\documentclass[11pt]{article}
\usepackage{stackengine}
\stackMath

\usepackage{pifont}

\usepackage{dsfont}
\usepackage{upgreek}

\usepackage{amsmath,amssymb,graphicx}
\usepackage[linktoc=all,
        hidelinks,
        colorlinks=true,
        pdfstartview=FitV,
        linkcolor=ColorLink,
        citecolor=ColorCite,
        urlcolor=ColorURL,
        hyperindex=true,
        hyperfigures=false%
    ]{hyperref}
\usepackage{cite}

\usepackage{caption}
\usepackage{slashed}
\usepackage[makeroom]{cancel}

\DeclareFontFamily{U}{matha}{\hyphenchar\font45}
\DeclareFontShape{U}{matha}{m}{n}{
      <5> <6> <7> <8> <9> <10> gen * matha
      <10.95> matha10 <12> <14.4> <17.28> <20.74> <24.88> matha12
      }{}
\DeclareSymbolFont{matha}{U}{matha}{m}{n}

\DeclareMathSymbol{\oleft}{2}{matha}{"68}
\DeclareMathSymbol{\oright}{2}{matha}{"69}

\DeclareFontFamily{U}{mathb}{\hyphenchar\font45}
\DeclareFontShape{U}{mathb}{m}{n}{
<-6> mathb5 <6-7> mathb6 <7-8> mathb7
<8-9> mathb8 <9-10> mathb9
<10-12> mathb10 <12-> mathb12
}{}
\DeclareSymbolFont{mathb}{U}{mathb}{m}{n}
\DeclareMathSymbol{\llcurly}{\mathrel}{mathb}{"CE}
\DeclareMathSymbol{\ggcurly}{\mathrel}{mathb}{"CF}

\usepackage{multicol,longtable}
\usepackage[dvipsnames]{xcolor}

\usepackage[T1]{fontenc}
\usepackage{hyphenat}
\usepackage{amsfonts}
\usepackage{mathrsfs}
\usepackage{mathdots}
\definecolor{darkred}{rgb}{0.65,0.15,0}
\definecolor{newgreen}{rgb}{0.2,0.62,0.14}
\definecolor{darkgreen}{rgb}{0.42, 0.46, 0.14}
\definecolor{darkcerulean}{rgb}{0.03, 0.27, 0.49}
\definecolor{oucrimsonred}{rgb}{0.45, 0.0, 0.0}
\hypersetup{pdfborder={0 0 0},colorlinks=true,urlcolor=darkcerulean,citecolor=darkcerulean,linkcolor=oucrimsonred,linktocpage=true}
\usepackage{graphicx}

\usepackage{array}
\newcolumntype{P}[1]{>{\centering\arraybackslash}p{#1}}
\usepackage{lscape}
\usepackage{multirow}

\usepackage{empheq}

\usepackage[normalem]{ulem}

\newcommand{\be}{\begin{equation}}
\newcommand{\ee}{\end{equation}}
\newcommand{\bea}{\setlength\arraycolsep{2pt} \begin{eqnarray}}
\newcommand{\eea}{\end{eqnarray}}

\newcommand{\ord}[1]{{\scriptscriptstyle (#1)}}

\newcommand{\seco}{\hspace{0.2mm};\hspace{0.2mm}}
\newcommand{\vardbtilde}[1]{\tilde{\raisebox{0pt}[0.92\height]{$\tilde{#1}$}}}

\newcommand{\ints}{\mathds{Z}}

\newcommand{\cE}{\mathcal{E}}
\newcommand{\cT}{\mathcal{T}}

\newcommand{\cL}{\mathcal{L}}
\newcommand{\cB}{\mathcal{B}}

\newcommand{\lb}{\left[}
\newcommand{\rb}{\right]}

\newcommand{\mf}[1]{{\mathfrak{#1}}}

\newcommand{\ta}{{\tilde\alpha}}
\newcommand{\tb}{{\tilde\beta}}

\newcommand{\tU}{\tilde{U}}

\newcommand{\wa}{{\widehat\alpha}}

\makeatletter

\@addtoreset{equation}{section}
\makeatother

\newcommand{\Tr}{\mathrm{Tr}}

\newcommand{\rd}{\mathrm{d}}
\newcommand{\rx}{\mathrm{x}}

\newcommand{\rY}{\mathrm{Y}}

\DeclareRobustCommand{\sscdots}{%
  \vbox{%
    \baselineskip=0.125\normalbaselineskip
    \hbox{.\hspace{-0.25mm}.\hspace{-0.25mm}.}%
    \kern-0.2\baselineskip
  }%
}

\usepackage[smalltableaux,centertableaux,centerboxes]{ytableau}
\ytableausetup{mathmode,boxsize=2.3em,centerboxes}

\makeatletter
\newcommand{\tvdots}{\raisebox{0.20\boxdim@YT}{\ensuremath{\vdots}}}
\newcommand{\tddots}{\raisebox{0.20\boxdim@YT}{\ensuremath{\ddots}}}
\makeatother

\begin{document}

\begin{flushright} CPHT-RR003.022026 \end{flushright}
\vspace{8mm}

\begin{center}

{\LARGE \bf \sc
Higher dualities in $E_{11}$ exceptional field theory
}\\[5mm]

\vspace{6mm}

\normalsize
{\large  Guillaume~Bossard${}^{1}$, Nicolas~Boulanger${}^{2}$, and Josh~O'Connor${}^{2\dagger}$}

\vspace{8mm}

\href{mailto:guillaume.bossard@polytechnique.edu}{\texttt{guillaume.bossard@polytechnique.edu}}
\;\;
\href{mailto:nicolas.boulanger@umons.ac.be}{\texttt{nicolas.boulanger@umons.ac.be}}\\
\href{mailto:josh.o'connor@umons.ac.be}{\texttt{josh.o'connor@umons.ac.be}}

\vspace{8mm}

${}^1${\it Centre de Physique Th\'eorique, CNRS,  Institut Polytechnique de Paris\\
91128 Palaiseau cedex, France}
\vskip 1 em
${}^2${\it Physique de l’Univers, Champs et Gravitation, Universit\'e de Mons -- UMONS\\
Place du Parc 20, 7000 Mons, Belgium}
\vspace{20mm}

\hrule

\vspace{5mm}

\begin{tabular}{p{14cm}}
It was proposed that there exists an $E_{11}$-invariant formulation of eleven-dimensional supergravity in which the propagating fields of the theory are realised through an infinite tower of higher duals.
We show how to derive the propagating equations of the higher dual fields within $E_{11}$ exceptional field theory at the linearised level.
Starting from the pseudo-Lagrangian, we construct parent actions for all higher gradient dual fields that are associated with the three-form, the six-form, and the dual graviton.
We show that the resulting Euler--Lagrange equations constrain the St\"{u}ckelberg fields to be pure curls, ensuring that the higher duals propagate the same physical degrees of freedom as the original supergravity fields.
The additional St\"{u}ckelberg fields, which are not predicted by the tensor hierarchy algebra, are shown to play a specific role as sources for the Labastida tensors of the higher dual fields.
\end{tabular}

\vspace{6mm}

\hrule

\end{center}

\vfill
\noindent{\footnotesize $\dagger$ FRIA grantee of the Fund for Scientific Research – FNRS, Belgium.}
\vspace{-10mm}
\null

\thispagestyle{empty}

\setcounter{page}{2}
\pagenumbering{arabic}

\newpage

\hrule

\setcounter{tocdepth}{2}
\tableofcontents

\null\hrule

\newpage

\section{Introduction}
\label{sec:introduction}

Eleven-dimensional supergravity \cite{Cremmer:1978km} compactified on a torus $T^d$ admits a hidden exceptional symmetry \cite{Cremmer:1978ds,Cremmer:1979up,Marcus:1983hb,Nicolai:1987kz}.
Observing that the rank of the algebra increases with each additional circle, thereby constructing the Dynkin diagram of the $E_d$ exceptional series, it was anticipated that exceptional Kac--Moody algebras should play a prominent role in supergravity \cite{Julia:1980gr,Julia:1982gx}.
These hidden symmetries are only unveiled after one dualises the appropriate degrees of freedom.
One finds the hidden Ehlers $\mathrm{SL}(2,\mathds{R})$ symmetry in pure gravity 
compactified down to three dimensions \cite{Ehlers:1957}.
This generalises to a wide range of gravitational theories 
\cite{Marcus:1983hb,Breitenlohner:1987dg,deWit:1992psp}.
In two dimensions, the group of symmetries enhances to the infinite-dimensional 
Geroch group \cite{Julia:1981wc,Nicolai:1987kz,Geroch:1970nt,Geroch:1972yt,Nicolai:1998gi}.
The system of equations is then integrable and the moduli space of solutions can be 
described as a coset  \cite{Breitenlohner:1986um,Katsimpouri:2012ky}.\footnote{The coset is identified as the moduli space of formal solutions that typically include conical singularities. See also \cite{Cesaro:2024ipq,Cesaro:2025msv} for generalisations beyond the usual toroidal compactification.}
However, the reduction on a circle down to one time-like dimension does not lead to a 
realisation of the hyperbolic Kac--Moody group \cite{Feingold:1983} as a symmetry of the 
moduli space of solutions \cite{Nicolai:1991kx}. 

The possibility of a hidden hyperbolic Kac--Moody $E_{10}$ symmetry group was revived with the 
observation that the Belinsky--Khalatnikov--Lifshitz dynamics near a spacelike 
singularity can be described as a free particle in the Weyl chamber of the 
$\mf{e}_{10}$ algebra \cite{Damour:2000hv,Damour:2001sa}.
This led to the proposal that the entire dynamics of eleven-dimensional supergravity 
could be recovered from the motion of a relativistic particle on the $E_{10}$ symmetric 
space \cite{Damour:2002cu,Damour:2002et}.

From a more conservative perspective, it was also observed that the degrees of freedom of maximal supergravity on $S^7$ 
are organised in terms of $E_7$ representations \cite{deWit:1981sst}.
This led to the idea that one may find the $E_d$ duality symmetry directly in eleven 
dimensions, but only the $\mathrm{SU}(8)$ invariance could be realised without 
extending the set of internal coordinates  \cite{deWit:1986mz,deWit:2000wu}.
First attempts to incorporate generalised coordinates were proposed in \cite{Duff:1990hn} 
from the membrane worldvolume perspective and in \cite{Siegel:1993xq} from the target 
space perspective in string theory.

It was proposed by West that the Kac--Moody algebra $\mathfrak{e}_{11}$ 
can be realised as a symmetry in eleven dimensions \cite{West:2001as} such that 
the dynamics would be determined by an $E_{11}$-invariant first-order duality 
equation for fields valued in the (relevant) $E_{11}$ symmetric space \cite{West:2011mm},\footnote{See \cite{marquisgroups,Keurentjes:2004bv,Bossard:2021ebg} for a precise definition of the symmetric space.} depending on infinitely many coordinates in the irreducible highest weight module $R(\Lambda_1)$ \cite{West:2003fc}. 
The evidence comes from the decomposition of $\mathfrak{e}_{11}$ with respect to its 
$\mathfrak{gl}(11)$ subalgebra -- that one calls level decomposition -- in 
which the $\mathfrak{e}_{11}$-invariant duality equation 
includes by assumption the duality between the three-form and the six-form 
of eleven-dimensional supergravity.
Besides the bosonic fields of eleven-dimensional supergravity, 
i.e.~the usual metric and three-form field, as well as the dual six-form, West's non-linear 
realisation requires an infinite set of dual potentials at higher levels 
\cite{Riccioni:2006az,Boulanger:2012df,West:2014qoa,Boulanger:2015mka}.
Using the first components of the fields and their generalised derivatives, 
it was shown that the non-linear equation for the three-form gauge field transforms 
under $E_{11}$ into the Einstein equation \cite{Tumanov:2015yjd,Tumanov:2016abm,West:2016xro,Tumanov:2017whf}, 
provided that one makes appropriate assumptions and that the fields eventually depend only on the eleven geometric coordinates.
This construction generalises to a wide set of theories 
\cite{Kleinschmidt:2003mf,Englert:2003py,Englert:2003zs,Englert:2004ph,Riccioni:2007hm}, 
including pure Einstein gravity in four dimensions 
\cite{Glennon:2020qpt,Boulanger:2022arw}.

A different line of research was pursued through the study of supersymmetric 
compactifications in string theory.
The concept of double geometry \cite{Hitchin:2003cxu} led to the development of double 
field theory with manifest $\mathrm{O}(d,d)$ invariance 
\cite{Hull:2007zu,Hull:2009mi,Hull:2009zb,Hohm:2013bwa}.
This introduced the very important concept of section constraint, imposing that the 
fields cannot depend locally on both the geometric coordinates and the T-dual 
string-winding coordinates.
This reflects that the radius $R$ and its T-dual $\frac{\alpha'}{R}$ cannot both 
be large at the same time, so one must decide which of the two is large before 
assuming that there is a local description of the effective field theory.
This construction was further generalised to exceptional $E_d$ groups with the 
introduction of exceptional geometry 
\cite{PiresPacheco:2008qik,Grana:2009im,Coimbra:2011ky}, that describes 
compactifications of eleven-dimensional supergravity and type II supergravities 
in terms of $K(E_d)$ structures, while making the exceptional invariance of the 
reduced gauged supergravity manifest \cite{Aldazabal:2013mya}.
A formulation including the membrane-winding coordinates appeared in  
\cite{Berman:2010is}, which required the introduction of an $E_d$-covariant generalisation of the section constraint so 
that the fields only depend locally on geometrical coordinates 
\cite{Coimbra:2011ky,Berman:2012vc}.
There are two inequivalent maximal solutions to the section constraint, one leading to 
eleven-dimensional supergravity, the other to type IIB supergravity in ten dimensions.
As a consequence, any choice of section, i.e.~the set of geometric coordinates the 
fields depend on, breaks the $E_d$ symmetry of the potential terms
to a $\mathrm{GL}(d)$ or $\mathrm{GL}(d{-}1)$ subgroup.
A complete formulation of eleven-dimensional supergravity with manifest $E_6$ 
invariance was introduced in \cite{Hohm:2013pua}, with internal local coordinates 
in the ${\bf 27}$ irreducible representation.
Exceptional field theories with $E_d$ invariance for $d=6,7,8$ were worked 
out in \cite{Hohm:2013vpa,Hohm:2013uia,Hohm:2014fxa} -- see also the reviews 
\cite{Berman:2020tqn,Samtleben:2025fta}.
Affine Kac--Moody exceptional diffeomorphisms were then introduced in 
\cite{Bossard:2017aae}, leading eventually to the formulation of $E_9$ exceptional 
field theory \cite{Bossard:2018utw,Bossard:2021jix}.

Revisiting West's conjecture from the perspective of exceptional field theory, it was 
shown that the $E_{11}$ section constraint introduced in \cite{West:2012qm} must be 
imposed, and the duality equation conjectured in \cite{West:2001as} should be defined 
for a field strength in the embedding tensor representation of $E_{11}$ 
\cite{Bossard:2017wxl}.
The embedding tensor representation of $E_{11}$ is an integrable module that is 
neither highest nor lowest weight.
It is only defined as a component of an infinite-dimensional tensor hierarchy 
superalgebra that includes and extends $\mathfrak{e}_{11}$ \cite{Bossard:2017wxl}.
The tensor hierarchy algebra was introduced in \cite{Palmkvist:2013vya} as a 
modification of the Borcherds superalgebra \cite{Henry-Labordere:2002xau}, such that 
its grade $-1$ component is the embedding tensor representation.
Defining a non-linear field strength in the embedding tensor representation, a 
duality equation invariant under generalised diffeomorphisms was introduced in \cite{Bossard:2017wxl}.
The major advantage of using an $E_{11}$ representation to encode the field strengths is that one can compute the $\mathrm{GL}(11)$-covariant components of the duality equation in the level decomposition systematically.

The field strength includes St\"{u}ckelberg fields that make most 
duality equation components tautological.
In order to have a complete description of the dynamics, the duality equation must therefore be supplemented by the non-linear Einstein field equation.
A pseudo-Lagrangian that transforms as a density under $E_{11}$ generalised diffeormorphisms was 
introduced in \cite{Bossard:2021ebg} from which the necessary Einstein equation follows as 
an Euler--Lagrange equation.\footnote{By the definition of a pseudo-Lagrangian, its Euler--Lagrange equations must be completed by a duality equation to obtain all the 
field equations that define a classical theory.}
This pseudo-Lagrangian is compatible with the duality equation in the sense that almost all of the Euler--Lagrange equations are the duality equations themselves or integrability conditions thereof.
Together, the pseudo-Lagrangian and the duality equation determine the entire dynamics in a $E_{11}$-covariant manner.
In order to obtain a true Lagrangian from which all the equations of $E_{11}$ exceptional field 
theory follow, one must add to the pseudo-Lagrangian an appropriate term that is quadratic in the 
duality equation components \cite{Bossard:2021ebg}.

$E_{11}$ exceptional field theory was shown to reproduce eleven-dimensional 
supergravity and $E_8$ exceptional field theory for the corresponding choices of 
section \cite{Bossard:2021ebg}.
One of the most striking features of this formulation of supergravity is 
that it not only describes the usual bosonic degrees of freedom but also 
infinitely many dual fields at higher levels, without introducing new
degrees of freedom.
In this paper we shall describe how it determines the dynamics of all propagating 
dual fields as conjectured in \cite{Riccioni:2006az}.

\vskip 5mm

At this point, it is worth pausing for a brief historical detour to recall how 
the notion of duality first developed in field theory.
The notion of electric-magnetic duality is very old, dating back to Heaviside 
\cite{heaviside1892xi}, and later beautifully extended by Dirac with his proposal for 
the existence of magnetic monopoles \cite{dirac1931monopole}, the existence of which 
would explain the quantisation of electric charge.
On the mathematical side, electric-magnetic duality involves the Hodge dual of 
Faraday's two-form field strength.
Staying at the level of field equations, i.e.~on-shell, electric-magnetic duality 
exchanges the field strength with its Hodge dual.
In other words, the vacuum Maxwell equations and the Bianchi identity for the Faraday 
tensor are exchanged under Hodge dualisation of the field strength. 
On-shell duality was further extended to higher-spin fields by Hull in 
\cite{Hull:2001iu}, then in \cite{Bekaert:2002dt} and \cite{deMedeiros:2002qpr}; see 
e.g.~\cite{Danehkar:2018yjp} for a review.

The free theory for a Maxwell $p$\,-form potential in dimension $D$ can equivalently be 
described in terms of a Maxwell $(D{-}p{-}2)$-form field in the same dimension, and the 
duality between their corresponding actions can be obtained from a parent action that 
contains both fields, as was explained in the introduction of \cite{Cremmer:1979up}.
This duality works at the action level, i.e.~off-shell.
That there might be infinitely many ways to represent the same dynamics in terms of gauge 
fields with extra sets of $D-2$ antisymmetric indices in addition to the indices carried 
by the original gauge field was apparently first observed by Siegel -- see ref.~16 
of \cite{Hull:2001iu}.

It took some time \cite{Boulanger:2012df} before the parent action techniques could be 
adapted to account for a spacetime covariant description of off-shell duality implying a 
potentially infinite tower of higher dual gauge fields.
As explained in \cite{Boulanger:2012df,Boulanger:2012mq}, the gauge potentials foreseen 
by Siegel result from the Hodge dualisation of a mere gradient acting on the original 
gauge field and its descendants.
For this reason, and for more clarity in the terminology, when the original gauge field 
is dualised on an empty set of indices along the lines of 
\cite{Boulanger:2012df,Boulanger:2012mq}, we shall refer to the resulting field as a 
\emph{higher dual} or \emph{gradient dual} field.
For example, the first higher dual counterpart to the Maxwell vector field is given, in 
$D$ dimensions, by an irreducible mixed-symmetry $\mathrm{GL}(D,\mathds{R})$ gauge field whose 
associated Young diagram is of type $[D-2,1]$ with two columns:~the first of height 
$D-2\,$, the second of height one.
In the case of the dual graviton in eleven space-time dimensions, it was shown in 
\cite{Boulanger:2012df} that there is an infinite tower of higher duals with Young 
symmetry types $[9,9,\ldots,9,8,1]$, therefore giving support to the claims made in \cite{Riccioni:2006az}.
Off-shell, higher gradient duality was further studied in 
\cite{Boulanger:2015mka,Bergshoeff:2016ncb,Boulanger:2020yib,Boulanger:2022arw,Boulanger:2024lwk}.

A non-linear action for gravity and dual gravity with an extra St\"uckelberg field 
was worked out in \cite{Boulanger:2008nd}.
The non-linearities however only concern the graviton, with the usual diffeomorphism 
algebra as the only non-abelian contribution to the full gauge algebra of the theory.
It was shortly thereafter confirmed in \cite{Bergshoeff:2009zq} that such extra fields 
beyond the set of fields in $\mathfrak{e}_{11}$ cannot be eliminated from the dual gravity sector.
A non-linear parent action for supergravity including the dual graviton was derived in $E_{11}$ 
exceptional field theory \cite{Bossard:2021ebg}, which reproduces the parent action of  
\cite{Boulanger:2008nd,Bergshoeff:2009zq} in eleven dimensions.
Most relevant to the present paper, $E_{11}$ exceptional field theory was also shown to provide a 
similar non-linear action for the first higher dual field in the three-form sector, namely the 
potential with irreducible Young symmetry $[9,3]$, and it was anticipated that parent 
actions for all the higher dual fields in the theory could be obtained \cite{Bossard:2021ebg}.
All these Lagrangians would include extra St\"uckelberg fields associated with the higher dual 
fields, such that the only component of the gauge algebra that is truly non-abelian is the algebra 
of spacetime diffeomorphisms, in line with \cite{Boulanger:2008nd,Bergshoeff:2009zq} and the no-go results of \cite{Bekaert:2002uh,Bekaert:2004dz}.

\vskip 5mm

In this paper we work out, using $E_{11}$ exceptional field theory, an infinite set of parent actions that describe all higher dual supergravity fields in eleven dimensions.
We work in the linearised approximation around a flat background.
While integrating out all the St\"{u}ckelberg fields reproduces eleven-dimensional supergravity, retaining a finite number allows one to construct parent actions whose Euler--Lagrange equations incorporate the corresponding duality equations.

We consider this an important result, 
i.e.~the derivation of parent action for all higher dual fields, that can straightforwardly be generalised to any number of spacetime dimensions, and this can be reverse-engineered to build parent actions for more general theories of gravity coupled to form fields.
Since the parent actions include the St\"{u}ckelberg fields, they can also in principle be promoted to full non-linear actions.

Writing the $E_{11}$ pseudo-Lagrangian would in principle have required the explicit form of the $E_{11}$-invariant tensors in a level decomposition, which would not have been feasible at arbitrarily high levels.
Remarkably, as we show in the present paper, the structure of the pseudo-Lagrangian, namely a sum of a topological 
term governing Chern--Simons-type couplings and a kinetic term, imposes consistency 
conditions that uniquely fix all the tensor contractions, without requiring explicit 
knowledge of their $E_{11}$ transformation properties.
Note that this requires the assumed existence and properties of the invariant bilinear form $\eta_{IJ}$ on the field strength representation.

We also prove that suitable integrability conditions force all St\"{u}ckelberg fields 
to be pure total derivatives in the linearised approximation, ensuring that the 
resulting duality equations correctly capture all the higher gradient dualities.
A key ingredient of the construction is an induction involving the Maxwell-like 
tensors (as coined in \cite{Campoleoni:2012th}) of the higher dual fields, 
which are second-order in derivatives and 
transform in the same representation as the fields themselves.
Specifically, the integrability condition for the St\"{u}ckelberg fields at 
level $\ell+3$ is equal to the Maxwell-like tensor at level $\ell$ which, in turn, 
can be expressed as a (non-local) double curl of fields at lower levels using a 
sourced Labastida equation \cite{Labastida:1986ft,Labastida:1987kw} 
for higher dual fields.

The presence of a source term in the Labastida equation reflects the fact that the equations of motion of higher dual fields cannot arise as Euler--Lagrange equations of an action that depends only on the dual field itself, in contrast to the case of conventional dual fields:~the action must necessarily involve all higher dual fields at lower levels in the same sector.

\paragraph{Outline of the paper.}

The plan for this paper is as follows.
We begin with a short summary of the notation and conventions that we use in this paper.
In Section~\ref{sec:linearised_duals} we review the notion of dualisation for linearised mixed-symmetry gauge fields, with an emphasis on higher gradient dualities that are baked in the structure of $E_{11}$ exceptional field theory.
We provide a lightning review of the $E_{11}$ theory in Section~\ref{sec:review_E11}, covering the tensor hierarchy algebra that underlies the theory and the core ingredients that lead to non-linear dynamics.
Lagrangians for the linearised gravity and three-form sectors featuring towers of higher dual fields are worked out in Section~\ref{sec:Lagrangian}, from which the second-order equations of motion and first-order duality equations are 
obtained as Euler--Lagrange equations.

In Section~\ref{sec:GradientDuals} we prove using the covariant equations of motion that all the constrained fields appearing in the propagating field equations, including those that are not found inside the tensor hierarchy algebra, are first derivatives of extra fields.
We also show that the covariant equations of motion (more precisely, the components of the equations that do not follow from nilpotency of the tensor hierarchy differential) can be expressed as an $E_{11}$-covariant Ricci-flat equation, where the Ricci tensor is defined in terms of the Maxwell tensor for mixed-symmetry fields previously considered in \cite{Skvortsov:2007kz,Campoleoni:2012th}.
Near the end, in Section~\ref{sec:E11dualityrels}, we work out in more detail the differential gauge transformations of the theory, and we show that the first-order duality equations can be obtained directly from the usual covariant form of the duality equations.
A summary of our results and an outlook for future research is given in Section~\ref{sec:conclusion}.

\section{Notation and conventions}

Throughout, we write $\approx$ to denote an on-shell equality, or more precisely an Euler--Lagrange equation that follows from some Lagrangian.
We work in the `mostly plus' convention for the flat metric $\eta=\mathrm{diag}(-1,+1,\dots,+1)$\,.

Mixed-symmetry fields have components that are denoted in a specific 
way:~commas are used to separate columns of antisymmetric indices 
within the same $\rm{GL}(D,\mathds{R})$-irreducible component of a 
(possibly reducible) 
field, while semicolons are used to separate indices associated 
with different $\rm{GL}(D,\mathds{R})$-irreducible components.
This is the antisymmetric convention for fields that have mixed-symmetry Young 
diagrams, as in \cite{Fulton:2004uyc}.
For the convenience of the reader, we refer 
to Appendix~\ref{Appendix:Schur-Weyl} 
for a brief review of Schur--Weyl duality and mixed-symmetry Young diagrams.
Indices in the same row of a Young tableau are symmetrised first, 
and the indices in each column are then antisymmetrised.
We also use a shorthand for diagrams that have several columns of 
the same height.
If a diagram $\rY$ has a total of 
$s\coloneqq w_1+\dots+w_B$ columns, including $w_1$ columns of height $h_1$, 
$w_2$ columns of height $h_2<h_1$, and so on, then it can be written as 
\begin{equation}
    \rY = [\underbrace{h_1,\ldots,h_1}_{w_1}, \underbrace{h_2, \ldots, h_2}_{w_2}, \ldots, 
    \underbrace{h_B, \ldots, h_B}_{w_B}]=[h_1^{w_1},\dots,h_B^{w_B}] \,,\quad h_1>h_2>\ldots h_B>0 \,.
\end{equation}
This makes 
explicit the decomposition of $\rY$ into $B$ vertical \emph{blocks}, 
where block $I$ $(1\leqslant I\leqslant B)$ is a rectangular Young subdiagram 
of height $h_I$ and width $w_I\,$. 
A totally symmetric field with $s$ indices thus has partition $[1^s]$ while a 
$p$-form field has partition $[p]$.
We denote by $s(\rY)$ the number of columns in $\rY$.
The diagram $\overline{\rY}$ is obtained from $\rY$ by adding one box 
at the bottom of every column, or stated differently, 
by duplicating the first row of $\rY$.

Consider $\rY=[h_1,\dots,h_s]$ and $\tilde{\rY}=
[\tilde{h}_1,\dots,\tilde{h}_{\tilde{s}}]$\,.
We say that $\rY$ is \emph{well-included} in $\tilde{\rY}$ if $\rY$ is obtained from 
$\tilde{\rY}$ by removing a single box located both at the end of a row and at 
the end of a column, or stated differently, by removing the 
lower-right box of a vertical block of $\rY$. There are thus $B$ inequivalent
well-included diagrams in $\rY$, if the latter possesses $B$ blocks. 
In particular $h_i=\tilde{h}_i-\delta_{i,k}$ for all $i$ and a fixed $k$ 
such that $\tilde{h}_{k}-1\geqslant\tilde{h}_{k+1}$ (with $\tilde{h}_{\tilde{s}+1}=0$ 
by convention).
We will denote the fact that $\rY$ is well included in $\tilde{\rY}$ by $\rY\prec\tilde{\rY}$.
For example, $[2,1]$ is well-included in $[3,1]$, $[2,2]$, and also $[2,1,1]$, but not in $[3,2]$.

We will denote the tensor product representation of two irreducible representations of Young diagrams $\rY$ and $\tilde{\rY}$ by separating the two corresponding partitions by a semicolon:
\begin{equation}
    [h_1,\dots,h_n\,;\tilde{h}_1,\dots,\tilde{h}_m] \;\equiv\; [h_1,\dots,h_n] \otimes [\tilde{h}_1,\dots,\tilde{h}_m] \,.
\end{equation}
In the same way that commas are used to separate indices associated with different columns in the Young tableau of an irreducible representation, we use semicolons to separate indices for which there is no Young symmetry projection.

When it is cumbersome to write out all the indices of a field, we use a shorthand notation to indicate the tensor structure.
For instance, $A_3$ denotes a three-form field $A_{a_1a_2a_3}=A_{[a_1a_2a_3]}$ and $\phi_{1,1,1,1}$ denotes a totally symmetric rank-four field $\phi_{a,b,c,d}=\phi_{(a,b,c,d)}$\,.
Indices sitting inside the same antisymmetric block (i.e.~column) share the same letter, and such blocks can be expressed concisely as $B_{a[4],b[2]}=B_{a_1a_2a_3a_4,b_1b_2}$\,.
For fields with several blocks (columns) of indices of the same size, we often use an additional shorthand.
An irreducible field $C_{4^n}=C_{4,\dots,4}$ with $n$ blocks of four indices can be written as $C_{a^i[4]|_{i=1}^{n}}=C_{a^1[4],a^2[4],\dots,a^n[4]}$\,.

A reducible field does not satisfy these over-antisymmetrisation constraints between different irreducible components.
For example, the reducible field $\chi_{10\seco8,1}=\chi_{a[10];b[8],c}$ satisfies only one such constraint $\chi_{a_1\dots a_{10};[b_1\dots b_8,c]}=0$ in its second component.
There is no constraint between the first component of ten indices and the second component of nine indices.

For example, the reducible field $\Phi_{4,1\seco3,2,2}$ is in the following tensor product representation:
\ytableausetup{boxsize=1em}
\begin{equation}
    [4,1\,;3,2,2] \;\equiv\;\; [4,1] \otimes [3,2,2] \;\equiv\;\;
    \ytableaushort{\null\null,\null,\null,\null}
    \;\otimes\;
    \ytableaushort{\null\null\null,\null\null\null,\null}
\end{equation}
With indices written explicitly, this field is $\Phi_{a[4],b;c[3],d^1[2],d^2[2]}=\Phi_{a[4],b;c[3],d^i[2]_{i=1}^{2}}$\,.

Lastly, we sometimes use angled brackets to denote a projection onto a 
particular diagram.
The projection of a field $h_{a[8];b}$ transforming in the reducible 
representation $[8\,;1]=[8]\otimes[1]$ onto the $[8,1]$ is written as 
$h_{\langle a[8];b\rangle}$\,.
This notation suppresses its explicit expression $h_{a[8];b}-h_{[a[8];b]}$ 
where the totally antisymmetric part $[9]$ is subtracted.
Note that $\phi_{\langle\,\dots\rangle}$ does not indicate the diagram onto 
which the field $\phi$ is being projected, so this will be explained in the 
text.
We also sometimes make use of a shorthand $(\,\dots)_\rY$ to denote a 
projection of a tensor onto the symmetry type denoted by $\rY$.
Lastly, $\Tr_{i,j}$ denotes a single trace on indices associated with the 
$i^\mathrm{th}$ and $j^\mathrm{th}$ columns.
For example, $(\Tr_{2,3})^3E_{6,5,4}$ is shorthand for 
$E_{a[6],b[5],c[4]}(\eta^{bc})^3=E_{a[6],d[3]b[2],}{}^{d[3]}{}_c$\,.

\paragraph{Differential operators.}

Consider mixed-symmetry tensor fields whose corresponding Young diagrams have $s$ columns, and introduce the exterior derivative
\begin{equation}
   \rd_i = \rd_ix^\mu \frac{\partial}{\partial x^\mu}\,,\qquad 
    i\in\{1,\ldots,s\}\,,
\end{equation} 
acting on the $i^\text{th}$ column and thus adding one box to it; see for example Appendix~A of \cite{Bekaert:2006ix} and references therein.

Greek indices refer to a holonomic basis, whereas Latin 
indices refer to a non-holonomic basis.  
At this stage we regard the field (in the antisymmetric convention for Young 
tableaux) as reducible, so that the differential operators 
$\{\rd_i\}_{i=1,\ldots,s}$ act as the de Rham operator in each of their 
associated columns.
In particular, these operators do not necessarily send an irreducible 
tensor into another irreducible tensor. It is only the case for $\rd_1$. 
These differential operators satisfy $\rd_i^2=0$ and can be chosen to obey 
$\rd_i\rd_j=\rd_j\rd_i$ for all $i\neq j$ so that for the total differential 
$\rd\coloneqq\sum_{i=1}^s\rd_i$ one has $\rd^{s+1}=0$\,.
The operators $\rd_i\,$, $i=1, \ldots, s\,$, naturally act on the 
space of multiforms 
\begin{equation}
    \wedge_{[s]}(V)\,=\,
    \underbrace{\wedge(V^*) \odot \ldots \odot \wedge(V^*)}_{s~\text{factors}} \quad
    = \bigoplus_{(h_1,\ldots,h_s)\in \mathbb{N}^s}\wedge_{[s]}^{h_1, \ldots, h_s}(V^*) \,, \qquad
    V \cong \mathds{R}^D
\end{equation}
having a fixed number of $s$ columns. 
The symbol $\odot$ stands for the symmetric tensor product. 
For example, when $s=1$, a $p\,$-form at a point of $\mathds{R}^D$ 
belongs to the exterior algebra $\wedge(V^*)\,$.\footnote{Note 
that multiforms belong to the space dual to  
$\bigoplus'_{\boldsymbol{\ell}}A^{\boldsymbol{\ell}}(\mathds{R}^D)$ 
defined in Appendix~\ref{Appendix:Schur-Weyl} (see eq.~\eqref{multiform2}), 
where the prime indicates that the direct sum is restricted to all multi-indices $\boldsymbol{\ell}$ such that 
$\sum_{i=1}^D \ell_i = s\,$.} In coordinates 
$x^\mu$ with $\mu\in\{0,\dots,D-1\}$
on $\mathds{R}^D$, the multiform algebra is presented by the 
(anti)commutation relations 
\begin{equation}
    {\rm d}_ix^\mu \,{\rm d}_jx^\nu - (-1)^{\delta_{ij}}{\rm d}_jx^\nu \,
    {\rm d}_ix^\mu = 0\,,\qquad
    i,j\in\{1, \ldots, s\} \,,
\end{equation}
where the wedge and/or symmetric product symbols are omitted. 
For more details, we refer to Appendix~A of \cite{Bekaert:2006ix}.

The Hodge star operator acting on the $i^\text{th}$ column is labelled 
$\star_i$\,.
It is also convenient to define \cite{Bekaert:2002dt,Bekaert:2003az} 
the curvature operator $K\coloneqq\prod_{i=1}^s\rd_i$ that acts by taking a 
curl on every column of indices.
In other words, the curvature of a field $\phi_{h_1,\dots,h_s}$ with Young 
diagram $[h_1,\dots,h_s]$ is
\begin{equation}
\label{eq:curvature_K}
    K[\phi]_{h_1+1,\dots,h_s+1} = \rd_1 \cdots \rd_s \phi_{h_1,\dots,h_s} \,.
\end{equation}
If we instead want to consider $s-1$ curls, it is useful to define
\begin{equation}
\label{eq:almost_curvatures}
    D^{s-1}_{(i)}\coloneqq\rd_1\cdots\rd_{i-1}\rd_{i+1}\cdots\rd_s\,,\quad 
    i\in\{1, \ldots, s\} \,,
\end{equation}
where an exterior 
derivative will act on all but the $i^\text{th}$ column.
The superscript indicates the number of distinct individual curls.
These `almost-curvatures' can similarly be generalised, e.g.~$D^{s-2}_{(i,j)}$ takes a curl on all but the $i^\text{th}$ and $j^\text{th}$ columns, and so on.

\paragraph{\texorpdfstring{Generalised Poincar\'e lemma.}{Generalised Poincare lemma.}}

The main statement from \cite{Bekaert:2002dt}, 
that generalises the main results of 
\cite{Dubois-Violette:1999iqe,Dubois-Violette:2001wjr}  
(see also \cite{Olver1982differential,Olver1987Invariant}),
requires the definition of the differential operators 
$\{\rd^\ord{i}\}_{i=1,\dots,s}$ that amount to applying $\rd_i$ to any tensor $T_{[h_1,\dots,h_s]}$ and then applying the Young projector to obtain
\begin{equation}
    \rd^\ord{i} (T_{[h_1,\dots,h_i,\dots,h_s]})
    = (\rd_i T)_{[h_1,\dots,h_{i-1},h_i+1,h_{i+1},\dots,h_s]} \,, 
\end{equation}
in the irreducible representation associated with 
$[h_1,\dots,h_{i-1},h_i+1,h_{i+1},\dots,h_s]$, if it is a partition.
However, if $[h_1,\dots,h_{i-1},h_i+1,h_{i+1},\dots,h_s]$ is not a 
partition then we define $\rd^\ord{i}$ to vanish by convention.
Importantly, one has $\prod_{i=1}^s\rd_i\equiv\prod_{i=1}^s\rd^\ord{i}$.

On a contractible domain (e.g.~$\mathds{R}^D$), the generalised Poincar\'e 
lemma \cite{Bekaert:2002dt} defines a differential multicomplex associated 
with mixed-symmetry gauge fields having $s$ columns in their diagrams and a 
differential operator that is nilpotent of order $s+1$\,, and then establishes 
the triviality of various cohomological groups in this differential 
multicomplex.
We will fortunately not need all the cohomological groups defined in 
\cite{Bekaert:2002dt}, only a couple of them that we describe in a 
simplified way as follows.

Concretely, let $T_{_\rY}$ be an irreducible tensor with $s$ columns 
of indices, i.e.~$\rY=[h_1^{w_1}, \ldots, h_B^{w_B}]$ 
with $\sum_{I=1}^Bw_I = s$. 
If $T_{_\rY}$ is $\rd_i$-closed for all $i\in\{1,\ldots, s\}$, 
then locally there exists a tensor $A_{_{\underline{\rY}}}$ 
with $\underline{\rY}=[(h_1-1)^{w_1}, \ldots, (h_B-1)^{w_B}]$,
such that 
$T_{_\rY}=K[ A_{ _{\underline{\rY}}} ]\coloneqq\rd_1\dots\rd_s A_{_{\underline{\rY}}}\,$.
Conversely, if $T_{_\rY}=\rd_1\dots\rd_n A_{_{\underline{\rY}}}$ 
for some $A_{_{\underline{\rY}}}$, then automatically 
$\rd_i T_{_\rY}=0$ for all $i\in\{1,\ldots, n\}$.
In short, one has locally
\begin{equation}
    \rd_i T_{_\rY} = 0 \quad \forall~i \quad \Longleftrightarrow
    \quad T_{_\rY}=K[ A_{ _{\underline{\rY}}} ]
    =\rd_1\dots\rd_s A_{_{\underline{\rY}}}\,.
\end{equation}
For $p\,$-forms ($s=1$) this is the local equivalence 
$\rd \omega_{[p]}=0\Longleftrightarrow \omega_{[p]}=\rd \lambda_{[p-1]}$\,, 
while in linearised gravity (for $s=2$) one has 
$\rd_1R_{[2,2]}=0=\rd_2R_{[2,2]}\Longleftrightarrow R_{[2,2]}=\rd_1\rd_2A_{[1,1]}\,$.

The generalised Poincar\'e lemma also implies that, locally, a vanishing curvature forces the potential to be pure gauge, while a pure-gauge potential has vanishing curvature.
In equations, one has the following local equivalence:
\begin{equation}
    K[T_{_\rY}] \coloneqq \rd_1 \dots \rd_s\, T_{_\rY} = 0 \quad \Longleftrightarrow \quad 
    T_{_\rY} = \sum_{\rY_i\prec\rY} \big( \rd_i \lambda_{_{\rY_i}} \big)_\rY 
    \,,
    \label{puregauge}
\end{equation}
where we recall that the notation $\rY_i\prec\rY$ means that 
$\rY_i$ is well-included into $\rY$.  
We also recall that the operator $\rd_i$ is the exterior derivative acting on the 
$i^\text{th}$ column, the column from which one removed a cell from 
$\rY$ to get $\rY_i$.

Another useful case involves the almost-curvature operators $D^{s-1}_{(i)}$ as defined in \eqref{eq:almost_curvatures}.
From the definition $\rd = \sum_{i=1}^s \rd_i\,$ and 
the commutativity $\rd_i \,\rd_j = \rd_j \,\rd_i$ (that is trivially 
satisfied when $i=j$), one obtains the identity 
$\rd^{s-1}=(s-1)\sum_{i=1}^s D^{s-1}_{(i)}\,$.
Therefore, the $s$ cocycle conditions $D^{s-1}_{(i)}T_{_\rY}=0\,$ for all 
$i\in\{1, \ldots, s\}$ imply that $T_{_\rY}$ is $\rd^{s-1}$-closed.
The Poincaré lemma of \cite{Bekaert:2002dt} tells us that 
$\rd^{s-1}$-closure implies $\rd^2$-exactness, in the sense 
that 
\begin{equation}
\label{2-derivative_coboundary}
    D^{s-1}_{(i)}T_{_\rY}=0\quad\forall ~ i\in\{1,\ldots,s\}
    \quad \Longleftrightarrow \quad T_{_\rY}=\sum_{\rY_{ij}\llcurly\rY}\rd_i\rd_jU_{_{\rY_{ij}}}\,,
\end{equation}
where the notation $\rY_{ij}\llcurly\rY$ means
(i) that $\rY_{ij}$ is such that $|\rY_{ij}|+2 = |\rY|\,$, 
obtained from $\rY$ by removing two boxes from it, 
in the $i^\text{th}$ and $j^\text{th}$ columns, if the result still 
is a diagram, and (ii) that in defining $\rY_{ij}$ from $\rY$, one
excludes the diagrams obtained by removing two boxes (or cells) 
in the \emph{same} column of $\rY$ (supposing the result is still a diagram).
This second condition on the definition of the relation
$\rY_{ij}\llcurly\rY$ that we call \emph{double inclusion} is 
natural, since by the nilpotency of each of the $s$ operators 
$\rd_i\,$, $i\in\{1,\ldots,s\}\,$, one may as well restricts $i< j$
in the sum appearing in \eqref{2-derivative_coboundary}, therefore 
excluding the cases where $i=j$. 
When using the notation $\rY_{ij}\llcurly\rY$, we say that the diagram 
$\rY_{ij}$ is \emph{doubly well-included} in $\rY$.

Finally, the generalised Poincar\'e lemma also gives the local 
following equivalence
\begin{equation}
\label{cocycle_2}
    \rd_i \rd_j\, T_{_\rY} = 0\quad \forall~~i,j\in\{1,\ldots,s\} 
    \quad\Longleftrightarrow\quad 
    T_{_\rY} = \sum_{I=1}^B D^{s-1}_{(p_I)} C_{_{\rY_I}}\,,\quad 
    p_I \coloneqq  1+\sum_{J=1}^{I-1} w_J\,,
\end{equation}
where the index $I\in\{1,\ldots,B\}$ runs over all the vertical blocks of 
the Young diagram $\rY$, 
and where the Young diagram $\rY_I$ characterising the tensor 
$C_{_{\rY_I}}$ in \eqref{cocycle_2} is obtained from 
$\rY$ by cutting off the top row of $\rY$, then adding a single cell 
at the bottom of the first column of the $I^{\text{th}}$ block of the 
resulting Young diagram.

\section{Linearised dynamics of gradient dual fields}
\label{sec:linearised_duals}

\subsection{Electromagnetic duality vs.~gradient duality}

Suppose that we have an irreducible massless bosonic field $\phi$ with Young diagram $[h_1,\dots,h_s]$\,.
There are two well-known ways to dualise this field.
Electromagnetic duality singles out one of the $s$ columns of antisymmetric indices, 
the $i^\text{th}$ column let's say, and creates from $\phi$ a dual field $\tilde{\phi}$ 
for which the $i^\text{th}$ column in the diagram is replaced by one of height $D-2-h_i$\,.
For example, in $D$ space-time dimensions the Maxwell $p$-form $A_{a_1\dots a_p}$ is dual 
to a $(D-p-2)$-form $\tilde{A}_{a_1\dots a_{D-p-2}}$\,, and the Fierz--Pauli field 
$h_{ab}=h_{(ab)}$ is dual to the \emph{dual graviton}:~a mixed-symmetry gauge field 
$\tilde{h}_{D-3,1}\equiv\tilde{h}_{a_1\dots a_{D-3},b}$ that satisfies the 
over-antisymmetrisation constraint $\tilde{h}_{[a_1\dots a_{D-3},b]}= 0$\,.

This works by first taking a curl of $\phi$ on a chosen block of indices, Hodge dualising this block, and defining this as the curl of the dual potential.
For the $p$-form $A_p$ we have the $(p+1)$-form field strength $F_{p+1}=\rd A_p$\,, Hodge dualised into the dual field strength $\tilde{F}_{D-p-1}\approx\star F_{p+1}$\,, and this defines the dual field via $\tilde{F}_{D-p-1}=\rd\tilde{A}_{D-p-2}$\,.
As usual, the equations of motion and the Bianchi identities of the gauge potential and its dual are exchanged as
\begin{equation}
    \begin{rcases}
        \partial^aF_{ab_1\dots b_p} \approx 0\;\;\\
        \partial_{[a_1}F_{a_2\dots a_{p+2}]} = 0\;\;
    \end{rcases}
    \quad\Longleftrightarrow\quad
    \begin{cases}
        \;\;\partial_{[a_1}\tilde{F}_{a_2\dots a_{D-p}]} = 0\\
        \;\;\partial^a\tilde{F}_{ab_1\dots b_{D-p-2}} \approx 0
    \end{cases}
\end{equation}

The second method of dualising a field concerns the higher gradients of the field strengths.
In the case of the Maxwell field $A_a$ that was worked out explicitly in \cite{Boulanger:2015mka}, we know that the field strength $F_{ab}=2\,\partial_{[a}A_{b]}$ and all its gradients
\begin{equation}
\label{eq:Maxwellgradients}
    F^{\ord{n}}_{ab,c_1,\dots,c_n} \coloneqq\; \partial_{c_1} \dots \partial_{c_n} F_{ab} \;\sim\; 
    \ytableausetup{smalltableaux}\ytableaushort{\null\null\cdots\null,\null}
\end{equation}
are needed to describe this propagating field around a given point in space-time 
\cite{Boulanger:2015mka}.
Gradients at higher order appear further along the Taylor expansion of the original field $A_a$ 
around a point in space-time, and so they contribute to a description of $A_a$ at longer distances.
The first such gradient $F_{2,1}\coloneqq\partial_{\langle1}F_{2\rangle}=(\partial_1F_2)_{[2,1]}$ 
is used to  express the on-shell value of the curvature 
of the dual potential:
\begin{equation}
\label{eq:Maxwell_dual}
    K_{a_1\dots a_{D-1},b_1b_2}[\tilde{A}_{D-2,1}] \approx
    \varepsilon_{a_1\dots a_{D-1}}{}^c \partial_c F_{b_1b_2} \,,
\end{equation}
where the dual potential $\tilde{A}_{D-2,1}$ satisfies 
$\tilde{A}_{[a_1\dots a_{D-2},b]}=0$\,.
We have
\begin{equation}
    K_{a_1\dots a_{D-1},b_1b_2}[\tilde{A}_{D-2,1}] \coloneqq 2(D-1)\,\partial_{a_1} \partial_{b_1} \tilde{A}_{a_2\dots a_{D-1},b_2} \,.
\end{equation}
Recall that indices with the same letter are implicitly antisymmetrised.

The equation of motion for the original field is the usual Maxwell equation 
$\partial^aF_{ab}\approx0$ and the Bianchi identity is $\partial_{a_1}F_{a_2a_3}\approx0$\,.
Together, they are equivalent to the Lorentz irreducibility properties of 
$F_{2,1}=F_{a[2],b}\coloneqq\partial_{\langle b}F_{a_1a_2\rangle}$\,.
Under \eqref{eq:Maxwell_dual}, these two equations are exchanged for a pair of constraints on the 
dual curvature $K_{D-1,2}=K_{a[D-1],b[2]}$ as
\begin{equation}
    \begin{rcases}
        \Tr(F_{2,1}) \approx 0\;\;\\
        F_{2,1}\text{ is $\mathrm{GL}(D)$ irreducible}\;\;
    \end{rcases}
    \quad\Longleftrightarrow\quad
    \begin{cases}
        \;\;K_{D-1,2}\text{ is $\mathrm{GL}(D)$ irreducible}\\
        \;\;\Tr^2(K_{D-1,2}) \approx 0
    \end{cases}
\end{equation}
where the $\mathrm{GL}(D)$ irreducibility of $F_{2,1}$ and $K_{D-1,2}$ are 
over-antisymmetrisation constraints
\begin{align}
    F_{[a_1a_2,b]} &= 0 \,,&
    K_{[a_1\dots a_{D-1},b_1]b_2} &= 0 \,.
\end{align}

Gradients are equivalent to curls on empty blocks of indices, so gradient duality 
generalises electromagnetic duality.
The on-shell relationship between electromagnetic dual fields and these gradient dual 
fields was studied in \cite{Chatzistavrakidis:2019bxo}. 
Actions for dual gravity were constructed in \cite{West:2001as,West:2002jj,Boulanger:2003vs} 
with a non-linear parent action in \cite{Boulanger:2008nd} featuring a St\"uckelberg-type extra 
field.
Actions for all higher dual gravity fields were worked out in 
\cite{Boulanger:2012df,Boulanger:2020yib,Boulanger:2022arw,Boulanger:2024lwk} where it was 
observed that extra fields in addition to the new dual fields are needed to ensure that the 
propagating degrees of freedom are those of the original graviton.
It is also possible to dualise the graviton on both its indices, leading to what can 
be called the double-dual graviton field whose on-shell and off-shell formulations 
were studied, respectively, in 
\cite{Henneaux:2019zod} and \cite{Boulanger:2012df,Boulanger:2012mq,Boulanger:2020yib}.

Hodge dualising $F^{\ord{n}}_{ab,c_1,\dots,c_n}$ on all gradient indices leads to an 
infinite tower of higher gradient dual fields $A_{(D-2)^n,1}=A_{D-2,\dots,D-2,1}$ each of 
which has $n$ columns of height $D-2$ in its Young diagrams.\footnote{Gradient 
dualities 
are called higher dualities (the original terminology), 
or exotic dualities, in the literature.}
Their curvatures $K_{D-1,\dots,D-1,2}$ 
are constrained on-shell to satisfy the equation
\begin{equation}
    K_{a^1[D-1],\dots,a^n[D-1],b[2]}[A_{(D-2)^n,1}] \approx \varepsilon_{a^1[D-1]}
    {}^{c_1} \dots \varepsilon_{a^n[D-1]}{}^{c_n} \partial_{c_1} \dots \partial_{c_n} 
    F_{b[2]} \,.
\end{equation}
As a result, adjacent curvatures are on-shell 
related to each other by
\begin{equation}
    K_{a[D-1],b^1[D-1],\dots,b^n[D-1],c[2]} = 
    \approx
    \varepsilon_{a[D-1]}{}^d 
    \,\partial_d K_{b^1[D-1],\dots,b^n[D-1],c[2]} \,.
\end{equation}
Similarly, just as the Maxwell equation $\partial^aF_{ab}\approx0$ and 
Bianchi identity $\partial_{[a}F_{bc]}=0$ are swapped under the usual duality 
relation, the equation of motion for the field $A_{(D-2)^{n+1},1}$ is 
equivalent to the integrability condition for the gradient of $K_{(D-1)^n,2}$ 
and the Bianchi identities for $A_{(D-2)^{n+1},1}$ are divergence-free 
constraints for $K_{(D-1)^n,2}$ that themselves follow from the equations of 
motion for the field $A_{(D-2)^{n},1}$\,.

Gauge transformations for the higher duals $A_{D-2,\dots,D-2,1}$ were 
worked out in \cite{Boulanger:2015mka} and we will present them again here.
The curvatures are defined as the curl on all blocks of indices
\begin{equation}
    K_{a^1[D-1],\dots,a^n[D-1],b[2]} = 2(D-1)^n \partial_{a^1} \dots 
    \partial_{a^n} \partial_{b} A_{a^1[D-2],\dots,a^n,[D-2],b} \,,
\end{equation}
where repeated indices are taken to be implicitly antisymmetrised.
Using the generalised Poincar\'e lemma \cite{Bekaert:2002dt} one finds 
that the gauge transformation that leaves $K_{(D-1)^n,2}$ invariant is
\begin{multline}
\label{eq:gauge_transfo_Maxwell}
    \delta_\lambda A_{a^1[D-2],\dots,a^n[D-2],b} 
    = n (D-2) \partial_{\langle a^n} \lambda_{a^1[D-2],
    \dots,a^{n-1}[D-2],a^n[D-3],b\rangle} \\
    + \partial_{\langle b} \lambda_{a^1[D-2],\dots,a^n[D-2]\rangle} \,,
\end{multline}
with two parameters $\lambda_{(D-2)^{n-1},D-3,1}$ and $\lambda_{(D-2)^n}$\,.
Recall that angled brackets in the subscript denotes the projection onto the 
$[(D-2)^n,1]=[D-2,\dots,D-2,1]$ diagram in this case, i.e.~the symmetry type of 
$A_{(D-2)^n,1}$ on the left-hand side.

The gauge parameters have Young diagrams that are \emph{well-included} in the 
diagram of the gauge field; see Section~\ref{sec:introduction} for the 
precise definition.
For example, the parameters of the field $A_{D-2,D-2,1}$ have diagrams 
$[D-2,D-3,1]$ and $[D-2,D-2]$\,.

In general, the gauge transformation of a field $\phi_\rY$ with Young 
diagram $\rY$ is given by
\begin{equation}
    \delta \phi_\rY = \sum_{\rY_i\prec\rY} \big( \rd_i \lambda_{\rY_i} \big)_{\rY} \,,
\end{equation}
where we are summing over each gauge parameter $\lambda_{\rY_i}$ 
whose diagram $\rY_i$ is well-included in $\rY$ with one empty box in the 
$i^\text{th}$ column.
A curl is taken on each column with the missing box and this is then projected 
back onto the $\rY$ diagram.
For example, the gauge transformation \eqref{eq:gauge_transfo_Maxwell} 
can be written as
\begin{equation}
    \delta A_{(D-2)^n,1} = \big( \rd_n \lambda_{(D-2)^{n-1},D-3,1} 
    + \rd_{n+1} \lambda_{(D-2)^n} \big)_{[(D-2)^n,1]} \,.
\end{equation}

The curvature $K[\phi]_{\overline{\rY}}$ is obtained from $\phi_{\rY}$ by 
applying an exterior derivative to every column of indices, the diagram 
$\overline{\rY}$ being thus obtained from $\rY$ by adding 
a row of length $s(\rY)$ on top of $\rY$, since the partial derivatives 
commute.
As a result, by the generalised Poincar\'e lemma, the curvature is a gauge 
invariant quantity:
\begin{equation}
    \delta K[\phi]_{\overline{\rY}} = K[\delta \phi]_{\overline{\rY}} = 0 \,.
\end{equation}

\subsection{Linearised gradient dual fields and their equations}

Maximal supergravity in eleven dimensions features two bosonic fields: the graviton $h_{1,1}=h_{(ab)}$ and the three-form field $A_{3}=A_{a_1a_2a_3}$\,.
Their magnetic duals are, respectively, the dual graviton $h_{8,1}=h_{a_1\dots a_8,b}$ and the six-form $A_{6}=A_{a_1\dots a_6}$\,.
These fields transform with $\mathrm{GL}(11)$ diagrams
\begin{align}
    &h_{1,1}\;\sim\;
    \ytableaushort{
    \null\null
    }\hspace{-3mm}&
    &A_{3}\;\sim\;
    \ytableaushort{
    \null,
    \null,
    \null
    }&
    &A_{6}\;\sim\;
    \ytableaushort{
    \null,
    \null,
    \null,
    \null,
    \null,
    \null
    }&
    &h_{8,1}\;\sim\;
    \ytableaushort{
    \null\null,
    \null,
    \null,
    \null,
    \null,
    \null,
    \null,
    \null
    }
\end{align}
Gradient dualities lead to infinite towers of higher dual three-forms, six-forms, and gravitons:
\begin{align}
    &A_{9,\dots,9,3}\;\sim\;
    \ytableaushort{
    \null\sscdots\null\null,
    \null\sscdots\null\null,
    \null\sscdots\null\null,
    \null\sscdots\null,
    \null\sscdots\null,
    \null\sscdots\null,
    \null\sscdots\null,
    \null\sscdots\null,
    \null\sscdots\null
    }&
    &A_{9,\dots,9,6}\;\sim\;
    \ytableaushort{
    \null\sscdots\null\null,
    \null\sscdots\null\null,
    \null\sscdots\null\null,
    \null\sscdots\null\null,
    \null\sscdots\null\null,
    \null\sscdots\null\null,
    \null\sscdots\null,
    \null\sscdots\null,
    \null\sscdots\null
    }&
    &h_{9,\dots,9,8,1}\;\sim\;
    \ytableaushort{
    \null\sscdots\null\null\null,
    \null\sscdots\null\null,
    \null\sscdots\null\null,
    \null\sscdots\null\null,
    \null\sscdots\null\null,
    \null\sscdots\null\null,
    \null\sscdots\null\null,
    \null\sscdots\null\null,
    \null\sscdots\null
    }
\end{align}
The curvature tensors obtained by taking curls on all columns 
of indices are related to each other algebraically.
For example, field strengths $F_4$ and $F_7$ for the three-form $A_3$ and six-
form $A_6$ are related to each other via the first-order duality equation
\begin{equation}
    F_{a_1\dots a_7}\approx 
    \frac{1}{4!} \varepsilon_{a_1\dots a_7}{}^{b_1\dots b_4} F_{b_1\dots b_4} \,,
\end{equation}
from which the usual equations of motion follow.
We also have a duality equation
\begin{equation}\label{92vs22}
    K_{9,2} \approx 
    \star_1 K_{2,2} 
    \quad\Longleftrightarrow\quad K_{a_1\dots a_9,b_1b_2} 
    \approx \frac{1}{2}\,
    \varepsilon_{a_1\dots a_9}{}^{c_1c_2} K_{c_1c_2,b_1b_2} 
\end{equation}
relating the graviton $h_{1,1}$ and dual graviton $h_{8,1}$ whose 
curvatures are
\begin{align}
    K_{2,2} &= \rd_1 \rd_2 h_{1,1} \,,&
    K_{9,2} &= \rd_1 \rd_2 h_{8,1} \,.
\end{align}
Integrating \eqref{92vs22} leads to a first-order duality equation 
for gravity and dual gravity.

The three-form $A_3$ and six-form $A_6$ are also related to the higher 
duals $A_{9,3}$ and $A_{9,6}$ via 
\begin{align}
    K_{a[10],b[4]} &\approx \varepsilon_{a[10]}{}^c 
    \,\partial_c F_{b[4]} \,,&
    K_{a[10],b[7]} &\approx \varepsilon_{a[10]}{}^c \,
    \partial_c F_{b[7]} \,,
\end{align}
where the curvatures $K_{10,4}$ and $K_{10,7}$ are 
\begin{align}
    K_{10,4} &= \rd_1 \rd_2 A_{9,3} \,,&
    K_{10,7} &= \rd_1 \rd_2 A_{9,6} \,.
\end{align}
Similarly, curvatures $K_{10^n,4}$ and $K_{10^n,7}$ for 
the higher duals $A_{9^n,3}$ and $A_{9^n,6}$ are related by
\begin{align}
\begin{aligned}
\label{FormHigherDuality} 
    K_{a[10],b^1[10],\dots,b^n[10],c[4]} &\approx
    \varepsilon_{a[10]}{}^e \,\partial_e 
    K_{b^1[10],\dots,b^n[10],c[4]} \,,\\
    K_{a[10],b^1[10],\dots,b^n[10],c[7]} &\approx\varepsilon_{a[10]}{}^e \,\partial_e 
    K_{b^1[10],\dots,b^n[10],c[7]} \,.
\end{aligned}
\end{align}
The Maxwell equation $\partial^aF_{abcd}\approx0$ and the Bianchi identity 
$\partial_{[a}F_{bcde]}=0$ for the three-form are exchanged under these 
relations for equations of motion and Bianchi identities for all higher dual 
three-forms, the six-form, and all higher dual six-forms.

The curvatures $K_{10^n,9,2}$ for the higher dual gravitons 
$h_{9^n,8,1}=h_{9,\dots,9,8,1}$ are related by
\begin{equation}
\label{GravityHigherDuality}
    K_{a[10],b^1[10],\dots,b^n[10],c[9],d[2]} \approx
    \varepsilon_{a[10]}{}^e 
    \,\partial_e K_{b^1[10],\dots,b^n[10],c[9],d[2]} \,,
\end{equation}
for all $n$\,.
Working on shell, the $\mathrm{GL}(11)$ irreducibility properties of one 
curvature are exchanged for trace constraints on another.
In this way, starting from the Ricci-flat equation of motion and the Bianchi 
identity for the graviton, the linearised equations of motion for all dual 
gravity fields in the theory were worked out \cite{Boulanger:2024lwk}.

The equations of motion for the higher dual fields take the form of higher 
trace constraints on the curvatures.
For example, the equation of motion for the first higher dual three-form 
$A_{9,3}$ is a fourth trace constraint on its curvature
\begin{equation}
\label{eq:Tr4_K10,4}
    K_{a[6]b[4],}{}^{b[4]} \approx 0
    \qquad\Longleftrightarrow\qquad
    \Tr^4K_{10,4} \approx 0 \,.
\end{equation}
Similarly, the equations for $A_{9^n,3}$ at higher levels with curvatures 
$K_{10^n,4}$ are given by
\begin{align}
    K_{a^1[10],\dots,a^i[10],\dots,a^j[10],\dots,a^n[10],b[4]} \big(\eta^{a^ia^j}\big)^{10} &\approx 0 \,,&
    K_{a^1[10],\dots,a^i[10],\dots,a^n[10],b[4]} \big(\eta^{a^ib}\big)^{4} &\approx 0 \,,& 
\end{align}
for all $i$ and $j\,$.
These higher trace constraints can be expressed more concisely as
\begin{align}
\label{eq:Tr10K_Tr4K}
    (\Tr_{i,j})^{10} K_{10^n,4} &\approx 0 \,,&
    (\Tr_{i,n+1})^{4} K_{10^n,4} &\approx 0 \,,&
    1\leqslant i<j\leqslant n \,,& 
\end{align}
where $\Tr_{i,j}$ denotes a single trace on the $i^\mathrm{th}$ and 
$j^\mathrm{th}$ columns.
Rather than there being one second-order equation of motion, we have several 
independent higher-order equations of motion.
Together, for each $n$\,, these equations are those required for the higher 
dual gauge potentials to propagate the correct degrees of freedom 
\cite{Bekaert:2002dt,Bekaert:2003az}.
For completeness, the equations of motion for the dual six-form fields 
$A_{9^n,6}$ are given by
\begin{align}
\label{eq:Tr10K_Tr7K}
    (\Tr_{i,j})^{10} K_{10^n,7} &\approx 0 \,,&
    (\Tr_{i,n+1})^{7} K_{10^n,7} &\approx 0 \,,&
    1\leqslant i<j\leqslant n \,,& 
\end{align}
and the equations of motion for the higher dual gravitons 
$h_{9^n,8,1}\equiv h_{9,\dots,9,8,1}$ are given by
\begin{align}
\begin{aligned}
    (\Tr_{i,j})^{10} K_{10^n,9,2} &\approx 0 \,,&\qquad
    (\Tr_{i,n+1})^{9} K_{10^n,9,2} &\approx 0 \,,\\
    (\Tr_{i,n+2})^{2} K_{10^n,9,2} &\approx 0 \,,&\qquad
    (\Tr_{n+1,n+2}) K_{10^n,9,2} &\approx 0 \,,
\end{aligned}
\quad\qquad 1\leqslant i<j\leqslant n\,.
\label{eq:Tr10K_Tr9K_Tr2K}
\end{align}
Note that a tensor in the irreducible representation of partition $[10^n,4]$ 
(or respectively $[10^n,7]$ or $[10^n,9,2]$) is automatically a trace, such 
that the irreducibility as an $SO(1,10)$ tensor is achieved by the above 
conditions, while a traceless condition would set it to zero.
This is in contrast with Young diagrams with columns of height at most nine, 
for which irreducibility requires all the traces to vanish.
This property singles out the gradient dual from more conventional dual fields.

\section{\texorpdfstring{Review of $E_{11}$ exceptional field theory}{Review of E11 exceptional field theory}}
\label{sec:review_E11}

In this section we review $E_{11}$ exceptional field theory following \cite{Bossard:2021ebg}.
This theory allows one to formulate (the bosonic sector of) both eleven-dimensional supergravity and type IIB supergravity in a way that makes the hidden exceptional symmetries $E_d$ appearing in Kaluza--Klein truncations manifest.
In this discussion we shall concentrate on the linearised free field equations for all the propagating higher dual fields.

\subsection{Algebraic preliminaries}
 
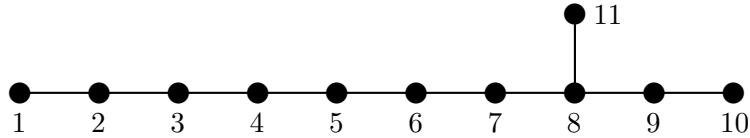
\begin{figure}[h]
\centering
\begin{picture}(300,50)
\thicklines
\multiput(10,10)(30,0){10}{\circle*{8}}
\put(10,10){\line(1,0){270}}
\put(220,40){\circle*{8}}
\put(220,10){\line(0,1){30}}
\put(7,-5){$1$}
\put(37,-5){$2$}
\put(67,-5){$3$}
\put(97,-5){$4$}
\put(127,-5){$5$}
\put(157,-5){$6$}
\put(187,-5){$7$}
\put(217,-5){$8$}
\put(247,-5){$9$}
\put(275,-5){$10$}
\put(227,36){$11$}
\end{picture}
\caption{\label{fig:e11dynk}{\sl Dynkin diagram of $E_{11}$ with labelling of 
nodes used in the text.}}
\end{figure}

Let us first introduce some notation.
We denote by $R(\lambda)$ the irreducible highest weight module with highest weight $\lambda = \sum_{i=1}^{11} n_i \Lambda_i$ (where $n_i \in \mathds{N}$) and we use the convention of Figure~\ref{fig:e11dynk} for the simple roots $\alpha_i$ with $(\Lambda_i,\alpha_j) = \delta_{ij}$\,.

The fields are functions of a subset of the generalised coordinates 
$z^{\mathcal{M}}$ valued in the module $R(\Lambda_1)$ such that any 
field or pair of fields satisfy the section constraint:
\begin{align}
\begin{aligned}
\label{eq:SC}
    \kappa_{\upalpha\upbeta} \,T^{\upalpha \mathcal{P}}{}_\mathcal{M} 
    \,T^{\upbeta \mathcal{Q}}{}_\mathcal{N} \,\partial_\mathcal{P}   
    \partial_\mathcal{Q} \phi(z)
    &= \frac12 \partial_\mathcal{M} \partial_\mathcal{N}\phi(z) \,,\\
    \kappa_{\upalpha\upbeta} \,T^{\upalpha \mathcal{P}}{}_\mathcal{M} 
    \,T^{\upbeta \mathcal{Q}}{}_\mathcal{N}\, 
    \partial_\mathcal{P} \phi_1(z)  \,\partial_\mathcal{Q} \phi_2(z)
    &= -\frac12 \partial_\mathcal{M} \phi_1(z)\, \partial_\mathcal{N}\phi_2(z) 
    + \partial_\mathcal{N} \phi_1(z) \,\partial_\mathcal{M}\phi_2(z) \,,
\end{aligned}
\end{align}
with $\kappa_{\upalpha\upbeta}$ the Killing form on $\mathfrak{e}_{11}$ and 
$T^{\upalpha \mathcal{M}}{}_\mathcal{N}$ the representation matrices of 
$\mathfrak{e}_{11}$ in $R(\Lambda_1)$.
This constraint can be defined equivalently as the condition that the quadratic 
differential
\begin{equation}
    \rd \phi_1(z) \otimes \rd \phi_2(z) \Big|_{R_\mathrm{sec}} = 0 \,,
\end{equation}
vanishes in the section module 
$R_\mathrm{sec}\coloneqq\big(R(\Lambda_1)\otimes 
R(\Lambda_1)\big)\ominus\big(R(2\Lambda_1)\oplus R(\Lambda_2)\big)$\,.
One shows that the maximal hyperplanes of solutions to the section constraint 
correspond either to the eleven or ten coordinates of 11D supergravity or type 
IIB supergravity, respectively \cite{Bossard:2015foa}.
Throughout this section, all the equalities will be understood to 
hold up to terms that vanish on section. 

In this paper we will study the linearised equations of motion of 
eleven-dimensional supergravity, and we assume accordingly that the 
fields only depend on the eleven coordinates $x^m$ among the infinitely 
many generalised coordinates
\begin{equation}
    z^\mathcal{M}= \big( x^m, y_{m[2]}, y_{m[5]}, y_{m[7];n}, y_{m[8];n[3]}, 
    y_{m[9];n,p}, \dots \big) \,.
\end{equation}
This solution to the section constraint is associated with the level 
decomposition with respect to the Cartan subgroup generator 
$H_{11}=(\Lambda_{11},H)$ 
where $H=\sum_{i=1}^{11}H_i\,\alpha_i$ is the vector of Cartan subalgebra elements
such that $\mathfrak{e}_{11}$ decomposes into eigenspaces of 
$\mathrm{ad}_{H_{11}}$ with eigenvalue $\ell$ as 
\begin{equation}
\label{LevelE}
    \mathfrak{e}_{11} = \bigoplus_{\ell \in \mathds{Z}} \mathfrak{g}_\ell = 
    \ldots \oplus (\mathfrak{gl}_{11})_0 \oplus [3]_1 \oplus [6]_2 \oplus 
    [8,1]_3 \oplus \big( [9,3] \oplus [10,1,1] \oplus [11,1] \big)_4 \oplus 
    \ldots
\end{equation}
and ${R(\Lambda_1)}$ decomposes as
\begin{equation}
    {R(\Lambda_1)} = \overline{[1]}_{\frac32} \oplus {[2]}_{\frac52} \oplus 
    {[5]}_{\frac72} \oplus [7\,;1]_{\frac92} \oplus \big( {[8\,;3]}\oplus 
    [9\,;1,1] \big)_{\frac{11}2} \oplus \ldots
\end{equation}
The fields only depend on the coordinates $x^m$ of lowest level 
$\ell=\frac32$\,.
Note that the components in $\mathfrak{e}_{11}$ of positive level 
$\ell\geqslant1$ have $3\ell$ covariant indices and the components of level 
$\ell\geqslant\frac52$ in ${R(\Lambda_1)}$ have $3\ell-\frac{11}2$ covariant 
indices.
This justifies using the columns of eleven indices to keep track of the 
$H_{11}$ level.
The type IIB solution is identically described using the level decomposition 
with respect to $H_9=(\Lambda_9,H)$ \cite{Schnakenburg:2001he}.

The fields of the theory include in particular a coset representative 
$\mathcal{V}\in E_{11}/K(E_{11})$ where $K(E_{11})$ is the involutive 
subgroup invariant under the Cartan involution \cite{Keurentjes:2004bv}, 
that acts on the Cartan--Weyl basis as
\begin{equation}
    \theta(H_i) = -H_i \,,\qquad
    \theta(E_\alpha) = -(-1)^{(\Lambda_1,\alpha)} E_{-\alpha} \,.
\end{equation}
This involution is defined such that 
$K(E_{11})\cap\mathrm{GL}(11)=\mathrm{SO}(1,10)$ for the 
group $\mathrm{GL}(11)\subset E_{11}$ that commutes with $H_{11}$\,.
Moreover, $\theta$ acts on the representation matrices as
\begin{equation}
    \theta( T^{\alpha M}{}_N) = -\eta^{MP} T^{\alpha Q}{}_P \eta_{QN} = -
    \eta^{\alpha\gamma} \kappa_{\gamma\beta}  T^{\beta M}{}_N \,,
\end{equation}
where we denote by $M,N,\dots$ and $\alpha,\beta,\dots$ the indices 
transforming under local $K(E_{11})$ and by $\mathcal{M},\mathcal{N},\dots$ and 
$\upalpha,\upbeta,\dots$ the indices transforming under rigid $E_{11}$\,, 
and
where $\eta_{\alpha\beta}$ and $\eta_{MN}$ are non-degenerate 
$K(E_{11})$-invariant bilinear forms on $\mf{e}_{11}$ and the 
$R(\Lambda_1)$ module, 
respectively. 
The existence of these non-degenerate, $K(E_{11})$-invariant forms 
was proven in \cite{Kac:1990algebras}.
In particular, the restriction of $\eta_{MN}$ to $M=a$ and $N=b$ is the 
eleven-dimensional Minkowski metric $\eta_{ab}$\,.

One chooses the coset representative $\mathcal{V}$ inside the parabolic 
subgroup of the algebra $\bigoplus_{\ell\geqslant 0} \mathfrak{g}_\ell$ 
according to \eqref{LevelE}.
For example, on a field satisfying the eleven-dimensional section constraint,
\begin{equation}
\label{11DSection}
    \mathcal{V}^{-1 \mathcal{N}}{}_M \partial_\mathcal{N} \phi 
    = \sqrt{e} ( e_a{}^m \partial_m \phi,0,0,0,\dots ) \,.
\end{equation}
The definition of $\mathcal{V}$ in a parabolic subgroup ensures generally 
that $  \mathcal{V}^{-1 \mathcal{N}}{}_M \partial_\mathcal{N} \phi$ admits a 
finite level expansion in the corresponding parabolic decomposition.
One defines the projected, right-invariant 
Maurer--Cartan one-form components
\begin{equation}\label{JMQ}
    J_M = J_{M \alpha} T^\alpha = \mathcal{V}^{-1 \mathcal{N}}{}_M \big( 
    \partial_\mathcal{N} \mathcal{V}\; \mathcal{V}^{-1} - \theta \big(   
    \partial_\mathcal{N} \mathcal{V}\; \mathcal{V}^{-1} \big) \big) \,.
\end{equation}
In the linearised approximation we have
\begin{equation}
    J_{M \alpha} \sim  \partial_M \phi_\alpha \,,
\end{equation}
where $\phi\in\mathfrak{e}_{11}\ominus K(\mf{e}_{11})$ has components
\begin{equation}
    \phi_\alpha = \big( h_{ab}=h_{(ab)}\,, A_{a[3]}\,, A_{a[6]}\,, h_{a[8],b}\,, A_{a[9],b[3]}\,, B_{a[10],b,c}\,, C_{a[11],b}\,, \dots \big)
\end{equation}
and satisfies 
\begin{equation}
\label{etakappa}
    \eta^{\alpha\beta} \phi_\beta = \kappa^{\alpha\beta} \phi_\beta \,.
\end{equation}

In addition to the coset fields $\mathcal{V}$, the theory includes 
constrained fields $\chi_M{}^{\tilde{\alpha}}$, $\zeta_M{}^{{\Lambda}}$, and 
$\zeta_M{}^{\tilde{\Lambda}}$ that themselves all satisfy the section 
constraint on their $M$ index.
For the eleven-dimensional solution to the section constraint we consider in 
this paper, they are therefore only non-zero for the level $\ell=\frac32$ 
indices $M=a$\,.
In order to define these fields it is convenient to introduce the tensor 
hierarchy algebra  $\cT(\mf{e}_{11})$~\cite{Bossard:2017wxl}.
For a given simple Lie algebra $\mathfrak{g}$ one can define an associated 
graded superalgebra $\mathcal{T}(\mathfrak{g})$ that encodes all the group 
theoretical structures of gauged supergravity inside a single algebra 
\cite{Palmkvist:2013vya}.
Similar algebraic structures were first proposed in~\cite{Henry-Labordere:2002xau,Cremmer:1997ct,Cremmer:1998px,Henneaux:2010ys}.
We refer to~\cite{Cederwall:2021ymp} for a precise definition of 
$\cT(\mathfrak{g})$ when $\mathfrak{g}$ is a hyperbolic Kac--Moody algebra.
The tensor hierarchy algebra $\cT(\mf{e}_{11})$ admits a $\ints$-grading
\begin{equation}
\label{eq:Tdec}
    \cT(\mf{e}_{11}) = \bigoplus_{p\in\ints} \cT_p
\end{equation}
consistent with the Grassmann $\ints_2$-grading, such that $\mf{e}_{11}$ is the maximal simple subalgebra of $\cT_0$\,.
Each $\cT_p$ is by construction a $\cT_0$ module and therefore an $\mathfrak{e}_{11}$ module.
In this way, the algebraic identities of $\cT(\mf{e}_{11})$ provide useful $E_{11}$-invariant tensors that allow to define $E_{11}$ exceptional field theory.
One important property is that $\cT_0$ is not the direct sum of irreducible $\mathfrak{e}_{11}$ modules, but includes an indecomposable component 
\begin{equation}
    \cT_0 = \mf{e}_{11} \oleft L(\Lambda_2) \oplus L''(\Lambda_{10}) \,,
\end{equation}
where we have introduced the symbol $\oleft$ to denote an indecomposable 
sum $M_1\oleft M_2$\,, meaning that $M_1\subset M_1\oleft M_2$ is a proper 
submodule while $M_2\cong M_1\backslash(M_1\oleft M_2)$ is a quotient.
We write the  generators as 
\begin{equation}
    t^\wa = (t^\alpha,t^\ta) \in \mf{e}_{11} \oleft L(\Lambda_2) \,.
\end{equation}
They satisfy the commutation relations 
\begin{align}
\label{eq:T0CR}
    \bigl[ t^\alpha, t^\beta \bigr] &= f^{\alpha\beta}{}_\gamma \,t^\gamma 
    \,,&
    \lb t^\alpha, t^\ta \rb &= -T^{\alpha\ta}{}_\tb\, t^\tb 
    - K^{\alpha\ta}{}_\beta \,t^\beta \,, 
\end{align}
where $T^{\alpha\ta}{}_\tb$ are representation matrices of $\mf{e}_{11}$ 
while $K^{\alpha\ta}{}_{\tb}$ is a Lie algebra cocycle responsible for the 
indecomposability of the module.
It is known from \cite{Bossard:2017wxl,Bossard:2021ebg} 
that $L(\Lambda_2)\supset R(\Lambda_2)$, and the (truncated) level 
decomposition of the tensor hierarchy algebra has not yet revealed the 
existence of any other irreducible module in $L(\Lambda_2)$.
It could well be that $L(\Lambda_2)=R(\Lambda_2)$ and we invite the reader to 
consider it is the case for all practical purposes.

In general we write 
$L(\lambda)=\bigoplus_{\lambda'\leqslant\lambda}R(\lambda^\prime)$ 
for a specific countable direct sum of highest weight modules such that 
$\lambda-\lambda'$ is in the positive cone of the simple roots.\footnote{These 
modules are integrable and should not be confused with Verma modules.}
We use $L(\lambda)$, $L'(\lambda)$, $L''(\lambda)$, and so on, to distinguish 
bounded weight modules of this kind with the same weight $\lambda$\,.
As we explained above, $L(\Lambda_2)$ is defined from the indecomposable 
component of $\cT_0$\,, and we define
\begin{equation}
    L(\Lambda_{10}) \coloneqq L''(\Lambda_{10}) \cap \big( R(\Lambda_1) 
    \otimes R(\Lambda_1) \big) \supset R(\Lambda_{10}) \oplus R(2\Lambda_3) 
    \,,
\label{L10}
\end{equation}
with associated generators $t^{\Lambda} \in L(\Lambda_{10})$.

Another important property of the tensor hierarchy algebra is that 
\begin{align}
\begin{aligned}
    \mathcal{T}_1 &= L(\Lambda_1) \supset R(\Lambda_1) \oplus \big( 
    R(\Lambda_1+\Lambda_{10}) \oplus R(\Lambda_{11}) \oplus R(\Lambda_1 + 
    2\Lambda_3) \big) \,, \label{T1def} \\
    \mathcal{T}_2 &= L'(\Lambda_{10}) \supset R(\Lambda_{10}) \oplus \big( 
    R(2\Lambda_3) \oplus R(\Lambda_2 + \Lambda_{10}) \oplus R(\Lambda_1 + 
    \Lambda_{11}) \oplus R(\Lambda_2 + 2\Lambda_3) \big) \,,
\end{aligned}
\end{align}
and one can define a differential complex of fields valued in 
$\mathcal{T}(\mathfrak{e}_{11})$ with the exterior derivative
\begin{align}
    \rd : \cT_p &\rightarrow \cT_{p+1} \,,&
    \phi(z) &\mapsto \rd \phi(z) = \mathrm{ad}(P^{M}) \partial_{M} \phi(z) \,,
\label{dTH}
\end{align}
where $P^{M}\in \mathcal{T}_1$ are the generators of 
$\mathcal{T}_1\cap R(\Lambda_1)$ and $\mathrm{ad}(\cdot)$ 
is the adjoint action in $\mathcal{T}(\mathfrak{e}_{11})$.
The Jacobi identity ensures that
\begin{equation}
    \rd^2 \phi(z) = \Pi_{\Lambda'}{}^{MN} {\rm ad}(P^{\Lambda'}) \partial_{M} \partial_{N} \phi(z) \,,
\end{equation}
where $\Pi_{\Lambda'}{}^{MN}$ are Clebsch--Gordan coefficients of 
$L'(\Lambda_{10})\cap\big(R(\Lambda_1) \otimes R(\Lambda_1)\big)
\hookrightarrow R(\Lambda_1) \otimes R(\Lambda_1)$ 
and where $P^{\Lambda'}\in\mathcal{T}_2$\,, such that $\rd^2\phi(z)=0$ on 
fields satisfying the section constraint.

The tensor hierarchy algebra admits an anti-involution 
that we will denote by~$\ddagger$ such that 
$\cT_p=\cT_{-2-p}^*$\,.
In particular, $\cT_{-1}\cong\cT_{-1}^*$ is a 
symplectic representation of $\mf{e}_{11}$ and so it admits an $E_{11}$-invariant 
non-degenerate antisymmetric bilinear form $\Omega_{IJ}$\,.
Moreover, it was conjectured in reference \cite{Bossard:2017wxl} that $\cT_{-1}$ is 
equipped with a $K(E_{11})$-invariant non-degenerate symmetric bilinear form 
$\eta_{IJ}$\,, or equivalently that the Cartan involution defines a 
representation equivalent to the dual representation such that
\begin{equation}
    (T^{\alpha I}{}_J)^\ddagger = -\eta^{IK} T^{\alpha L}{}_K \eta_{LJ} \,.
\end{equation}
The level decomposition of $\cT_{-1}$ under finite-dimensional subalgebras of 
$\mf{e}_{11}$ has not unveiled any proper submodule of $\cT_{-1}$\,.
If $\cT_{-1}$ were an irreducible representation, $\eta_{IJ}$ would indeed exist 
and be unique.\footnote{The most general possible structure of the module 
$\cT_{-1}\cong M_{k}$ for some (possibly infinite) integer $k$ 
is such that 
$M_{n+1} = M_{n,+} \oleft M_{n} \oleft M_{n,-}$\,, 
with $M_0$ fully reducible and $M_{n,+}\cong M_{n,-}^*$ fully reducible 
for all $0\leqslant n\leqslant k$\,.
The module $M_{n,+}$ must be the direct sum of bounded lowest and highest 
weight modules. The existence of $\eta_{IJ}$ requires that 
$M_{n,+}\cong M_{n,-}$\,.}
We refer to \cite{Bossard:2017wxl} for the exhaustive list of evidence in 
favour of this conjecture.

In order to define the theory one needs to introduce constrained fields 
$\zeta_M{}^{{\Lambda}}$ and $\zeta_M{}^{\tilde{\Lambda}}$ for the entire 
section module $R_\mathrm{sec}$ and we define accordingly $L(\Lambda_4)$ 
such that  
\begin{equation}
    R_\mathrm{sec} \coloneqq \big( R(\Lambda_1) \otimes R(\Lambda_1) \big) 
    \ominus \big( R(2\Lambda_1) \oplus R(\Lambda_2) \big) = L(\Lambda_{10}) 
    \oplus L(\Lambda_4) \,,
\end{equation}
where $L(\Lambda_{10})$ was defined in \eqref{L10}.
The field $\zeta_M{}^{{\Lambda}}$ valued in $L(\Lambda_{10})$ is predicted by 
the tensor hierarchy algebra $\mathcal{T}(\mathfrak{e}_{11})$ while the other 
field $\zeta_M{}^{\tilde{\Lambda}}$ valued in $L(\Lambda_4)$ must be included 
to ensure that the equations of motion are gauge invariant.

\subsection{\texorpdfstring{Duality equations and the $E_{11}$ pseudo-Lagrangian}{Duality equations and the E11 pseudo-Lagrangian}}

As usual in exceptional field theory, the generalised diffeomorphisms are 
defined for a vector field $\xi^M\in R(\Lambda_1)$.
They act in the linearised approximation as
\begin{equation}
\label{eq:dxiJ}
    \delta_\xi \phi{}^\alpha = T^{\alpha N}{}_P \big( \partial_N \xi^P + 
    \eta_{NQ}\, \eta^{PR}\, \partial_R \xi^Q \big) \,.
\end{equation}
The first term is identical to the gauge transformation first proposed in 
\cite{West:2014eza}.
The constrained fields transform as
\begin{align}
\begin{aligned}
\label{eq:chiGT}
    \delta_\xi \chi_M{}^\ta &= T^{\tilde\alpha N}{}_P \big( 
    \partial_M\partial_N \xi^P + \eta_{NQ} \eta^{PR} \partial_M \partial_R 
    \xi^Q \big) + \Pi^{\tilde\alpha}{}_{QP} \eta^{NQ} \partial_M \partial_N 
    \xi^P +  T^{\tilde\alpha N}{}_{\tilde{P}} \partial_M\partial_N 
    \alpha^{\tilde{P}}\,,\\
    \delta_\xi \zeta_M{}^\Lambda &= \Pi^{{\Lambda}}{}_{QP} \eta^{NQ} \partial_M 
    \partial_N \xi^P +  T^{\Lambda N}{}_{\tilde{P}} \partial_M\partial_N 
    \alpha^{\tilde{P}} \,, \qquad\qquad \delta_\xi \zeta_M{}^{\tilde{\Lambda}} 
    = \Pi^{\tilde{\Lambda}}{}_{QP} \eta^{NQ} \partial_M \partial_N \xi^P \,,
\end{aligned}
\end{align}
where $\Pi^\ta{}_{MN}$\,, $\Pi^\Lambda{}_{MN}$\,, and 
$\Pi^{\tilde{\Lambda}}{}_{MN}$ are the Clebsch--Gordan coefficients for the 
embeddings of $L(\Lambda_2)\oplus L(\Lambda_{10})\oplus L(\Lambda_4)$ into 
$R(\Lambda_1)\otimes R(\Lambda_1)$.\footnote{If $L(\Lambda_2)$ was strictly 
bigger than $R(\Lambda_2)$ then $\Pi^\ta{}_{MN}$ would only be non-zero for 
$\tilde{\alpha}$ valued in $R(\Lambda_2)$\,.}
Additional parameters $\alpha^{\tilde{P}}\in L(\Lambda_1)\ominus R(\Lambda_1)$ 
were not considered in \cite{Bossard:2021ebg}, but it follows from 
\eqref{T1def} and \eqref{dTH} that they are also gauge symmetries of the 
theory.\footnote{One has the algebraic identity 
$( C^{IM}{}_{\tilde{\alpha}} T^{\tilde{\alpha}N}{}_{\tilde{P}} + C^{IM}{}_{\Lambda} T^{\Lambda N}{}_{\tilde{P}} ) \partial_M \partial_N \alpha^{\tilde{P}} = 0$ 
and the constrained fields $\chi^{\tilde{\alpha}}$ and $\zeta^\Lambda$ only 
appear in the duality equation and the pseudo-Lagrangian through 
$C^{IM}{}_{\tilde{\alpha}} \chi_M{}^{\tilde{\alpha}} + C^{IM}{}_{\Lambda} \zeta_M{}^{\Lambda}$, 
so it is clearly a gauge symmetry.}

It appears that the algebra of generalised diffeormorphisms only closes on 
fields satisfying the section constraint up to additional ancillary gauge 
transformations \cite{Hohm:2014fxa,Bossard:2021ebg}.
These additional transformations may be interpreted as St\"uckelberg 
shifts that can be used in order to eliminate higher dual fields, see for 
example \cite{Boulanger:2008nd,Bossard:2021ebg,Hohm:2018qhd}.
Their explicit action on the constrained fields have not been worked 
out and we shall not attempt to describe them in this paper.

The field strength 
\begin{equation}
\label{eq:FStemp}
    F^I =  C^{IM}{}_\alpha\, \partial_M \phi^\alpha 
    + C^{IM}{}_{\tilde{\alpha}}\, \chi_M{}^{\tilde{\alpha}} 
    + C^{IM}{}_{\Lambda}\, \zeta_M{}^\Lambda 
    + C^{IM}{}_{\tilde{\Lambda}}\, \zeta_M{}^{\tilde{\Lambda}}
\end{equation}
is defined with the structure coefficients of the tensor hierarchy algebra 
$[\cT_1,\cT_{-2}]\rightarrow\cT_{-1}$\,, except for 
${C}^{IM}{}_{\tilde{\Lambda}}$ which is an $E_{11}$-invariant tensor 
determined by the identity 
\begin{equation}
\label{Master}
    \Omega_{IJ} C^{JM}{}_\wa T^{\wa N}{}_P \equiv \eta_{IJ}\, \eta_{PQ} \Big( 
    \eta^{\ta\tb} C^{JQ}{}_\ta \Pi_\tb{}^{MN} + \eta^{\Lambda\Xi} 
    C^{JQ}{}_{\Lambda} \Pi_{\Xi}{}^{MN} + \eta^{\tilde\Lambda\tilde\Xi} 
    C^{JQ}{}_{\tilde\Lambda} \Pi_{\tilde\Xi}{}^{MN} \Big) \,.
\end{equation}
The definition \eqref{eq:FStemp} 
for the field strength is suggested by the tensor hierarchy algebra 
differential \eqref{dTH} that gives 
\begin{equation}
    \rd \big( \phi^\alpha \bar{t}_\alpha + X^{\tilde{\alpha}} 
    \bar{t}_{\tilde{\alpha}} + X^{\Lambda} \bar{t}_{\Lambda} \big) 
    = \Big( C^{IM}{}_\alpha \partial_M \phi^\alpha + C^{IM}{}_{\tilde{\alpha}} 
    \partial_M X^{\tilde{\alpha}} + C^{IM}{}_{\Lambda} \partial_M X^\Lambda 
    \Big) t_I \,,
\end{equation}
where $\bar{t}_\alpha=-(t^\alpha)^\ddagger$, 
$\bar{t}_{\tilde\alpha}=-(t^{\tilde\alpha})^\ddagger$, and 
$\bar{t}_{\Lambda}=-({t}^{\Lambda})^\ddagger$ are the generators of 
$\cT_{-2}=\big(\cT_0\big)^\ddagger$.
The field strength does not transform as a tensor under generalised 
diffeomorphisms:~its linearised gauge transformation gives a non-covariant 
contribution
\begin{align}
\begin{aligned}
\label{nablaF}
    \delta_\xi F^I = \Big( C^{IM}{}_{\wa} T^{\wa R}{}_Q \eta^{QN}\eta_{RP} 
    + C^{IM}{}_{\tilde\alpha} \Pi^{\tilde\alpha}{}_{QP} \eta^{QN} \hspace{50mm} 
    \\
    {} + C^{IM}{}_\Lambda \Pi^\Lambda{}_{QP}\eta^{QN} 
    + C^{IM}{}_{\tilde{\Lambda}} \Pi^{\tilde{\Lambda}}{}_{QP}
    \eta^{QN} \Big) \partial_M \partial_N \xi^P \,.
\end{aligned}
\end{align}
Nonetheless, the duality equation 
\begin{equation}
\label{eq:DR}
    \cE_I \coloneqq \eta_{IJ} F^J - \Omega_{IJ} F^J \approx 0
\end{equation}
\emph{is} gauge invariant, as follows from the identity \eqref{Master} 
that was proved in \cite{Bossard:2017wxl}.\footnote{It was proved that 
$C^{IM}{}_\Lambda$ is indeed given by the structure coefficients of the tensor 
hierarchy algebra $\cT(\mf{e}_{11})$ for the $L(\Lambda_{10})$ index $\Lambda$ 
restricted to $R(\Lambda_{10})\subset L(\Lambda_{10})$\,. If it were not true 
for the full $L(\Lambda_{10})$ module, one should simply redefine accordingly 
which irreducible representation is in $L(\Lambda_{10})$ 
(for which $C^{IM}{}_\Lambda$ is given by the structure coefficients 
of the tensor hierarchy algebra) and which is in $L(\Lambda_4)$\,.}

Using the section constraint, most of the duality equations are tautological 
and can be solved in terms of the constrained fields 
$\chi_a{}^{\tilde{\alpha}}$, $\zeta_a{}^{\Lambda}$, and 
$\zeta_a{}^{\tilde{\Lambda}}$.
In order to determine the dynamics of the theory, one introduces a 
pseudo-Lagrangian $\cL_{E_{11}}$ that transforms as a density under 
generalised diffeomorphisms.
We refer to \cite{Bossard:2021ebg} for the explicit form of $\cL_{E_{11}}$ 
and here we shall only review a few of its established properties, assuming 
that the bilinear form $\eta_{IJ}$ exists.
Considering fields that only depend on the eleven coordinates at level 
$\ell=\frac32$\,, the pseudo-Lagrangian $\cL_{E_{11}}$ reduces to the sum of 
the eleven-dimensional supergravity Lagrangian $\cL^{\scalebox{0.6}{sugra}}$ 
(restricted to the bosonic sector) plus the sum of the squares of the duality 
equations, up to a total derivative:
\begin{equation}
\label{priorcLc}
    \cL_{E_{11}} = \cL^{\scalebox{0.6}{sugra}} - \frac{1}{4} 
    \sum_{\ell\,\geqslant\frac12} \eta^{I_{(\ell)}J_{(\ell)}} 
    \mathcal{E}_{I_{(\ell)}} \mathcal{E}_{J_{(\ell)}} + \partial(\,\dots) \,.
\end{equation}
It turns out that the Euler-Lagrange equations of $\cL_{E_{11}}$ give 
formal infinite sums at the non-linear level, and one must instead consider 
the Euler--Lagrange equations for the Lagrangians 
\begin{equation}
\label{cLc}
    \cL_c = \cL_{E_{11}} + \frac{1}{4} \sum_{\ell\,\geqslant\frac12} 
    (1-c_\ell)\, \eta^{I_{(\ell)}J_{(\ell)}} \mathcal{E}_{I_{(\ell)}} 
    \mathcal{E}_{J_{(\ell)}} \,,
\end{equation}
where $c_\ell=0$ for all $\ell>\ell_0$ with a fixed $\ell_0\geqslant\frac12$ 
to get well-defined non-linear equations.
Setting every $c_\ell$ to zero leads to the supergravity Einstein and Maxwell 
equations in eleven dimensions.
For all choices of $c_\ell$ the Euler--Lagrange equations for the constrained 
fields are given by linear combinations of the duality equations.

The linearised Euler--Lagrange equations for the $E_{11}$ fields $\phi^\alpha$ 
are finite and can be defined directly from $\mathcal{L}_{E_{11}}$.
The Euler-Lagrange derivative of $\cL_{E_{11}}$ 
with respect to $\phi^\alpha$ transforms (indecomposably) under 
$\mathfrak{e}_{11}$ into itself and (the derivative of) the Euler-Lagrange derivative 
of $\cL_{E_{11}}$ with respect to 
$\chi_M{}^{\tilde{\alpha}}$ so that
\begin{align}
\begin{aligned}
    \delta \cL_{E_{11}} =  \Big( \widehat{\mathcal{E}}_\alpha 
    + \frac{1}{2} C^{IM}{}_\alpha \partial_M \mathcal{E}_I \Big) 
    \delta \phi^\alpha - \frac{1}{2} C^{IM}{}_{\tilde{\alpha}} \mathcal{E}_I 
    \delta \chi_M{}^{\tilde{\alpha}} 
    \hspace{50mm} \\
    {}- \frac{1}{2} C^{IM}{}_{{\Lambda}} \mathcal{E}_I 
    \delta \zeta_M{}^{{\Lambda}} - \frac{1}{2} C^{IM}{}_{\tilde{\Lambda}} 
    \mathcal{E}_I \delta \zeta_M{}^{\tilde{\Lambda}}
\end{aligned}
\end{align}
defines the covariant equation\footnote{Note also that one has 
$\widehat{\mathcal{E}}^\alpha = \kappa^{\alpha\beta}\,\widehat{\mathcal{E}}_\beta = 
\eta^{\alpha\beta}\,\widehat{\mathcal{E}}_\beta$ due to \eqref{etakappa}. There is a sign misprint for the 
$\zeta$ term in Eq. (5.20) of \cite{Bossard:2021ebg}.}
\begin{align}
\begin{aligned}
\label{LEN}
    \widehat{\mathcal{E}}^\alpha \coloneqq \frac12 \big( \eta^{\alpha\beta} 
    + \kappa^{\alpha\beta} \big) \Omega_{IJ} C^{IM}{}_\beta 
    \Big( C^{JN}{}_\ta \partial_{[M} \chi_{N]}{}^\ta +C^{JN}{}_\Lambda 
    \partial_{[M} \zeta_{N]}{}^\Lambda +C^{JN}{}_{\tilde\Lambda} 
    \partial_M \zeta_N{}^{\tilde\Lambda} \Big) 
    \hspace{10mm} \\
    {} - \frac12 \big( \eta^{\alpha\gamma} + \kappa^{\alpha\gamma} \big) 
    \bigg( \frac12 \eta^{MN} \kappa_{\gamma\beta} 
    - T_{\beta}{}^M{}_P T_\gamma{}^N{}_Q\,\eta^{PQ} \bigg) 
    \partial_M \partial_N \phi^\beta \approx 0 \,.
\end{aligned}
\end{align}

Assuming the fields only depend on the usual eleven coordinates $x^m$, 
it is convenient to reduce \eqref{LEN} level-by-level.
For $\ell=0$ one finds the linearised Einstein equation for 
$\phi_{\alpha_{(0)}} \equiv h_{ab}$ and for $\ell\geqslant1$ one 
obtains\footnote{The $\cT_{-1}(\mf{e}_{11})$ symplectic form 
$\Omega_{I_{(\ell_1)}J_{(\ell_2)}}$ and the $\mf{e}_{11}$ Killing form 
$\kappa^{\alpha_{(\ell_1)}\beta_{(\ell_2)}}$ are non-zero when 
$\ell_1+\ell_2=0$ while $\eta^{\alpha_{(\ell_1)}\beta_{(\ell_2)}}$ and 
$\eta^{I_{(\ell_1)}J_{(\ell_2)}}$ are non-zero when $\ell_1=\ell_2$\,.}
\begin{multline}
\label{LENexpand}
    \Omega_{I_{(-\ell-3/2)}J_{(\ell+3/2)}} 
    C^{I_{(-\ell-3/2)}a}{}_{\alpha_{(-\ell)}} 
    \\
    \times \Big( C^{J_{(\ell+3/2)}b}{}_{\tb_{(\ell+3)}} \partial_{[a} 
    \chi_{b]}{}^{\tb_{(\ell+3)}} + C^{J_{(\ell+3/2)}b}{}_{\Lambda_{(\ell+3)}} 
    \partial_{[a} \zeta_{b]}{}^{\Lambda_{(\ell+3)}} + C^{J_{(\ell+3/2)}b}
    {}_{\tilde{\Lambda}_{(\ell+3)}} \partial_{a} \zeta_{b}
    {}^{\tilde{\Lambda}_{(\ell+3)}} \Big) 
    \\
    \approx \Big(\eta^{ab} \eta_{\alpha_{(-\ell)}\beta_{(-\ell)}} 
    - T_{\beta_{(-\ell)}}{}^{\! a}{}_{P_{(\ell+3/2)}} 
    T_{\gamma_{(-\ell)}}{}^{\! b}
    {}_{Q_{(\ell+3/2)}}\,\eta^{P_{(\ell+3/2)}Q_{(\ell+3/2)}} \Big) 
    \partial_a \partial_b \phi^{\beta_{(-\ell)}} \,, 
\end{multline}
where indices with subscript $(\ell+\mu)$ denote the restriction 
to level $\ell+\mu$ which always corresponds to a finite set of 
$\mathrm{GL}(11)$ irreducible representations.
The right-hand side gives the second-order field equation for the three-form at 
$\ell=1$ and the second-order field equation for the six-form at $\ell=2$\,.
As we shall discuss in more detail in Section \ref{sec:Lagrangian}, the above
discussion justifies the definition of the generalised Maxwell-like tensor
\begin{equation}
\label{RicciEqua}
    M[\phi]^\alpha \coloneqq \big( \eta^{\alpha\gamma} + \kappa^{\alpha\gamma} \big) \bigg( \frac12 \eta^{ab} \kappa_{\gamma\beta} - T_{\beta}{}^a{}_P T_\gamma{}^b{}_Q\, \eta^{PQ} \bigg) \partial_a \partial_b \phi^\beta \,.
\end{equation}
For $\ell\geqslant1$ we shall see that it indeed gives the Maxwell-like tensors 
as defined in \cite{Campoleoni:2012th}, while 
for $\ell=0$ it actually gives the linearised Einstein tensor 
$M[h]^{ab}=-2R[h]^{ab}+\eta^{ab}R[h]$.
For $\ell\geqslant3$ the Maxwell-like tensor does not vanish on-shell, but we will show in this paper that
\begin{equation}
\label{LinearisedRicciU}
    M[\phi]^{\alpha_{(\ell)}} \approx \kappa^{\alpha_{(\ell)}\beta_{(-\ell)}} \Omega_{I_{(-\ell-3/2)}J_{(\ell+3/2)}} C^{I_{(-\ell-3/2)}a}{}_{\beta_{(-\ell)}} C^{J_{(\ell+3/2)}a}{}_{\tilde{\Lambda}_{(\ell+3)}} \partial_a \partial_b U^{\tilde{\Lambda}_{(\ell+3)}}
\end{equation}
for all components associated with propagating fields, where 
$U^{\tilde{\Lambda}_{(\ell+3)}}$ depends (non-locally) on the $E_{11}$ fields 
$\phi^\alpha$
at lower levels $\ell'\leqslant\ell$\,.
As we will show, one can actually decompose the covariant equation 
\eqref{LEN} into the following equation for the constrained fields
\begin{equation}
\label{BianchiE11}
    \mathcal{B}^\alpha \coloneqq \frac12 \Big( \eta^{\alpha\beta} + \kappa^{\alpha\beta} \Big) \Omega_{IJ} C^{IM}{}_\beta \Big( C^{JN}{}_\ta \partial_{[M} \chi_{N]}{}^\ta + C^{JN}{}_\Lambda \partial_{[M} \zeta_{N]}{}^\Lambda \Big) = 0 \,,
\end{equation}
and the generalised Ricci-flat equation of motion
\begin{equation}
\label{eq:generalisedRicci}
    R^\alpha \coloneqq M[\phi]^\alpha - \big( \eta^{\alpha\gamma} + \kappa^{\alpha\gamma} \big) \Omega_{IJ} C^{Ia}{}_{\gamma} C^{Jb}{}_{\tilde{\Lambda}} \partial_{a} \zeta_{b}{}^{\tilde{\Lambda}} \approx 0 \,,
\end{equation}
where $\zeta_a{}^{\tilde\Lambda}$ will shortly be identified with 
$\partial_aU^{\tilde\Lambda}$.

Since the linearised gauge transformations of the constrained fields in 
\eqref{eq:chiGT} are total derivatives, one finds that the gauge 
transformation of $\cB^\alpha$ in \eqref{BianchiE11} vanishes 
automatically such that the generalised Ricci-flat equation is 
independently gauge invariant.
For any solution of \eqref{BianchiE11} and \eqref{eq:generalisedRicci}, restricting to the 11D solution $\partial_M=(\partial_a,0,\dots)$ of the section constraint,
we will see for the components of the constrained fields associated with propagating $E_{11}$ fields that we necessarily have
\begin{align}
\label{LinearChi}
    \chi_M{}^\ta &= \partial_M X^\ta \,,& 
    \zeta_M{}^\Lambda &= \partial_M X^\Lambda \,,&
    \zeta_M{}^{\tilde{\Lambda}} &= \partial_M {U}^{\tilde{\Lambda}} \, . 
\end{align}
At this stage, where the duality equations are not yet considered, 
the fields $X^{\tilde{\alpha}}$ and $X^\Lambda$ are arbitrary, while   
$U^{\tilde{\Lambda}}$ depends non-locally on $\phi^\alpha$ via 
\eqref{LinearisedRicciU}.
As we shall describe in more detail, it follows that the covariant equation \eqref{LEN} and the duality equation \eqref{eq:DR} describe the linearised dynamics of all gradient dual fields.

However, since the components of $\widehat{\mathcal{E}}_\alpha\approx0$ are not themselves the Euler--Lagrange equations of any Lagrangian, it is useful to relate them to the Euler--Lagrange equations of the Lagrangians $\cL_c$ in \eqref{cLc}.
After computation \cite{Bossard:2021ebg}, it appears that one has 
\begin{align}
\begin{aligned}
    \delta \cL_c &= \widehat{\mathcal{E}}_\alpha \delta \phi^\alpha + \frac{1}{2} \sum_{\ell\,\geqslant\,\frac12} c_{\ell} \, \eta_{I_{(\ell)}J_{(\ell)}} C^{I_{(\ell)} m}{}_\alpha \partial_m  F^{J_{(\ell)}} \delta \phi^\alpha \\
    &{}\hspace{22mm} + \frac{1}{2} \sum_{\ell\,\geqslant\,\frac12} (2-c_{\ell}) \, C^{I_{(-\ell)}m}{}_\alpha \partial_m \Big( \eta_{I_{(-\ell)}J_{(-\ell)}} F^{J_{(-\ell)}} - \Omega_{I_{(-\ell)}J_{(\ell)}} F^{J_{(\ell)}} \Big) \delta \phi^\alpha \\
    & = \delta \mathcal{L}^{\scalebox{0.6}{sugra}} + \frac{1}{2} \sum_{\ell\,\geqslant\,\frac12} c_\ell \, \eta_{I_{(\ell)}J_{(\ell)}} C^{I_{(\ell)}m}{}_\alpha \partial_m F^{J_{(\ell)}} \delta \phi^\alpha \\
    &{}\hspace{30mm} + \frac{1}{2} \sum_{\ell\,\geqslant\,\frac12} c_\ell \, C^{I_{(-\ell)}m}{}_\alpha \partial_m \Big( \Omega_{I_{(-\ell)}J_{(\ell)}} F^{J_{(\ell)}} -\eta_{I_{(-\ell)}J_{(-\ell)}} F^{J_{(-\ell)}} \Big) \delta \phi^\alpha \, . 
\end{aligned}
\end{align}
As a result, we find that for different choices of $c_\ell$ one can recover 
all the required equations as Euler--Lagrange equations.

Using \eqref{LinearChi} together with the duality equations and the property 
that the constrained fields only appear at positive level $\ell\geqslant 3$\,, 
one obtains the linearised duality equations 
\begin{align}
\begin{aligned}
\label{eq:DR_fixed}
    &C^{I_{(\ell+3/2)}a}{}_{\alpha_{(\ell+3)}} \partial_a 
    \phi^{\alpha_{(\ell+3)}} + C^{I_{(\ell+3/2)}a}{}_{\ta_{(\ell+3)}} 
    \partial_a X^{\ta_{(\ell+3)}} + C^{I_{(\ell+3/2)}a}{}_{\Lambda_{(\ell+3)}} 
    \partial_a X^{\Lambda_{(\ell+3)}} \\
    &\hspace{20mm} + C^{I_{(\ell+3/2)}a}{}_{\tilde{\Lambda}_{(\ell+3)}} 
    \partial_a U^{\tilde{\Lambda}_{(\ell+3)}} \approx \eta^{I_{(\ell+3/2)} K} 
    \Omega_{KJ_{(-\ell-3/2)}} C^{J_{(-\ell-3/2)}a}{}_{\alpha_{(-\ell)}} 
    \partial_a \phi^{\alpha_{(-\ell)}} \, , 
\end{aligned}
\end{align}
which imply the duality equations \eqref{FormHigherDuality} and \eqref{GravityHigherDuality} for the propagating higher dual fields, 
upon acting on it with appropriately projected derivatives, so as to 
eliminate the $X$ and $U$ fields.

\subsection{\texorpdfstring{$\mathrm{GL}(11)$ field representations and duality equations}{GL(11) field representations and duality equations}}
\label{GL11E11}

To describe the dynamics it is necessary to solve the section constraint.
In this paper we consider only the eleven-dimensional supergravity solution, 
in which case it is relevant to decompose $E_{11}$ with respect to its 
$\mathrm{GL}(11)$ subgroup.
One defines accordingly the $\mathrm{GL}(1)$ level $\ell$ associated to the 
weight $(\Lambda_{11},H)$ as was introduced in \eqref{LevelE}.
The low level components of the field strength $F=F^It_I\in\cT_{-1}$ were 
determined in \cite{Bossard:2017wxl} as
\begin{equation}
    \dots \,;
    \underbrace{F_{1}{}^6\,,F^{4,1}}_{\ell\,=\,-\frac72} \,;
    \underbrace{F_{1}{}^3\,,F^{1,1}}_{\ell\,=\,-\frac52} \,;
    \underbrace{F_{2}{}^1}_{\ell\,=\,-\frac32} \,;
    \underbrace{F_{4}}_{\ell\,=\,-\frac12} \,;
    \underbrace{F_{7}}_{\ell\,=\,\frac12} \,;
    \underbrace{F_{9\seco1}}_{\ell\,=\,\frac32} \,;
    \underbrace{F_{10\seco3}\,,F_{11,1,1}}_{\ell\,=\,\frac52} \,;
    \underbrace{F_{10\seco6}\,,F_{11,4,1}}_{\ell\,=\,\frac72} \,; \dots
\end{equation}
Recall that we use a shorthand for $\mathrm{GL}(11)$ indices, 
where a semicolon separates the irreducible components of a reducible 
field and a comma separates antisymmetric 
columns of indices within the same irreducible 
component.
The symplectic form $\Omega_{IJ}$ on $\cT_{-1}$ features in the duality 
equation \eqref{eq:DR} as a pairing between field strengths at grades $\ell$ 
and $-\ell$ and involves the Levi--Civita  tensor in eleven dimensions.

The level $\ell=\frac12$ component of the duality equation \eqref{eq:DR} is the familiar $F_7= \star F_4$ equation, and its level $\ell=\frac32$ component is the dual graviton duality equation.
In our shorthand notation we write them as
\begin{align}
\label{Duality12}
    &\mathcal{E}_{a[7]} \coloneqq F_{a[7]} - \frac{1}{24} \varepsilon_{a[7]}{}^{b[4]} F_{b[4]} \approx 0 \,,&
    &\cE_{a[9];b} \coloneqq F_{a[9];b} + \varepsilon_{a[9]}{}^{c_1c_2} \partial_{c_1} h_{c_2b} \approx 0 \,,
\end{align}
where the graviton, three-form, six-form, and dual graviton field strengths are
\begin{align}
\label{eq:F2^1_F4_F7_F9;1}
    F_{a[2]}{}^b &= 2\,\partial_a h_a{}^b \,,&
    F_{a[4]} &= 4\,\partial_a A_{a[3]} \,,&
    F_{a[7]} &= 7\,\partial_a A_{a[6]} \,,&
    F_{a[9];b} &= 9\,\partial_a h_{a[8],b} + \chi_{b;a[9]} \,.
\end{align}

The higher dual fields associated with the three-form, six-form, and dual 
graviton are $A_{9^n,3}$\,, $A_{9^n,6}$\,, and $h_{9^n,8,1}$\,.
They are found inside $\mf{e}_{11}$ at levels $3n+1$\,, $3n+2$\,, 
and $3n+3$\,, respectively.
We denote by $n\in\mathbb{Z}^+$ the number of 
columns
of nine indices in each such field.
Their associated field strengths in $\cT_{-1}$ are
\begin{equation}
    \underbrace{F_{10\seco9^{n-1},3}}_{\ell\,=\,3n\,-\,\frac12} \,;
    \underbrace{F_{10\seco9^{n-1},6}}_{\ell\,=\,3n\,+\,\frac12} \,;
    \underbrace{F_{10\seco9^{n-1},8,1}}_{\ell\,=\,3n\,+\,\frac32} \,.
\end{equation}
There is precisely one of these fields, and hence field strengths, for each value of $n$\,.
Using $\rx$ to denote $[3]$, $[6]$, or $[8,1]$ depending on the sector, we find that the field strength $F_{10\seco9^{n-1},\,\rx}$ for a higher dual field $\phi_{9^n,\,\rx}$ is:
\begin{equation}
\label{eq:fieldstrength}
    F_{a[10]\seco b^i[9]_{i=1}^{n-1},c[\rx]} = 10\,\partial_a \phi_{a[9],b^i[9]_{i=1}^{n-1},c[\rx]} + \chi_{a[10]\seco b^i[9]_{i=1}^{n-1},c[\rx]} \,,
\end{equation}
and $\chi_{a[10]\seco b^i[9]_{i=1}^{n-1},c[\rx]}$ are the components of 
\begin{equation}
\label{chichi}
    \chi^I \coloneqq C^{IM}{}_{\tilde{\alpha}}\, \chi_M{}^{\tilde{\alpha}} 
    + C^{IM}{}_{\Lambda}\, \zeta_M{}^\Lambda 
    + C^{IM}{}_{\tilde{\Lambda}}\, \zeta_M{}^{\tilde{\Lambda}} \,, 
\end{equation}
for which we do not introduce yet another letter to avoid proliferation. 
Defining $\ell(\rx)$ as the $\mathrm{GL}(11)$ level of the three-form, six-
form, and dual graviton, 
\begin{equation}
\label{eq:ell(x)}
    \ell([3]) \coloneqq 1 \,, \qquad
    \ell([6]) \coloneqq 2 \,, \qquad
    \ell([8,1]) \coloneqq 3 \,,
\end{equation}
we have that the duality equation 
\begin{equation}
\label{eq:dualityequation}
    \cE_{a[10];b^i[9]_{i=1}^{n-1},c[\rx]} \coloneqq F_{a[10];b^i[9]_{i=1}^{n-1},c[\rx]} - \varepsilon_{a[10]}{}^d \partial_d \phi_{b^i[9]_{i=1}^{n-1},c[\rx]} \approx 0 \,, 
\end{equation}
is a component at level $\ell=3n+\ell(\rx)-\frac32$ of $\cE_I\approx0$ in \eqref{eq:DR}.

In addition, there is an infinite number of field strengths associated with fields that contain at least one block of ten antisymmetric indices.
The first two are $B_{10,1,1}$ at level four and $B_{10,4,1}$ at level five, and the associated field strengths are $F_{11,1,1}$ and $F_{11,4,1}$\,.
Unlike the propagating fields, the fields with at least one column of ten or eleven indices have a multiplicity that grows non-polynomially with the level.
There is no closed form expression for their multiplicity \cite{Kleinschmidt:2003jf}.

To study these equations systematically, it is useful to consider the tensor hierarchy algebra.
We already saw that $\cT(\mf{e}_{11})$ admits a bi-grading $\bigoplus_{p,\ell}\big(\cT_p\big)_\ell$\,.
Using instead the $q$-grading \cite{Bossard:2017wxl}
\begin{equation}
    \cT(\mathfrak{e}_{11}) = \bigoplus_{p\in\mathds{Z}} \cT_p =  \bigoplus_{q\in\mathds{Z}} W_q \,,
\end{equation}
where
\begin{equation}
   W_q \coloneqq \bigoplus_{p\in\mathds{Z}} \big( \cT_p \big)_{\ell\,=\,q-\frac32p} \,,
\end{equation}
one computes that $W_0$ is the superalgebra $W(11)$ of superdiffeomorphisms in eleven Grassmann variables $\vartheta_m$\,.
Moreover, each component $W_q$ admits a finite $p$-grading and splits into a finite set of superfields $\Phi(\vartheta)$\,.
In this decomposition, the exterior derivative \eqref{dTH} takes the simple form 
\begin{equation}
    \rd = \frac{\partial\;}{\partial\vartheta_a} \frac{\partial\;}{\partial x^a} \,.
\label{dsuper}
\end{equation}

The components of $\mathcal{T}_0$ at level $\ell\geqslant1$ are those 
of $\mathfrak{e}_{11}$ at the same level, and 
for each level $\ell\geqslant1$ there exists a superfield 
$\mathcal{B}_{a[11]}{}^{\alpha{(\ell)}}(\vartheta)$ with lowest 
components
\begin{align} \label{All9!}
    &\cB_{a[11];}{}^{\alpha_{(\ell)}}(0) = B_{a[11];}{}^{\alpha_{(\ell)}} \in \big(\cT_0\big)_{\ell} \,,
    &F_{a[10];}{}^{\alpha_{(\ell)}} \in \big(\cT_{-1}\big)_{\ell+\frac32} \,,& &\phi_{a[9];}{}^{\alpha_{(\ell)}} \in \big(\cT_{-2}\big)_{\ell+3} \,.
\end{align}
The tensor hierarchy algebra exhibits a duality $\big(\cT_p\big)_\ell\cong\big((\cT_{-2-p})_{-\ell}\big)^*$ about $p=-1$\,.
As a result, there is a dual superfield with highest components
\begin{align}
    &\phi^{\alpha_{(-\ell)}} \in \big(\cT_{-2}\big)_{-\ell} \,,&
    &F_a{}^{\alpha_{(-\ell)}} \in \big(\cT_{-1}\big)_{-\ell-\frac32} \,,&
    &B_{a[2]}{}^{\alpha_{(-\ell)}} \in \big(\cT_0\big)_{-\ell-3} \,.
\end{align}
From this property one can conclude that the components of $F^{I_{\scalebox{0.6}{$(-\ell-3/2)$}}}$ for $\ell\geqslant1$ always include a field $F_a{}^{\alpha_{(-\ell)}}$ in the tensor product representation, with one covariant index and $3\ell$ contravariant indices, while all the other components of $F^{I_{\scalebox{0.6}{$(-\ell-3/2)$}}}$ only carry $3\ell-1$ contravariant indices.
One can always fix conventions so that the other components vanish for fields satisfying the section constraint, and $F_a{}^{\alpha_{(-\ell)}}=\partial_a\phi^{\alpha_{(-\ell)}}$ holds, consistently with the action of the differential \eqref{dsuper} and the Bianchi identity $B_{a[2]}{}^{\alpha_{(-\ell)}}=0$ with 
\begin{equation}
\label{eq:Bianchi2alpha}
    B_{a[2]}{}^{\alpha_{(-\ell)}} = 2\,\partial_{[a_1} F_{a_2]}{}^{\alpha_{(-\ell)}} \,.
\end{equation}
In particular it follows that field strength at level $\ell<-\frac32$ are simply gradients $F_1{}^{9^n,\,\rx}=\partial_1\phi^{9^n,\,\rx}$ for the propagating fields and vanish otherwise.
One finds for example that $F^{1,1}$ and $F^{4,1}$ only depend on higher level derivatives (that vanish in the eleven-dimensional solution to the section constraint) and can only be non-trivial in non-geometric backgrounds \cite{Bossard:2017wxl}.

By duality, $F^{I_{\scalebox{0.6}{$(\ell+3/2)$}}}$ for $\ell\geqslant 1$ must include a field $F_{a[10];}{}^{\alpha_{(\ell)}}$ with $3\ell+10$ covariant indices, and every other component (with at least one column 
of height eleven) vanishes using the duality equations.
Despite the fact that we have a very poor knowledge of the set of irreducible representations appearing in the $\mathrm{GL}(11)$ decomposition of $\mathfrak{e}_{11}$ at high levels, the structure of the tensor hierarchy algebra and the equations $F_{a[10];}{}^{\alpha_{(\ell)}}=0$ strongly suggest that all the non-propagating $E_{11}/K(E_{11})$ fields (with at least one column of height ten) are pure gauge and can be disregarded in the linearised dynamics.
We describe how it should work at low levels in  Appendix~\ref{sec:CountingIR}. 

Iteratively, one deduces that the only propagating fields $\phi^{\alpha_{(\ell)}}$ for $\ell\geqslant 1$ are in the irreducible $\mathrm{GL}(11)$ representations $[9^{\lfloor\frac{\ell-1}{3}\rfloor},\rx]$ \cite{Damour:2002cu,Nicolai:2003fw,Riccioni:2006az} with $\ell=3\lfloor\frac{\ell-1}{3}\rfloor+\ell(\rx)$ for $\ell(\rx)$ defined in \eqref{eq:ell(x)}.
Using this notation, one finds that the field 
$\phi_{a[9];}{}^{\alpha_{(\ell)}}\in\big(\cT_{-2}\big)_{\ell+3}$ 
at level $\ell=3n+\ell(\rx)$ with $n\geqslant 1$ includes all the irreducible representations in the tensor product $[9]\otimes[9^n,\rx]$\,.
The component $[9^{n+1},\rx]$ belongs to $\mathfrak{e}_{11}$ while all the others belong to $L(\Lambda_2)\oplus L(\Lambda_{10})$\,.
Indeed, one checks that
\begin{align}
\label{eq:R2L10}
    R(\Lambda_2) \oplus R(\Lambda_{10}) \supset \null&
    \big([9]\big)_3
    \oplus \big( [10,2] \oplus [11,1] \big)_4
    \oplus \big( [10,5] \oplus [11,4] \big)_5 \nonumber\\
    & \oplus \big( [10,7,1] \oplus [10,8] \oplus [11,6,1] \oplus [11,7] \big)_6 \nonumber\\
    &\hspace{-30mm} \oplus \bigoplus_{n\geqslant1} \big( [10,9^{n-1},8,3] \oplus [10,9^n,2] \oplus [11,9^{n-1},7,3] \oplus [11,9^{n-1},8,2] \oplus [11,9^n,1] \big)_{4+3n} \nonumber\\
    &\hspace{-30mm} \oplus \bigoplus_{n\geqslant1} \big( [10,9^{n-1},8,6] \oplus [10,9^n,5] \oplus [11,9^{n-1},7,6] \oplus [11,9^{n-1},8,5] \oplus [11,9^n,4] \big)_{5+3n} \nonumber\\
    &\hspace{-30mm} \oplus \bigoplus_{n\geqslant1} \big( [10,9^{n-1},8,8,1] \oplus [10,9^n,7,1] \oplus [10,9^n,8] \\
    &\hspace{-2mm} \oplus [11,9^{n-1},8,7,1] \oplus [11,9^{n-1},8,8] \oplus [11,9^{n},6,1] \oplus [11,9^{n},7] \big)_{6+3n} \,.\nonumber
\end{align}
This defines the set of fields $\big\{X^{\ta},X^\Lambda\big\}$ contributing to 
the field strength $F_{10\seco9^n,\,\rx}$ at level $\ell\geqslant \frac52$\,.
More precisely, one checks that all the representations above including an eleven 
in the partition of $3\ell$ appear in both $R(\Lambda_2)$ and $R(\Lambda_{10})$ with 
positive multiplicities, and $F_{10\seco9^n,\,\rx}$ depends on a specific linear 
combination of the corresponding components of $X^{\ta}$ and $X^\Lambda$ in each 
irreducible representation.

The gauge transformation of the field $\phi_{9^n,\,\rx}$ gives the exterior derivative 
of gauge parameters in all the representations with Young diagrams well-included in 
$[9^n,\rx]$, and the gauge invariance of the duality equation therefore requires to 
include a field component of $X^{\tilde{\alpha}}$, $X^\Lambda$, or $U^{\tilde{\Lambda}}$ 
in the representations $[10]\otimes\rY$ for each well-included 
diagram $\rY\prec[9^n,{\rm x}]$ at level $\ell=3n+3+\ell(\rx)$, as was explained in the 
introduction.
Using that 
\begin{equation}
    [9] \otimes [\rY] \;\cong\;
    [9,\rY] \oplus \bigoplus_{\rY'\prec\rY} \big( [10] \otimes [\rY'] \big) \ominus \bigoplus_{\rY''\llcurly\rY} [11,\rY''] \,,
\end{equation}
where the sum over all the doubly well-included Young diagrams cancel the terms in $[10]\otimes[\rY']$ that are not compatible with the symmetry of $[9]$, one obtains that the remaining irreducible representations captured by the field $U^{\tilde{\Lambda}}$ are precisely those associated with the doubly well-included Young diagrams $\rY''\llcurly \rY$.
We find that this is compatible with  $U^{\tilde{\Lambda}}\in L(\Lambda_4)$ as this 
representation includes all doubly well-included Young diagram representations (up to 
the $[11]$ factor that simply accounts for the $\mathrm{GL}(1)$ weight)\footnote{This $\mathrm{GL}(1)$ is simply the $\mathds{R}^+$ factor of $\mathrm{GL}(11)\cong \mathds{R}^+\times\mathrm{SL}(11)$.}
\begin{align}
\label{eq:L4}
    R(\Lambda_{4}) \supset \null& \big([11,7]\big)_6 \oplus \big([11,8,2]\big)_7 \oplus \big([11,8,5]\big)_8 \oplus \big( [11,8,7,1] \oplus [11,8,8] \oplus [11,9,7] \big)_9 \nonumber\\
    & \oplus \bigoplus_{n\geqslant2} \big( [11,9^{n-2},8,8,3] \oplus [11,9^{n-1},8,2] \big)_{4+3n} \nonumber\\
    & \oplus \bigoplus_{n\geqslant2} \big( [11,9^{n-2},8,8,6] \oplus [11,9^{n-1},8,5] \big)_{5+3n} \\
    & \oplus \bigoplus_{n\geqslant2} \big( [11,9^{n-2},8,8,8,1] \oplus [11,9^{n-1},8,7,1] \oplus [11,9^{n-1},8,8] \oplus [11,9^n,7] \big)_{6+3n} \,.\nonumber
\end{align}
For example, the $[7]$ inside $[11,7]$ is doubly well-included in $[8,1]$ and the $[8,2]$ inside $[11,8,2]$ is doubly well-included in $[9,3]$.

We will show that the representations in \eqref{eq:L4} are precisely those that are required to satisfy the generalised Ricci-flat 
equations \eqref{eq:generalisedRicci}.
Note that these representations, together with those in \eqref{eq:R2L10}, 
all appear already in $R(\Lambda_2)$ (with multiplicity $\mu \ge 2$ when they appear in both \eqref{eq:L4} and \eqref{eq:R2L10}) as we show in equation \eqref{eq:R2Big} in 
Appendix~\ref{app:proof_of_spectrum}. If one were merely counting the number of fields in each irreducible representation of $\mathrm{GL}(11)$ required to write the linearised equations, one might conclude that it would have been sufficient to introduce a single constrained field $\chi_M{}^{\tilde{\alpha}}$ in $R(\Lambda_2)$.
However, the structure of the tensor hierarchy algebra and the properties of its 
differential \eqref{dsuper} are such that only the components of $X^{\tilde{\alpha}}$ 
and $X^\Lambda$ in the representations displayed in \eqref{eq:R2L10} can contribute to the field strength $F_{10\seco9^n,\,\rx}$\,.
As a result, we really do need the field components of $U^{\tilde{\Lambda}}$ in the 
representations shown in \eqref{eq:L4}.
We show in Appendix~\ref{sec:CountingIR} that one can also see the need for a field in the representation $R(\Lambda_4)$ if one writes all the field strengths for the non-dynamical fields.

The proof given in Appendix~\ref{app:proof_of_spectrum} relies on the branching of these 
$E_{11}$ modules under $\mathrm{SL}(2)\times E_9$ and the action of the Virasoro generator 
$L_1$ on the lowest weight modules of $E_9$.
The $\mathrm{GL}(11)$ modules depicted in \eqref{eq:R2L10} and \eqref{eq:L4} actually 
appear with a multiplicity growing exponentially with the level, but the components 
relevant to the description of $E_{11}$ exceptional field theory dynamics are 
specifically those for which the $\mathrm{GL}(9)$ highest weight vector is obtained 
from $L_1$.
This suggests that one may be able to understand the infinite sequences of irreducible 
$\mathrm{GL}(11)$ representations of the type $[9^n,\rY]$ from the action of 
$t^{\tilde{\alpha}}$ in the tensor hierarchy algebra.\footnote{Recall that 
$t^{\tilde{\alpha}}$ includes the Virasoro generator $L_1$  when branching 
$\mathcal{T}(\mathfrak{e}_{11})$ under $\mathcal{T}(\mathfrak{e}_9)$.}
By construction, $t^{\tilde{\alpha}}$ acts on $\mathcal{T}_0$\,, and the 
(appropriate) structure coefficients in $\mathcal{T}_0$ give linear maps
\begin{equation}
    L(\Lambda_2)\otimes \mathfrak{e}_{11} \rightarrow \mathfrak{e}_{11} \,, \quad
    L(\Lambda_2)\otimes L(\Lambda_2) \rightarrow L(\Lambda_2) \,, \quad
    L(\Lambda_{2})\otimes L(\Lambda_{10}) \rightarrow L(\Lambda_{10}) \,,
\end{equation} 
that should extend the action of $L_1$ on the $E_{9}$ modules to an action of 
$t_{a[9]} \in L(\Lambda_2)$.
This was analysed for $\mathcal{T}(\mathfrak{e}_{10})$ in \cite{Cederwall:2025muh}.
If true, this underlying algebraic structure might be useful to analyse the dynamics 
of the theory.

\section{Parent Lagrangians for higher gradient dual fields}
\label{sec:Lagrangian}

It was proposed in \cite{Bossard:2021ebg} that the $E_{11}$ pseudo-Lagrangian \eqref{cLc} with the appropriate choice of $c_\ell$ provides a parent Lagrangian for higher gradient dual fields in 11D supergravity.
In this section we write these parent Lagrangians explicitly in the (Gaussian) linearised approximation.
In order to find the explicit form of \eqref{cLc}, we need to compute the squared duality terms, i.e.~the squares of the left-hand sides $\cE_I$ of the duality equations.
For $\ell\geqslant1$ one finds
\begin{multline}
\label{dualitysquare}
    \eta^{I_{(\ell+\scalebox{0.5}{$\frac32$})}J_{(\ell+\scalebox{0.5}{$\frac32$})}} \mathcal{E}_{I_{(\ell+\scalebox{0.5}{$\frac32$})}} \mathcal{E}_{J_{(\ell+\scalebox{0.5}{$\frac32$})}}
    = \eta_{I_{(\ell+\scalebox{0.5}{$\frac32$})}J_{(\ell+\scalebox{0.5}{$\frac32$})}} F^{I_{(\ell+\scalebox{0.5}{$\frac32$})}} F^{J_{(\ell+\scalebox{0.5}{$\frac32$})}}
    + 2\,\Omega_{I_{(-\ell-\scalebox{0.5}{$\frac32$})}J_{(\ell+\scalebox{0.5}{$\frac32$})}} F^{I_{(-\ell-\scalebox{0.5}{$\frac32$})}} F^{J_{(\ell+\scalebox{0.5}{$\frac32$})}} \\
    + \Big( \eta^{ab} \eta_{\alpha_{(-\ell)}\beta_{(-\ell)}} - T_{\alpha_{(-\ell)}}{}^{\!a}{}_{P_{(\ell+3/2)}} T_{\beta_{(-\ell)}}{}^{\!b}{}_{Q_{(\ell+3/2)}}\,\eta^{P_{(\ell+3/2)}Q_{(\ell+3/2)}} \Big) \partial_a \phi^{\alpha_{(-\ell)}} \partial_b \phi^{\beta_{(-\ell)}} \,.
\end{multline} 
The structure coefficients of $E_{11}$ in the $\mathrm{GL}(11)$ level decomposition have only been computed up to level $\ell=4$ in \cite{Bossard:2021ebg}.
We will now show that the last line above is the Maxwell-like kinetic term (up to total derivatives).
Let us first argue why one should expect this.
One can always define the normalisation of the gauge parameters $\xi^M$ such that the linearised gauge transformation of the level $\ell\geqslant1$ fields
\begin{equation}
    \delta \phi_{\alpha_{(-\ell)}} = T_{\alpha_{(-\ell)}}{}^{\!a}{}_{P_{(\ell+3/2)}}  \partial_a \xi^{P_{(\ell+3/2)}}
\end{equation}
can be written for each corresponding representation with Young diagram $\rY$ as
\begin{equation}
    \delta \phi_\rY = \sum_{\rY_i\prec\rY} \rd_i \xi_{\rY_i} \,.
\end{equation}
To obtain this, one needs to prove that all the coefficients 
$T_{\alpha_{(-\ell)}}{}^{\!a}{}_{P_{(\ell+3/2)}}$ are indeed non-zero for the 
corresponding $\mathrm{GL}(11)$ irreducible representations.
We prove this is the case for all $\rY=[9^n,\rx]$ with $\rx\in\{[3],[6],[8,1]\}$ 
and $n\geqslant1$ in Appendix~\ref{app:proof_of_spectrum}.
We normalise the field $\phi^\alpha$ such that the invariant bilinear form 
$\eta_{\alpha\beta}$ has the canonical normalisation for all irreducible representations:
\begin{equation}
    \eta^{a^i[h_i]|_{i=1}^{s(\rY)},\,b^i[h_i]|_{i=1}^{s(\rY)}} = \frac{1}{\prod_{i=1}^{s(\rY)}h_i!} \prod_{i=1}^{s(\rY)} \prod_{j=1}^{h_i} \eta^{a^i_j b^i_j} \,.
\end{equation}
One finds that the operator 
\begin{equation}
    \delta \bar{\xi}{}^{P_{(\ell+3/2)}} = T_{\alpha_{(-\ell)}}{}^{\!a}{}_{Q_{(\ell+3/2)}} \eta^{P_{(\ell+3/2)}Q_{(\ell+3/2)}} \partial_a \phi^{\alpha_{(-\ell)}}
\end{equation}
can be written as a Hermitian conjugate, provided that the normalisation of 
$\eta^{P_{(\ell+3/2)}Q_{(\ell+3/2)}}$ is appropriate for each irreducible 
representation, such that
\begin{equation}
\label{MaxwellSection5}
    M[\phi]^{\alpha_{(-\ell)}} = \eta^{ab} \partial_a \partial_b \phi^{\alpha_{(-\ell)}} - \eta^{\alpha_{(-\ell)}\gamma_{(-\ell)}} T_{\gamma_{(-\ell)}}{}^{\!a}{}_{Q_{(\ell+\scalebox{0.5}{$\frac32$})}} \eta^{P_{(\ell+\scalebox{0.5}{$\frac32$})}Q_{(\ell+\scalebox{0.5}{$\frac32$})}} T_{\beta_{(-\ell)}}{}^{\!b}{}_{P_{(\ell+\scalebox{0.5}{$\frac32$})}} \partial_a \partial_b \phi^{\beta_{(-\ell)}}
\end{equation}
is the Maxwell-like tensor for propagating fields:\footnote{Note that we write 
Maxwell tensors for the $E_{11}$ fields at positive levels since at level zero we 
directly obtain the linearised Einstein tensor for the graviton rather than the Maxwell tensor in 
the covariant equation of motion.}
\begin{align}
    M[A]_{a^i[9]|_{i=1}^n,b[3]} & \coloneqq \Box A_{a^i[9]|_{i=1}^n,b[3]} - 9\sum_{i=1}^n \partial_{a^i} \partial^c A_{a^j[9]|_{j\ne i},a^i[8]c,b[3]} - 3\,\partial_{b} \partial^c A_{a^i[9]|_{i=1}^n,b[2]c} \,, \nonumber\\
    M[A]_{a^i[9]|_{i=1}^n,b[6]} & \coloneqq \Box A_{a^i[9]|_{i=1}^n,b[6]} - 9\sum_{i=1}^n \partial_{a^i} \partial^c A_{a^j[9]|_{j\ne i},a^i[8]c,b[6]} + 6\,\partial_{b} \partial^c A_{a^i[9]|_{i=1}^n,b[5]c} \,, \nonumber\\
    M[h]_{a^i[9]|_{i=1}^n,b[8],c} & \coloneqq \Box h_{a^i[9]|_{i=1}^n,b[8],c} - 9\sum_{i=1}^n \partial_{a^i} \partial^d h_{a^j[9]|_{j\ne i},a^i[8]d,b[8],c} \\
    &\hspace{50mm} + 8\,\partial_{b} \partial^d h_{a^i[9]|_{i=1}^n,b[7]d,c} - \partial_{c} \partial^d h_{a^i[9]|_{i=1}^n,b[8],d} \,.\nonumber
\end{align}
In order to rigorously prove this is the case, one could check the normalisation 
of the bilinear form $\eta^{P_{(\ell+3/2)}Q_{(\ell+3/2)}}$ for each irreducible 
$\mathrm{GL}(11)$ representation using the $\mathfrak{e}_{11}$ algebra, but one cannot do this 
to arbitrary high $\mathrm{GL}(1)$ level.
Assuming the normalisations are all as such, so that the last line in \eqref{dualitysquare} is indeed the Maxwell-like kinetic term, one determines the symmetric and skew bilinear forms $\eta_{I_{(\ell+\scalebox{0.5}{$\frac32$})}J_{(\ell+\scalebox{0.5}{$\frac32$})}}$ and $\Omega_{I_{(-\ell-\scalebox{0.5}{$\frac32$})}J_{(\ell+\scalebox{0.5}{$\frac32$})}}$.
As a non-trivial consistency check, we will find that the expression for $\Omega_{I_{(-\ell-\scalebox{0.5}{$\frac32$})}J_{(\ell+\scalebox{0.5}{$\frac32$})}}$ is indeed consistent with \eqref{BianchiE11}, 
i.e.~the following tensor hierarchy algebra Jacobi identities are satisfied
\begin{subequations}
\begin{align}
    \Omega_{I_{(-\ell-3/2)}J_{(\ell+3/2)}} C^{I_{(-\ell-3/2)}(a}{}_{\alpha_{(-\ell)}} C^{J_{(\ell+3/2)}b)}{}_{\beta_{(\ell+3)}} &= 0 \,,\\
    \Omega_{I_{(-\ell-3/2)}J_{(\ell+3/2)}} C^{I_{(-\ell-3/2)}(a}{}_{\alpha_{(-\ell)}} C^{J_{(\ell+3/2)}b)}{}_{\tb_{(\ell+3)}} &= 0 \,,\\
    \Omega_{I_{(-\ell-3/2)}J_{(\ell+3/2)}} C^{I_{(-\ell-3/2)}(a}{}_{\alpha_{(-\ell)}} C^{J_{(\ell+3/2)}b)}{}_{\Lambda_{(\ell+3)}} &= 0 \,.\label{AlgebraicBianchi}
\end{align}
\end{subequations}
Conversely, these Jacobi identities determine the normalisation of each irreducible component of $\Omega_{I_{(-\ell-\scalebox{0.5}{$\frac32$})}J_{(\ell+\scalebox{0.5}{$\frac32$})}}$ associated with the field strength $F_{a[10];b^i[9]|_{i=1}^n,c[{\rm x}]}$ (up to an overall normalisation) such that the last line of \eqref{dualitysquare} must be the Maxwell-like kinetic term.
This proves that \eqref{MaxwellSection5} is indeed the Maxwell-like tensor.

Since they do not propagate, and since we know very little about their 
representations and multiplicities, we disregard all the non-propagating 
fields in the following.
Equivalently, consider that one has integrated out all the corresponding 
constrained fields.

The Lagrangian $\cL_c$ in \eqref{cLc}, restricted to the propagating fields, decouples in the linearised approximation as the sum of three Lagrangians
\begin{equation}
    \cL_c = \cL_3 + \cL_6 + \cL_{8,1} + \partial (\,\dots) \,,
\end{equation}
up to a total derivative, with 
\begin{multline}
\label{eq:Lag_3}
    \mathcal{L}_3 = -\frac{1-\frac{c_{1/2}}{2}}{48} F_{a[4]} F^{a[4]} + \frac{1}{40\cdot3!} \sum_{n=0}^\infty \frac{c_{3n+5/2}}{(9!)^{n+1}} \bigg( (9n+2) \mathcal{E}_{a[10];b^i[9]|_{i=1}^n,c[3]} \\
    - 90\sum_{i=1}^n \mathcal{E}_{a[9]b^i;b^j[9]|_{j\ne i},b^i[8]a,c[3]} - 30\,\mathcal{E}_{a[9]c;b^i[9]|_{i=1}^n,c[2]a} \bigg) \mathcal{E}^{a[10];b^i[9]|_{i=1}^n,c[3]} \,, 
\end{multline}
for the three-form and its higher gradient duals, followed by
\begin{multline}
    \cL_6 = -\frac{c_{1/2}}{4\cdot7!} F_{a[7]} F^{a[7]} + \frac{1}{40\cdot6!} \sum_{n=0}^\infty \frac{c_{3n+7/2}}{(9!)^{n+1}} \bigg( (9n+5)\cE_{a[10];b^i[9]|_{i=1}^n,c[6]} \\
    - 90\sum_{i=1}^n \cE_{a[9]b^i;b^j[9]|_{j\ne i},b^i[8]a,c[6]} + 60\,\cE_{a[9]c;b^i[9]|_{i=1}^n,c[5]a} \bigg) \cE^{a[10];b^i[9]|_{i=1}^n,c[6]} \,,
\end{multline}
for the six-form and its duals, and lastly
\begin{multline}
    \cL_{8,1} = -\frac14 \big( \partial_a h_{bc} - 2\,\partial_b h_{ac} \big) \partial^a h^{bc}  + \frac14 \big( \partial_a h_b{}^b - 2\,\partial_b h_a{}^b \big) \partial^a h_c{}^c - \frac{c_{3/2}}{4\cdot8!} \cE_{ba[8];a} \cE^{a[9];b} \\
    + \frac1{40\cdot8!} \sum_{n=0}^\infty \frac{c_{3n+9/2}}{(9!)^{n+1}} \bigg( (9n+8)\cE_{a[10];b^i[9]|_{i=1}^n,c[8],d} - 90 \sum_{i=1}^n \cE_{a[9]b^i;b^j[9]|_{j\ne i},b^i[8]a,c[8],d} \\
    + 80\,\cE_{a[9]c;b^i[9]|_{i=1}^n,c[7]a,d} - 10\,\cE_{a[9]d;b^i[9]|_{i=1}^n,c[8],a} \bigg) \cE^{a[10];b^i[9]|_{i=1}^n,c[8],d} \, ,  \label{L81E2}
\end{multline}
for the metric, the dual graviton, and the higher dual gravitons.
The coefficients $c_k$ are all equal to one in the $E_{11}$-invariant pseudo-Lagrangian.
Note that, although  $\mathcal{L}_3$ and $\mathcal{L}_6$ decouple, they describe the same degrees of freedom 
due to the duality equation $F_7=\star F_4$ in \eqref{Duality12}.
This is the only duality equation that is not obtained as an Euler--Lagrange equation. 
In practice we shall either choose to describe the three-form and set $c_{1/2}=0$ or the six-form and set $c_{1/2} = 2$.\footnote{Note that $E_{11}$ exceptional field theory does not provide a parent Lagrangian for the standard electromagnetic duality $F_7=\star F_4$, unless one breaks manifest Lorentz invariance.
This contrasts with the cases of the other dualities, for which there is always a St\"{u}ckelberg field $\chi$\,.}

\subsection{Dual three-form Lagrangians}

In order to obtain a parent Lagrangian for higher gradient dual counterparts of the three-form, we decouple the six-form by setting $c_{1/2}=0$ and we fix $c_{3n+5/2}=2$ for $n\leqslant k$ and zero otherwise.
One then obtains, up to total derivatives, 
\begin{multline}
    \mathcal{L}^\ord{k+1}_3 = \frac{1}{12} \sum_{n=1}^k \frac{1}{(9!)^n} \partial^a A^{b^i[9]|_{i=1}^n,c[3]} \bigg( \partial_a A_{b^i[9]|_{i=1}^{n},c[3]} - 3\,\partial_c A_{b^i[9]|_{i=1}^n,c[2]a} \\
    \hspace{90mm} - 9 \sum_{i=1}^n \partial_{b^i} A_{b^j[9]|_{j\ne i},b^i[8]a,c[3]} \bigg) \\
    + \frac{1}{60} \sum_{n=0}^k \frac{\varepsilon^{a[11]}}{(9!)^{n+1}} A^{b^i[9]|_{i=1}^n,c[3]} \bigg( {-}\partial_a F_{a[10];b^i[9]|_{i=1}^n,c[3]} + 3\,\partial_{c} F_{a[10];b^i[9]|_{i=1}^n,c[2]a} \\
    \hspace{90mm} + 9 \sum_{i=1}^n \partial_{b^i} F_{a[10];b^j[9]|_{j\ne i},b^i[8]a,c[3]} \bigg) \\
    + \frac{1}{120} \sum_{n=0}^k \frac{1}{(9!)^{n+1}} \bigg( (9n+2) F_{a[10];b^i[9]|_{i=1}^n,c[3]} - 30F_{a[9]c;b^i[9]|_{i=1}^n,c[2]a} \\
    - 90\sum_{i=1}^n F_{a[9]b^i;b^j[9]|_{j\ne i},b^i[8]a,c[3]} \bigg) F^{a[10];b^i[9]|_{i=1}^n,c[3]} \,.
\end{multline}
According to the discussion in the introduction of this section, we observe indeed that the first two lines give the Maxwell-like kinetic term for the field $A_{b^i[9]|_{i=1}^{n},c[3]}$ which justifies a posteriori the definition \eqref{eq:Lag_3}.

The Euler--Lagrange equation for $A_3$ gives 
\begin{equation}
\label{eq:EOM_9,3}
    \frac{1}{10!} \varepsilon^{a[11]} \bigg( {-}\partial_a F_{a[10];c[3]}  + 3\,\partial_{c} F_{a[10];c[2]a} \bigg)\approx 0 \,,
\end{equation}
and the Euler--Lagrange equation for the $A_{9^n,3}$ field with $1\leqslant n\leqslant k$ gives
\begin{multline}
\label{eq:EOM_9,...,9,3}
    \frac{1}{10!} \varepsilon^{a[11]} \bigg( {-}\partial_a F_{a[10];b^i[9]|_{i=1}^n,c[3]} + 9\sum_{i=1}^n \partial_{b^i} F_{a[10];b^j[9]|_{j\ne i},b^i[8]a,c[3]} + 3\,\partial_{c} F_{a[10];b^i[9]|_{i=1}^n,c[2]a} \bigg) \\
    \approx M[A]_{b^i[9]|_{i=1}^n,c[3]} - \partial^a F_{a\langle b^1[9];b^i[9]|_{i=2}^n,c[3]\rangle} \,,
\end{multline}
while for $n=k+1$ one finds
\begin{equation}
    \partial^a F_{a\langle b^1[9];b^i[9]|_{i=2}^{k+1},c[3]\rangle} \approx 0 \,.
\end{equation}
The field strength $F_{a[10];b^i[9]|_{i=1}^n,c[3]}$ depends algebraically on the St\"{u}ckelberg field $\chi_{a[10];b^i[9]|_{i=1}^n,c[3]}$ according to the definition \eqref{eq:fieldstrength}.
Therefore, one can integrate $\chi_{a[10];b^i[9]|_{i=1}^k,c[3]}$ out to get the Lagrangian $\mathcal{L}^{\ord{k}}_3$.
Integrating out all the St\"{u}ckelberg fields gives the free field Lagrangian for the three-form gauge field.
It follows that the Euler--Lagrange equations include 
\begin{equation}
    \partial^a F_{a\langle b^1[9];b^i[9]|_{i=2}^{n},c[3]\rangle} \approx 0 \,,
\end{equation}
for all $n\leqslant k+1$\,.
Moreover, substituting the expression \eqref{eq:fieldstrength} 
for the field strength, one obtains that the potential $A_{a[9],b^i[9]|_{i=1}^n,c[3]}$ drops out in \eqref{eq:EOM_9,...,9,3} for all $0\leqslant n\leqslant k$, such that one gets 
\begin{multline}
\label{eq:Bianchi993}
    \partial_a \chi_{a[10];b^i[9]|_{i=1}^n,c[3]} - 9\sum_{i=1}^n \partial_{b^i} \chi_{a[10];b^j[9]|_{j\ne i},b^i[8]a,c[3]} - 3\,\partial_c \chi_{a[10];b^i[9]|_{i=1}^n,c[2]a} \\
    \approx \frac{1}{11} \varepsilon_{a[11]} M[A]_{b^i[9]|_{i=1}^n,c[3]} 
\end{multline}
for all $n\in\{1,\ldots,k\}$, and
\begin{equation}
\label{eq:EL_A3}
    \partial_a \chi_{a[10];c[3]} - 3\,\partial_c \chi_{a[10];c[2]a} 
    \approx0 \,,
\end{equation}
for $n=0$.
As we shall prove in the next section, this set of Bianchi-type equations 
implies that $\chi_{a[10];b^i[9]|_{i=1}^n,c[3]}$ is a total derivative so that 
$A_{a[9],b^i[9]|_{i=1}^n,c[3]}$ is indeed dual to $A_{c[3]}$ and it propagates 
the same degrees of freedom.
One checks that the relative coefficients between the $n+2$ terms in the left-hand 
side of \eqref{eq:Bianchi993} is uniquely determined such that this is the case, 
therefore proving that \eqref{MaxwellSection5} must indeed be the Maxwell-like 
tensor in $E_{11}$ exceptional field theory.

\paragraph{\texorpdfstring{The $k=0$ case.}{The k=0 case.}}

For illustration we review the simplest case of the first higher gradient dual, that was described at the non-linear level in \cite{Bossard:2021ebg}.
The linearised Lagrangian simplifies to
\begin{multline} \label{L31}
    \mathcal{L}_3^\ord{1} = \frac{1}{6\cdot10!} \varepsilon^{a[11]} A^{b[3]}  \Big( {-}\partial_a F_{a[10];b[3]} + 3\,\partial_{b} F_{a[10];b[2]a} \Big) \\
    + \frac{1}{12\cdot10!} \Big( 2F_{a[10];b[3]} - 30F_{a[9]b;b[2]a} \Big) F^{a[10];b[3]} \,,
\end{multline}
with independent fields $\{A_3\,,A_{9,3}\,,\chi_{10\seco3}\}$\,.
The Euler--Lagrange equation for $\chi_{a[10];b[3]}$ is
\begin{equation}
\label{eq:EOM_chi10;3}
    2F_{a[10];b[3]} - 30F_{a[9]b;b[2]a} + \varepsilon_{a[10]c}\,\partial^c A_{b[3]} - 3\,\varepsilon_{a[10]b}\,\partial^c A_{b[2]c} \approx 0 \,,
\end{equation}
and implies the duality equation
\begin{equation}
    \cE_{a[10];b[3]} \coloneqq F_{a[10];b[3]} - \varepsilon_{a[10]c}\,\partial^c A_{b[3]} \approx 0 \, .
\end{equation}
Eliminating $\chi_{a[10];b[3]}$ from the Lagrangian $\cL_3^{\ord{1}}$ yields by construction the Maxwell Lagrangian $\cL_3^{\ord0}=-\frac1{48}(F_4)^2$ for the three-form $A_3$\,.

If instead we keep $\chi_{10\seco3}$ in the Lagrangian, we find that the Euler--Lagrange equation of $A_3$ gives the Bianchi identity \eqref{eq:EL_A3} for $\chi_{10\seco3}$\,.
This Bianchi identity can be identified with the component $B_{a[11];b[3]}=0$ of \eqref{BianchiE11}, and it implies via the Poincar\'e lemma that 
\begin{equation}
\label{eq:chi10;3}
    \chi_{a[10];b[3]} = 3\,\partial_bX_{a[10];b[2]} + 10\,\partial_a \tilde{A}_{a[9],b[3]} \,,
\end{equation}
for a reducible field $X_{10 \seco 2}$\,.
Using a St\"{u}ckelberg transformation, one can absorb $\tilde{A}_{9,3}$ by shifting $A_{9,3}$\,. 
We shall explain the general proof for arbitrary levels in Section~\ref{sec:BI_sols_homog}.

Lastly, the Euler--Lagrange equation for $A_{9,3}$ is
\begin{equation}
    \partial^c F_{c\langle a[9];b[3]\rangle} = \Box A_{a[9],b[3]} - 9\,\partial_a \partial^c A_{a[8]c,b[3]} - 3\,\partial_b \partial^c A_{a[9],b[2]c} + \partial^c \chi_{c\langle a[9];b[3]\rangle} \approx 0 \,.
\end{equation}
If the three-form had already been integrated out, then the $A_{9,3}$ equation becomes
\begin{equation}
\label{eq:A9,3_X10;2}
    \Box A_{a[9],b[3]} - 9\,\partial_a \partial^c A_{a[8]c,b[3]} - 3\,\partial_b \partial^c A_{a[9],b[2]c} + 3\,\partial_{\langle b} \partial^c X_{ca[9];b[2]\rangle} \approx 0 \,.
\end{equation}
Taking a triple trace kills the $X_{10\seco2}$ field, and \eqref{eq:A9,3_X10;2} becomes the higher trace equation of motion $\Tr^4K_{10,4}\approx 0$ in \eqref{eq:Tr4_K10,4}.

As a side remark, after integrating out the three-form, one may consider the Lagrangian for $A_{9,3}$ and $X_{10\seco2}$ only.
By construction, the Euler--Lagrange equation for $X_{10\seco2}$ is the divergence of 
the Euler--Lagrange equation \eqref{eq:EOM_chi10;3} for $\chi_{10\seco3}$ on the $b$ index
\begin{equation}
    2\,\partial^c F_{a[10];b[2]c} - 20\,\partial^c F_{a[9]b;abc} - 10\,\partial^c F_{a[9]c;b[2]a} \approx 0 \,.
\end{equation}
This obviously does not depend on the three-form $A_3$, as it has been 
integrated out, and consistently with the fact that the divergence of 
\eqref{eq:EOM_chi10;3} on a $b$ index makes $A_3$ disappear from it. 
In total, the Euler--Lagrange equations give the (reducible) integrability condition of the duality equation:
\begin{equation}
    \partial^c F_{ca[9];b[3]} = \Box A_{a[9],b[3]} - 9\,\partial_a \partial^c A_{a[8]c,b[3]} + 3\,\partial_b \partial^c X_{ca[9];b[2]} \approx 0 \,.
\end{equation}

Before closing this section, let us compare our parent action with the one originally introduced in \cite{Boulanger:2015mka}.\footnote{Here we correct a typo that appears in equations (3.2.13) and (3.2.14) of the arXiv version of \cite{Boulanger:2015mka}, or (3.51) and (3.52) of the published version.}
They write the Lagrangian 
\begin{equation} 
    \mathcal{L}_{\scalebox{0.6}{BSW}} = - \frac{1}{12} \big( G_{a;b[3]} G^{a;b[3]} - 3\,G^{a;}{}_{ab[2]} G_{c;}{}^{cb[2]} \big) 
    - \frac{1}{6\cdot9!} \varepsilon^{a[11]} G_{a;b[3]} \partial_a Y_{a[9];}{}^{b[3]} \,,  
\end{equation}
where $Y_{a[9];b[3]}$ is a Lagrange multiplier whose Euler-Lagrange equation enforces $G_{a;b[3]}\approx\partial_aA_{b[3]}$\,.
Integrating out $G_{a;b[3]}$ instead, one obtains the Lagrangian 
\begin{equation} 
    \widetilde{\mathcal{L}}_{\scalebox{0.6}{BSW}}= - \frac{1}{12\cdot10!} \Big( G_{a[10];b[3]} G^{a[10];b[3]} - \frac{33}{8}  G_{a[10];ab[2]} G^{a[10];ab[2]} \Big)  \,,
\end{equation}
for the reducible field strength 
\begin{equation}
    G_{a[10];b[3]} = 10\,\partial_a Y_{a[9];b[3]} \,. 
\end{equation}
One finds perfect agreement with \eqref{L31} after integrating out the three-form, upon identifying
\begin{align}
    G_{a[10];b[3]} &= - 2 F_{a[10];b[3]} + 30 F_{a[9]b;b[2]a}\quad \Rightarrow \quad F_{a[10];b[3]} =\frac{5}{8} G_{a[10];b[3]} + \frac{15}{4} G_{a[9]b;b[2]a}  \,, \nonumber \\
    Y_{a[9];b[3]} &= A_{a[9],b[3]} + 3 X_{a[9]b;b[2]} + 27 X_{a[8]b[2];ab} \,.
\end{align} 
Indeed, substituting the expression for $F_{10\seco3}$ in the Lagrangian \eqref{L31}, one obtains
\begin{equation}
\label{L31bis}
    \mathcal{L}_3^\ord{1} =
    - \frac{1}{12\cdot 10! } \Big( G_{a[10];b[3]} G^{a[10];b[3]} - \frac{33}{8} G_{a[10];ab[2]} G^{a[10];ab[2]} \Big) + \frac{1}{6\cdot10!} \varepsilon^{a[11]} \partial_a A^{b[3]} G_{a[10];b[3]} \,.
\end{equation}
We find therefore that the Lagrangian \eqref{L31bis} obtained from $E_{11}$ exceptional field theory $\mathcal{L}[G_{10\seco3},A_3]$ is equivalent to the Lagrangian $\mathcal{L}[G_{1\seco3},Y_{9\seco3}]$ introduced in \cite{Boulanger:2015mka}, where the three-form is now the Lagrange multiplier enforcing $G_{a[10];b[3]}=10\,\partial_aY_{a[9];b[3]}$\,.

\subsection{Dual graviton Lagrangians}
\label{sec:Lagrangian_gravity}

In order to write a parent Lagrangian for higher dual gravitons, we fix $c_{n+3/2}=1$ for $n\leqslant k+1$ and zero otherwise.
This gives the Lagrangian
\begin{multline}
    \mathcal{L}_{8,1}^{\ord{k+1}} = -\frac{1}{4\cdot8!} \Big( \partial_b h_{c[8],a} \partial^a h^{c[8],b} + 2 \chi^{a[9];b} \partial_a h_{a[8],b} + \chi_{bc[8];a} \chi^{ac[8];b} + 2\,\varepsilon^{a[11]} h_a{}^b \partial_a \chi_{a[8]b;a} \Big) \\
    \hspace{-30mm} + \frac{1}{4\cdot8!} \sum_{n=1}^k \frac{1}{(9!)^n} \bigg( \partial_a h_{b^i[9]|_{i=1}^n,c[8],d} - 9\sum_{i=1}^n \partial_{b^i} h_{b^j[9]|_{j\ne i},b^i[8]a,c[8],d} \\
    \hspace{50mm} + 8\,\partial_c h_{b^i[9]|_{i=1}^n,c[7]a,d} - \partial_d h_{b^i[9]|_{i=1}^n,c[8],a} \bigg) \partial^a h^{b^i[9]|_{i=1}^n,c[8],d} \\
    + \frac1{20\cdot8!} \sum_{n=0}^k \frac{\varepsilon^{a[11]}}{(9!)^{n+1}} h^{b^i[9]|_{i=1}^n,c[8],d} \bigg( {-}\partial_a F_{a[10];b^i[9]|_{i=1}^n,c[8],d} + 9 \sum_{i=1}^n \partial_{b^i} F_{a[10];b^j[9]|_{j\ne i},b^i[8]a,c[8],d} \\
    \hspace{50mm} - 8\,\partial_c F_{a[10];b^i[9]|_{i=1}^n,c[7]a,d} + \partial_d F_{a[10];b^i[9]|_{i=1}^n,c[8],a} \bigg) \\
    + \frac1{40\cdot8!} \sum_{n=0}^k \frac{1}{(9!)^{n+1}} \bigg( (9n+8)F_{a[10];b^i[9]|_{i=1}^n,c[8],d} - 90\sum_{i=1}^n F_{a[9]b^i;b^j[9]|_{j\ne i},b^i[8]a,c[8],d} \\
    + 80F_{a[9]c;b^i[9]|_{i=1}^n,c[7]a,d} - 10F_{a[9]d;b^i[9]|_{i=1}^n,c[8],a} \bigg) F^{a[10];b^i[9]|_{i=1}^n,c[8],d} \,,
\end{multline}
where we have used \eqref{Duality12}, \eqref{eq:F2^1_F4_F7_F9;1}, and \eqref{eq:dualityequation}.

The Euler--Lagrange equation for $h_{ab}$ gives 
\begin{equation}
\label{eq:EL_metric}
    \varepsilon_{(b|}{}^{a[10]} \partial_a \chi_{a[8]|c);a} \approx 0 \,,
\end{equation}
and that for $h_{8,1}$ gives 
\begin{equation}
\label{h81EulerLagrange}
    \frac{1}{10!} \varepsilon^{a[11]} \bigg( {-}\partial_a F_{a[10];c[8],d}  - 8\,\partial_{\langle c} F_{a[10];c[7]a,d\rangle}
    + \partial_{\langle d} F_{a[10];c[8]\rangle,a} \bigg) 
    \approx M[h]_{c[8],d} - \partial^a F_{a\langle c[8],d\rangle} \,.
\end{equation}
The Euler--Lagrange equation for the $h_{9^n,8,1}$ field with $1\leqslant n\leqslant k$ is
\begin{multline} \label{h9981EulerLagrange}
    \frac{1}{10!} \varepsilon^{a[11]} \bigg( {-}\partial_a F_{a[10];b^i[9]|_{i=1}^n,c[8],d} + 9\sum_{i=1}^n \partial_{\langle b^i} F_{a[10];b^j[9]|_{j\ne i},b^i[8]a,c[8],d\rangle}  - 8\,\partial_{\langle c} F_{a[10];b^i[9]|_{i=1}^n,c[7]a,d\rangle} \\
    + \partial_{\langle d} F_{a[10];b^i[9]|_{i=1}^n,c[8]\rangle,a} \bigg) 
    \approx M[h]_{b^i[9]|_{i=1}^n,c[8],d}-\partial^a F_{a\langle b^1[9];b^i[9]|_{i=2}^n,c[8],d\rangle} \,,
\end{multline}
while for $n=k+1$ we obtain
\begin{equation}
\label{eq:Lag_8,1_EOM_n=k}
    \partial^a F_{a\langle b^1[9];b^i[9]|_{i=2}^{k+1},c[8],d\rangle} \approx 0 \,.
\end{equation}
Integrating out the St\"{u}ckelberg field $\chi_{a[10];b^i[9]|_{i=1}^k,c[8],d}$ gives $\mathcal{L}_{8,1}^{\ord{k}}$ such that, recursively,
\begin{equation}
\label{eq:Lag_9,...,8,1_EOM}
    \partial^a F_{a\langle b^1[9];b^i[9]|_{i=2}^{n},c[8],d\rangle} \approx 0
\end{equation}
is an Euler--Lagrange equation for all $n\leqslant k+1$ and so is
\begin{equation}
\label{eq:Lag_8,1_EOM}
    \partial^a F_{a\langle c[8];d\rangle} \approx 0\; . 
\end{equation}
It follows that the we have the equation 
\begin{multline}
\label{IntegrabilityChiGravity}
    \partial_a \chi_{a[10];b^i[9]|_{i=1}^n,c[8],d} - 9\sum_{i=1}^n \partial_{\langle b^i} \chi_{a[10];b^j[9]|_{j\ne i},b^i[8]a,c[8],d\rangle} + 8\,\partial_{\langle c} \chi_{a[10];b^i[9]|_{i=1}^n,c[7]a,d\rangle} \\
    - \partial_{\langle d} \chi_{a[10];b^i[9]|_{i=1}^n,c[8]\rangle,a}
    \approx \frac1{11} \varepsilon_{a[11]} M[h]_{b^i[9]|_{i=1}^n,c[8],d} \,.
\end{multline}
As we shall show in the next section, this set of Bianchi-type equations allows to prove that $\chi_{a[10];b^i[9]|_{i=1}^n,c[8],d}$ is a total derivative such that the field $h_{a[9],b^i[9]|_{i=1}^n,c[8],d}$ is indeed dual to $h_{c[8],d}$ and therefore propagates the same degrees of freedom as the metric field $h_{ab}$\,.
This proves a posteriori that the form of the `squared duality terms' in \eqref{L81E2} is indeed determined such that \eqref{AlgebraicBianchi} holds.

\paragraph{\texorpdfstring{The $k=0$ case.}{The k=0 case.}}

As before, we first consider the simplest example of a Lagrangian in the dual gravity sector which features a higher gradient dual field.
The independent fields in $\cL_{8,1}^{\ord{1}}$ are the (linearised) metric $h_{ab}$\,, the dual graviton $h_{8,1}$\,, the higher dual graviton $h_{9,8,1}$\,, and two reducible extra fields $\chi_{9\seco1}$ and $\chi_{10\seco8,1}$\,.
The Lagrangian is given by
\begin{multline}
    \mathcal{L}_{8,1}^{\ord{1}} = - \frac{1}{4\cdot8!} \partial_b h_{c[8],a} \partial^a h^{c[8],b} - \frac1{2\cdot8!} \chi^{a[9];b} \partial_a h_{a[8],b} - \frac{1}{4\cdot8!} \chi_{ba[8];a} \chi^{a[9];b} - \frac{1}{2\cdot8!} \varepsilon^{a[11]} h_a{}^b \partial_a \chi_{a[8]b;a} \\
    + \frac1{20\cdot8!\cdot9!} \varepsilon^{a[11]} h^{b[8],c} \Big( {-}\partial_a F_{a[10];b[8],c} - 8\,\partial_b F_{a[10];b[7]a,c} + \partial_c F_{a[10];b[8],a} \Big) \\
    + \frac1{40\cdot8!\cdot9!} \Big( 8F_{a[10];b[8],c} + 80F_{a[9]b;b[7]a,c} - 10F_{a[9]c;b[8],a} \Big) F^{a[10];b[8],c} \,.
\end{multline}
The Euler--Lagrange equations for $\chi_{9\seco1}$ and $\chi_{10\seco8,1}$ are equivalent to the duality equations
\begin{align}
\label{eq:DR_E9;1_E10;8,1}
    \cE_{a[9];b} &= F_{a[9];b} + \varepsilon_{a[9]}{}^{c[2]} \partial_c h_{cb} \approx 0 \,,&
    \cE_{a[10];b[8],c} &= F_{a[10];b[8],c} - \varepsilon_{a[10]}{}^d \partial_d h_{b[8],c} \approx 0 \,,
\end{align}
and integrating out these fields gives back the Fierz--Pauli 
Lagrangian $\cL_{8,1}^\ord{0}$\,.

Varying with respect to the metric field first  gives instead the Euler--Lagrange equation 
\eqref{eq:EL_metric} that imposes via the Poincar\'e lemma that
\begin{equation}
\label{eq:chi9;1}
    \chi_{a[9];b} = \partial_b X_{a[9]} + 10\,\partial_a \tilde{h}_{a[8],b} \,,
\end{equation}
where one can set $\tilde{h}_{8,1}=0$ such that one obtains the second order duality equation 
\begin{equation}
    2\,\partial_b \cE_{a[9];b} = 18\,\partial_a \partial_b h_{a[8],b} + 2\,\varepsilon_{a[9]}{}^{c[2]} \partial_c \partial_b h_{cb} \approx 0 \,,
\end{equation}
and the dual graviton wave equation 
\begin{equation}
\label{DualGraviton}
    \partial^a \cE_{a\langle b[8];c\rangle} - \partial_{\langle c} \cE_{b[8]\rangle a;}{}^a = M[h]_{b[8],c} - 8\,\partial_c \partial_b h_{b[7]a,}{}^a \approx 0 \,.
\end{equation}
The Euler--Lagrange equation for $h_{8,1}$ gives \eqref{h81EulerLagrange}, while the divergence of  the higher dual graviton duality equation \eqref{eq:DR_E9;1_E10;8,1} on the $a$ index gives \eqref{eq:Lag_8,1_EOM}, hence we obtain \eqref{IntegrabilityChiGravity} for $n=0$, i.e.
\begin{equation}
\label{eq:EL_h8,1_alt}
    \partial_a \chi_{a[10];b[8],c} + 8\,\partial_b \chi_{a[10];b[7]a,c} - \partial_c \chi_{a[10];b[8],a}
    \approx \frac1{11} \varepsilon_{a[11]} M[h]_{b[8],c}
    \approx \frac8{11} \varepsilon_{a[11]}\,\partial_c \partial_b h_{b[7]d,}{}^d \,,    
\end{equation}
where we have used the wave equation for the dual graviton in the second step.
We will show in Section~\ref{sec:BI_sols_inhomog} that this equation implies that 
$\chi_{a[10];b[8],c}$ is a total derivative 
\begin{equation}
\label{Chi1081} 
    \chi_{a[10];b[8],c} = 8\,\partial_b X_{a[10];b[7],c} + \partial_{\langle c} X_{a[10];b[8]\rangle} \,,
\end{equation}
for which the following seven-form irreducible component of the $X$ fields 
is determined by the trace of the dual graviton as 
\begin{equation}
\label{eq:EL_h8,1_alt3}
   \partial_b \partial_c \Big( X_{a[10];b[7]a} -X_{a[10];b[7],a}  \Big) \approx \frac8{11} \varepsilon_{a[11]} \, \partial_c \partial_b h_{b[7]a,}{}^a \,.
\end{equation}
According to the discussion of Section~\ref{GL11E11}, we conclude that this specific combination is the extra field $U^{\tilde{\Lambda}}\in L(\Lambda_4)$, that does not belong to the tensor hierarchy algebra, i.e.
\begin{equation}
    X_{a[10];b[7]a} - X_{a[10];b[7],a} = -\frac{8}{11} U_{a[11],b[7]} \,.
\end{equation} 
Equation \eqref{eq:EL_h8,1_alt3} shows that this $L(\Lambda_4)$ field is identified on-shell with (minus) the trace of the dual graviton up to a total derivative.

Using \eqref{Chi1081} in the higher duality equation \eqref{eq:DR_E9;1_E10;8,1}, one obtains the covariant higher duality equation
\begin{equation}
    18\,\partial_b \partial_c \cE_{a[10];b[8],c}
    = 180\,\partial_a \partial_b \partial_c h_{a[9],b[8],c} - 18\,\varepsilon_{a[10]}{}^d \partial_d \partial_b \partial_c h_{b[8],c} \approx 0 \,,
\end{equation}
showing that $h_{a[9],b[8],c}$ indeed propagates the same degrees of freedom as the metric field.

\section{Solving the Bianchi identities and Ricci-flat equations}
\label{sec:GradientDuals}

In order to show that the equations of motion imply the higher duality equations between the fields 
and their higher dual counterparts, we must use the generalised Ricci-flat equation 
\eqref{eq:generalisedRicci} and equation \eqref{BianchiE11} 
to demonstrate that the constrained fields are total derivatives.
Our strategy is to proceed by induction.
Assuming that the Maxwell tensor of the propagating field at level $\ell$ is a double curl, we solve equation  \eqref{LENexpand} for the constrained fields at level $\ell+3$\,.
Combined with the duality equation \eqref{eq:dualityequation}, this yields the covariant duality 
equations \eqref{GravityHigherDuality} and \eqref{FormHigherDuality} for the fields at level 
$\ell+3$\,.
We then show that these duality equations imply, in turn, that the Maxwell-like tensor of the field 
at level $\ell+3$ is itself a double curl.

To initiate the induction one uses the fact that the Maxwell tensors for the three-form and six-form 
are by definition the left-hand sides of their equations of motion:
\begin{align}
\label{MaxwellEq}
    M[A]_{a[3]} &= \partial^b F_{ba[3]} \approx 0 \,,&
    M[A]_{a[6]} &= \partial^b F_{ba[6]} \approx 0 \,.
\end{align}
For the dual graviton, only the Labastida tensor vanishes on-shell, which implies
that the Maxwell tensor is the double curl of its trace on shell
\begin{equation}
\label{DualGraMax}
    M[h]_{a[8],b} \approx 8\,\partial_a \partial_b h_{a[7]c,}{}^c \,.
\end{equation}

\subsection{Bianchi-type equations}
\label{sec:BI_sols_homog}

In this subsection we solve the covariant equation of motion \eqref{LENexpand} assuming \eqref{LinearisedRicciU} is satisfied up to level $\ell=3+3n$ for $n\geqslant0$, i.e.~that the Maxwell tensor is a double curl
\begin{equation}
\label{M=ddU}
    M[\phi]_{9^n,\,\rx} \approx \sum_{\rY_{i,j}\llcurly[9^n,\,\rx]} \big( \rd_i \rd_j \tU_{\rY_{i,j}} \big)_{[9^n,\,\rx]} \,,
\end{equation}
where $\phi_{9^n,\,\rx}$ is a propagating $E_{11}$ field.

\subsubsection*{\texorpdfstring{Level $\ell=4+3n$ for $n\geqslant 0$}{Level l=4+3n for n>-1}}

For the higher dual three-forms, the Maxwell tensor \eqref{M=ddU} 
reads\footnote{For $n=0$ one understands that $M_{c[3]}\approx 0$ according to 
\eqref{MaxwellEq}, and for $n=1$ the first line vanishes and we are left with 
$M_{b[9],c[3]}\approx27\,\partial_{b}\partial_{c}\tU_{b[8],c[2]}\approx27\,\partial_b\partial_cA_{b[8]}{}^d{}_{,c[2]d}+\varepsilon_{b[9]}{}^{de}\partial_dF_{c[3]e}$\,. 
Our notation is such that $\sum_{i\neq j}$ denotes a double sum, 
over all values of $i$ and $j$ from $1$ to $n$ with $i\ne j$\,. 
In our subscript $|_{k\neq i,j}$ we mean $k$ takes values from $1$ to $n$ except 
$i$ and $j$\,.}
\begin{multline}
\label{M=ddU,3form}
    M_{b[9]|_{i=1}^{n},c[3]} \approx 81 \sum_{i\ne j} \partial_{b^i} \partial_{b^j} \Big( \tfrac12\,\tU_{b^k[9]|_{k\neq i,j},b^i[8],b^j[8],c[3]} + \tfrac{3}{8}\,\tU_{b^k[9]|_{k\neq i,j},b^i[8]c,b^j[8],c[2]} \Big) \\
    + 27 \sum_{i=1}^n \partial_{b^i} \partial_{c} \tU_{b[9]|_{j\neq i},b^i[8],c[2]} \,,
\end{multline}
and the Bianchi-type equation for the field strength  $F_{10\seco9^n,3}$ is
\begin{multline}
\label{eq:HoBianchi3}
    \partial_a F_{a[10];b^i[9]|_{i=1}^n,c[3]} - 9 \sum_{i=1}^n \partial_{b^i} F_{a[10];b^j[9]|_{j\neq i},b^i[8]a,c[3]} - 3\,\partial_c F_{a[10];b^i[9]|_{i=1}^n,c[2]a} \\
    \approx \frac1{11} \varepsilon_{a[11]} \bigg( 81 \sum_{i\ne j}  \partial_{b^i} \partial_{b^j} \Big( \tfrac12\,\tU_{b^k[9]|_{k\neq i,j},b^i[8],b^j[8],c[3]} + \tfrac{3}{8}\,\tU_{b^k[9]|_{k\neq i,j},b^i[8]c,b^j[8],c[2]} \Big) \\
    + 27 \sum_{i=1}^n \partial_{b^i} \partial_{c} \tU_{b[9]|_{j\neq i},b^i[8],c[2]} \bigg) \,.
\end{multline}
In order to solve this equation, 
it is convenient to decompose the field strength into irreducible representations as 
$F_{10\seco9^n,3}=F_{10,9^n,3}+F_{11,9^{n-1},8,3}+F_{11,9^{n},2}$\,.
Equation \eqref{eq:HoBianchi3} does not trivially give zero on any of these representations, 
so the solution must at least be a first derivative.
We write the Ansatz for each $F$ as the sum of curls of fields, which by $\mathrm{GL}(11)$ 
covariance must be in irreducible representations associated with well-included Young diagrams 
according to
\begin{equation}
\label{FAnsatz} 
    F_\rY = \sum_{\rY_i\prec\rY} \big( \rd_i X_{\rY_i} \big)_\rY \,.
\end{equation}
One checks that the set of diagrams $\rY_i$ well-included in $[11,9^{n-1},8,3]$ and $[11,9^{n},2]$ are, respectively, the irreducible representations that appear in the reducible $[10\,;9^{n-1},8,3]$ and $[10\,;9^{n},2]$\,.
The diagrams well-included in $[10,9^n,3]$ are defined similarly as the irreducible representations inside $[9\,;9^n,3]$ except those with a column of height eleven.

We define accordingly the general Ansatz for $F_{10\seco9^n,3}$ as
\begin{multline}
\label{eq:Fansatz3}
    F_{a[10];b^i[9]|_{i=1}^n,c[3]} = 10\,\partial_a A_{a[9];b^i[9]|_{i=1}^n,c[3]} + 9 \sum_{i=1}^n \partial_{b^i} X^{\ord{1}}_{a[10];b^j[9]|_{j\ne i},b^i[8],c[3]} \\
    + \frac{21}{n+7} \partial_c X^{\ord{2}}_{a[10];b^i[9]|_{i=1}^n,c[2]} + \frac{27}{n+7} \sum_{i=1}^n \partial_{b^i} X^{\ord{2}}_{a[10];b^j[9]|_{j\ne i},b^i[8]c,c[2]} \,, 
\end{multline}
where $A_{a[9];b^i[9]|_{i=1}^n,c[3]}$ includes the irreducibles 
$[9^{n+1},3]\oplus[10,9^{n-1},8,3]\oplus[10,9^n,2]$, but we can assume it 
\emph{does not} include $[11,9^{n-1},7,3]\oplus[11,9^{n-1},8,2]\oplus[11,9^{n},1]$ 
as they are already included in $X^{\ord{1}}$ and $X^{\ord{2}}$.

By construction, the non-vanishing terms on the right-hand side of the Bianchi-type equation \eqref{eq:HoBianchi3} are double curls of fields 
in irreducible representations associated with Young diagrams that are each 
doubly well-included in $[11,9^n,3]$, i.e.~$[10,9^{n-1},8,3]$, $[10,9^n,2]$, 
$[11,9^{n-2},8,8,3]$, and $[11,9^{n-1},8,2]$.
We find accordingly that 
\begin{multline} \label{UwillSolve}
     \partial_a F_{a[10];b^i[9]|_{i=1}^n,c[3]} - 9 \sum_{i=1}^n \partial_{b^i} F_{a[10];b^j[9]|_{j\neq i},b^i[8]a,c[3]} - 3\,\partial_c F_{a[10];b^i[9]|_{i=1}^n,c[2]a} \\
     = -90 \sum_{i=1}^n \partial_a \partial_{b^i} A_{a[9];b^j[9]|_{j\ne i},b^i[8]a,c[3]} - 30\,\partial_a \partial_c A_{a[9];b^i[9]|_{i=1}^n,c[2]a} - 27 \sum_{i=1}^n \partial_{b^i} \partial_{c} U_{a[11],b[9]|_{j\neq i},b^i[8],c[2]} \\
     -81 \sum_{i\ne j} \partial_{b^i} \partial_{b^j} \Big( \tfrac12\,U_{a[11],b^k[9]|_{k\neq i,j},b^i[8],b^j[8],c[3]} + \tfrac{3}{8}\,U_{a[11],b^k[9]|_{k\neq i,j},b^i[8]c,b^j[8],c[2]} \Big) \,, 
\end{multline}
where the $[11,9^{n-2},8,8,3]$ and $[11,9^{n-1},8,2]$ irreducible fields are
\begin{subequations}
\label{Yall} 
\begin{multline}
\label{Y883}
    U_{a[11],b^i[9]|_{i=1}^{n-2},c^1[8],c^2[8],d[3]} \coloneqq  X^{\ord{1}}_{a[10];b^i[9]|_{i=1}^{n-2},c^1[8]a,c^2[8],d[3]} \\
    - \frac{3}{8} X^{\ord{1}}_{a[10];b^i[9]|_{i=1}^{n-2},c^1[8]d,c^2[8],d[2]a} + \big(c^1\leftrightarrow c^2\big) \,,
\end{multline}
\vspace{-8mm}
\begin{multline}
\label{Y982}
    U_{a[11],b^i[9]|_{i=1}^{n-1},c[8],d[2]} \coloneqq  X^{\ord{1}}_{a[10];b^i[9]|_{i=1}^{n-1},c[8],d[2]a} \\
    + \frac{8}{n+7} X^{\ord{2}}_{a[10];b^i[9]|_{i=1}^{n-1},c[8]a,d[2]} + \frac{2}{n+7} X^{\ord{2}}_{a[10];b^i[9]|_{i=1}^{},c[8]d,da} \,.
\end{multline}
\end{subequations}
They must be equal to $-\frac1{11} \varepsilon_{11}\tU$ up to total derivatives, while the components of $A_{a[9]\seco b^i[9]|_{i=1}^n,c[3]}$ in the irreducible representations $[10,9^{n-1},8,3]$ and $[10,9^{n},2]$ are necessarily curls themselves in order to solve \eqref{eq:HoBianchi3}.
Including one more curl in the Ansatz \eqref{eq:Fansatz3} for these fields, one checks that they can then be absorbed by a redefinition of $X^{\ord{1}}_{10\seco9^{n-1},8,3}$ and $X^{\ord{2}}_{10\seco9^{n},2}$\,, so we may as well set $A_{a[9]\seco b^i[9]|_{i=1}^n,c[3]}$ equal to the irreducible $A_{a[9],b^i[9]|_{i=1}^n,c[3]}$ field.
We find therefore precisely the field content predicted by $E_{11}$ exceptional field theory, 
where $A_{a[9],b^i[9]|_{i=1}^n,c[3]}$ belongs to $E_{11}$ and the constrained field $\chi$ 
must be a total derivative as
\begin{multline}
\label{eq:Chi=dX3}
    \chi_{a[10];b^i[9]|_{i=1}^n,c[3]} = 9 \sum_{i=1}^n \partial_{b^i} 
    X^{\ord{1}}{}_{a[10];b^j[9]|_{j\ne i},b^i[8],c[3]} \\
    + \frac{21}{n+7}\, \partial_c X^{\ord{2}}{}_{a[10];b^i[9]|_{i=1}^n,c[2]} + \frac{27}{n+7}\, 
    \sum_{i=1}^n \partial_{b^i} X^{\ord{2}}{}_{a[10];b^j[9]|_{j\ne i},b^i[8]c,c[2]} \,.
\end{multline}
The $U$ fields as defined in \eqref{Yall} in terms of $X^{\ord{1}}$ and $X^{\ord{2}}$ are recognised to be the additional fields in $L(\Lambda_4)$ that do not cancel out in the left-hand side of \eqref{LENexpand} as discussed in Section \ref{GL11E11}.

The general solution to equation \eqref{eq:HoBianchi3} for the $U$ fields,
taking into account the Ans\"atze \eqref{eq:Fansatz3} and \eqref{Yall},
according to the generalised Poincar\'e lemma, is given by
\begin{subequations}
\begin{multline}
    U_{11,9^{n-1},8,2} = -\frac1{11} \varepsilon_{11} \tU_{9^{n-1},8,2} + D^{n+1}_{(1)} Z_{11,8^{n-1},7,1} + D^{n+1}_{(2)} Z_{10,9,8^{n-2},7,1} \\
    + D^{n+1}_{(n+1)} Z_{10,8^n,1} + D^{n+1}_{(n+2)} Z_{10,8^{n-1},7,2} \,,
\end{multline}
\vspace{-8mm}
\begin{multline}
    U_{11,9^{n-2},8,8,3} = -\frac1{11} \varepsilon_{11} \tU_{9^{n-1},8,2}+D^{n+1}_{(1)} Z_{11,8^{n-2},7,7,2} + D^{n+1}_{(2)} Z_{10,9,8^{n-3},7,7,2} \\
    + D^{n+1}_{(n)} Z_{10,8^{n-1},7,2} + D^{n+1}_{(n+2)} Z_{10,8^{n-2},7,7,3} \,,
\end{multline}
\end{subequations}
where it is understood that the right-hand sides are projected onto the symmetry types of the left-hand sides.
We defined $D^{s-1}_{(i)}\coloneqq\rd_1\dots\rd_{i-1}\rd_{i+1}\dots\rd_s$ in equation \eqref{eq:almost_curvatures}.

This concludes the proof of \eqref{LinearisedRicciU} and \eqref{LinearChi} for all levels $\ell=4+3n$, assuming the Ricci-flat equation \eqref{M=ddU} is satisfied for some $\tU$ field for all levels $\ell=1+3n$\,.
The proof is identical for $\ell=5+3n$ so we shall directly proceed to the case of the higher dual graviton.

\subsubsection*{\texorpdfstring{Level $\ell=6+3n$ for $n\geqslant 0$}{Level l=6+3n for n>-1}}

The proof is in fact also identical for the higher dual gravitons, but involves more irreducible representations because the dual graviton is already in a mixed-symmetry irreducible representation.
In this case, \eqref{M=ddU} takes the explicit form
\begin{multline}
\label{HigherDGMaxwell}
    M_{b[9]|_{i=1}^{n},c[8],d} \approx \sum_{i\ne j} \partial_{b^i} \partial_{b^j} \Big( \tfrac12\,\tU_{b^k[9]|_{k\neq i,j},b^i[8],b^j[8],c[8],d} + 3\,\tU_{b^k[9]|_{k\neq i,j},b^i[8]c,b^j[8],c[7],d} \\
    + \frac{81}{88} \tU_{b^k[9]|_{k\neq i,j},b^i[8]d,b^j[8],c[8]} - \frac{9}{88} \tU_{b^k[9]|_{k\neq i,j},c[8]d,b^i[8],b^j[8]} + \frac{81}{20} \tU_{b^k[9]|_{k\neq i,j},b^i[8]d,b^j[8]c,c[7]} \Big) \\
    \hspace{-50mm} + \sum_{i=1}^n \partial_{b^i} \partial_{c} \Big( \tU_{b[9]|_{j\neq i},b^i[8],c[7],d} - \tU_{b[9]|_{j\neq i},b^i[8],c[7]d} \\
    \hspace{50mm} + \frac{27}{20} \tU_{b[9]|_{j\neq i},b^i[8]d,c[7]} - \frac{63}{20} \tU_{b[9]|_{j\neq i},b^i[8]c,c[6]d} \Big) \\
    + \sum_{i=1}^n \partial_{b^i} \partial_{d} \Big( \tU_{b[9]|_{j\neq i},b^i[8],c[8]} + \frac{9}{2} \tU_{b[9]|_{j\neq i},b^i[8]c,c[7]} \Big) + \partial_{c} \partial_{d} \tU_{b[9]|_{i=1}^{n},c[7]} \,.
\end{multline}
As for the higher dual three-forms, in order to solve the Bianchi identity 
\begin{multline}
\label{eq:HoBianchi8,1}
    \partial_a F_{a[10];b^i[9]|_{i=1}^n,c[8],d} - 9 \sum_{i=1}^n \partial_{b^i} F_{a[10];b^j[9]|_{j\neq i},b^i[8]a,c[8],d} \\
    +8\,\partial_c F_{a[10];b^i[9]|_{i=1}^n,c[7]a,d} - \partial_d F_{a[10];b^i[9]|_{i=1}^n,c[8],a} = \frac{1}{11} \varepsilon_{a[11]} M_{b[9]|_{i=1}^{n},c[8],d} \,,
\end{multline}
assuming \eqref{HigherDGMaxwell}, we take an Ansatz for each irreducible component of $F_{10\seco9^n,8,1}$ as a sum of curls \eqref{FAnsatz}.
One has $F_{10\seco9^n,8,1}=F_{10,9^n,8,1}+F_{11,9^{n-1},8,8,1}+F_{11,9^{n},7,1}+F_{11,9^{n},8}$\,.
Well-included diagrams in $[11,9^{n-1},8,8,1]$, $[11,9^{n},7,1]$, and $[11,9^{n},8]$ are, respectively, the irreducible representations appearing in the reducible $[10\,;9^{n-1},8,8,1]$\,, $[10\,;9^{n},7,1]$\,, and $[10\,;9^{n},8]$, while diagrams well-included in $[10,9^n,8,1]$ can be defined as the irreducible components of $[9\,;9^n,8,1]$ excluding those with a column of height eleven.
Accordingly, we write the Ansatz 
\begin{multline}
\label{eq:Fansatz8,1}
    F_{a[10];b^i[9]|_{i=1}^{n},c[8],d} \coloneqq 10\,\partial_a h_{a[9];b^i[9]|_{i=1}^{n},c[8],d}
    + 9\sum_{i=1}^n \partial_{b^i} X^{\ord{1}}_{a[10];b^j[9]|_{j\ne i},b^i[8],c[8],d} \\
    + 8\,\partial_{\langle c} X^{\ord{2}}_{a[10];b^i[9]|_{i=1}^{n},c[7],d\rangle} + \partial_{\langle d} X^{\ord{3}}_{a[10];b^i[9]|_{i=1}^n,c[8]\rangle} \,,
\end{multline}
where $h_{a[9];b^i[9]|_{i=1}^{n},c[8],d}$ includes $[9^{n+1},8,1]\oplus[10,9^{n-1},8,8,1]\oplus[10,9^n,7,1]\oplus[10,9^n,8]$ but \emph{does not} include $[11,9^{n-1},8,8]\oplus[11,9^{n-1},8,7,1]\oplus[11,9^{n},6,1]\oplus[11,9^{n},7]$ by definition.
We used the explicit projections
\begin{multline}
\label{eq:F10;9,...,9,8,1_X(2)}
    8\,\partial_{\langle c} X^{\ord{2}}{}_{a[10];b^i[9]_{i=1}^{n},c[7],d\rangle} 
    = \frac{16}{n+2}\, \partial_c X^{\ord{2}}{}_{a[10];b^i[9]_{i=1}^{n},c[7],d} \\
    + \frac{72}{n+2} \sum_{i=1}^n 
    \partial_{b^i} X^{\ord{2}}{}_{a[10];b^j[9]_{j\neq i},b^i[8]c,c[7],d} \,,
\end{multline} 
\vspace{-5mm}
\begin{multline}
\label{eq:F10;9,...,9,8,1_X(3)}
    \partial_{\langle d} X^{\ord{3}}{}_{a[10];b^i[9]_{i=1}^n,c[8]\rangle} = 
    \frac{80}{9(n+10)}\, \partial_{d} X^{\ord{3}}{}_{a[10];b^i[9]_{i=1}^n,c[8]} 
    - \frac{80}{9(n+10)} \,\partial_{c} X^{\ord{3}}{}_{a[10];b^i[9]_{i=1}^n,c[7]d} \\
    + \frac{9}{n+10}\, \sum_{i=1}^n \partial_{b^i} 
    X^{\ord{3}}{}_{a[10];b^j[9]_{j\neq i},b^i[8]d,c[8]} 
    - \frac{1}{n+10}\, \sum_{i=1}^n \partial_{b^i} 
    X^{\ord{3}}{}_{a[10];b^j[9]_{j\neq i},c[8]d,b^i[8]} \,.
\end{multline}
Once again, the non-vanishing terms in the Bianchi-type equation must be double 
curls of fields $\tU$ in irreducible representations associated with Young diagrams 
doubly well-included in $[11,9^n,3]$, and one finds 
\begin{multline}
\label{eq:BianchiCheck6}
    \partial_a F_{a[10];b^i[9]|_{i=1}^n,c[8],d} - 9 \sum_{i=1}^n \partial_{b^i} F_{a[10];b^j[9]|_{j\neq i},b^i[8]a,c[8],d} \\
    \hspace{70mm} +8\,\partial_c F_{a[10];b^i[9]|_{i=1}^n,c[7]a,d} - \partial_d F_{a[10];b^i[9]|_{i=1}^n,c[8],a} \\
    = -\sum_{i\ne j} \partial_{b^i} \partial_{b^j} \Big( \tfrac12\,U_{a[11],b^k[9]|_{k\neq i,j},b^i[8],b^j[8],c[8],d} + 3\,U_{a[11],b^k[9]|_{k\neq i,j},b^i[8]c,b^j[8],c[7],d} \hspace{30mm} \\
    + \frac{81}{88} U_{a[11],b^k[9]|_{k\neq i,j},b^i[8]d,b^j[8],c[8]} - \frac{9}{88} U_{a[11],b^k[9]|_{k\neq i,j},c[8]d,b^i[8],b^j[8]} + \frac{81}{20} U_{a[11],b^k[9]|_{k\neq i,j},b^i[8]d,b^j[8]c,c[7]} \Big) \\
    - \sum_{i=1}^n \partial_{b^i} \partial_c \Big( U_{a[11],b[9]|_{j\neq i},b^i[8],c[7],d} - U_{a[11],b[9]|_{j\neq i},b^i[8],c[7]d} \hspace{60mm} \\
    \hspace{60mm} + \frac{27}{20} U_{a[11],b[9]|_{j\neq i},b^i[8]d,c[7]} - \frac{63}{20} U_{a[11],b[9]|_{j\neq i},b^i[8]c,c[6]d} \Big) \\
    -\sum_{i=1}^n \partial_{b^i} \partial_d \Big( U_{a[11],b[9]|_{j\neq i},b^i[8],c[8]} + \frac{9}{2} U_{a[11],b[9]|_{j\neq i},b^i[8]c,c[7]} \Big) - \partial_{c} \partial_{d} U_{a[11],b[9]|_{i=1}^{n},c[7]} \\
    - 90 \sum_{i=1}^n\partial_a \partial_{b^i} h_{a[9];b^j[9]|_{j\neq i},b^i[8]a,c[8],d} + 80\,\partial_a \partial_c h_{a[9];b^i[9]|_{i=1}^n,c[7]a,d} - \partial_a \partial_d h_{a[9];b^i[9]|_{i=1}^n,c[8],a} \,,
\end{multline}
where all the irreducible $U$ fields are specific linear combinations of the reducible $X$ fields with respective partition $[11,9^{n-2},8,8,8,1]$, $[11,9^{n-1},8,7,1]$, $[11,9^{n-1},8,8]$, and $[11,9^{n},7]$\,:
\begin{subequations}
\label{tildeUdualGravity}
\begin{multline}
    \frac12 U_{a[11],b^i[9]|_{i=1}^{n-2},c^1[8],c^2[8],c^3[8],d}
    \coloneqq 27 X^{\ord{1}}_{a[10];b^i[9]|_{i=1}^{n-2},c^1[8]a,c^2[8],c^3[8],d} \\
    - \frac{27}{11} X^{\ord{1}}_{a[10];b^i[9]|_{i=1}^{n-2},c^1[8]d,c^2[8],c^3[8],a} + \text{cyc}(c^i) \,,
\end{multline}
\vspace{-5mm}
\begin{multline}
    U_{a[11],b^i[9]|_{i=1}^{n-1},c[8],d[7],e} \coloneqq 9 X^{\ord{1}}_{a[10];b^i[9]|_{i=1}^{n-1},c[8],d[7]e,a} - 72 X^{\ord{1}}_{a[10];b^i[9]|_{i=1}^{n-1},c[8],d[7]a,e} \\
    \hspace{-30mm} + \frac{216}{n+2} X^{\ord{2}}_{a[10];b^i[9]|_{i=1}^{n-1},c[8]a,d[7],e} - \frac{108}{5(n+2)} X^{\ord{2}}_{a[10];b^i[9]|_{i=1}^{n-1},c[8]e,d[7],a} \\
    - \frac{504}{n+2} X^{\ord{2}}_{a[10];b^i[9]|_{i=1}^{n-1},c[8]d,d[6]a,e} + \frac{252}{5(n+2)} X^{\ord{2}}_{a[10];b^i[9]|_{i=1}^{n-1},c[8]d,d[6]e,a} \,,
\end{multline}
\vspace{-5mm}
\begin{multline}
    U_{a[11],b^i[9]|_{i=1}^{n-1},c^1[8],c^2[8]} \coloneqq 9 X^{\ord{1}}_{a[10];b^i[9]|_{i=1}^{n-1},c^1[8],c^2[8],a} + \frac{44}{n+10} X^{\ord{3}}_{a[10];b^i[9]|_{i=1}^{n-1},c^1[8]a,c^2[8]} \\
    + \frac{44}{n+10} X^{\ord{3}}_{a[10];b^i[9]|_{i=1}^{n-1},c^2[8]a,c^1[8]} \,,
\end{multline}
\vspace{-5mm}
\begin{multline}
    U_{a[11],b^i[9]|_{i=1}^{n},c[7]} \coloneqq \frac{16}{n+2} X^{\ord{2}}_{a[10];b^i[9]|_{i=1}^{n},c[7],a} - \frac{80}{n+10} X^{\ord{3}}_{a[10];b^i[9]|_{i=1}^{n},c[7]a} \,, \hfill
\end{multline}
\end{subequations}
where ``cyc$(c^i)$'' in $\tilde{U}_{11,9^{n-2},8^3,1}$ denotes the cyclic permutations of the three blocks of $c^i[8]$ indices.

The terms that depend on $h_{9\seco9^n,8,1}$ involve the irreducible representations $[10,9^{n-1},8,8,1]$, $[10,9^n,7,1]$, and $[10,9^n,8]$, and do not vanish by themselves.
The only possible solution is to assume that the contributions in $[9\,;9^n,8,1]\ominus[9^{n+1},8,1]$ are themselves curls, in which case they can be absorbed in the $X$ fields by construction and we can therefore set them to zero without loss of generality.
We find therefore precisely the field content predicted by $E_{11}$ exceptional field theory where the constrained field $\chi$ must be a total derivative
\begin{multline}
\label{eq:Chi10981=dX}
    \chi_{a[10];b^i[9]|_{i=1}^{n},c[8],d} =  9\sum_{i=1}^n \partial_{b^i} 
    X^{\ord{1}}{}_{a[10];b^j[9]|_{j\ne i},b^i[8],c[8],d} \\
    + 8\,\partial_{\langle c} X^{\ord{2}}{}_{a[10];b^i[9]|_{i=1}^{n},c[7],d\rangle} 
    + \partial_{\langle d} X^{\ord{3}}{}_{a[10];b^i[9]|_{i=1}^n,c[8]\rangle} \,,
\end{multline}
and the $U$ components of $X^{\ord{i}}$ in equation \eqref{tildeUdualGravity} are recognised as the additional fields in $L(\Lambda_4)$ as discussed in Section~\ref{GL11E11}.
We again used the projections \eqref{eq:F10;9,...,9,8,1_X(2)} and \eqref{eq:F10;9,...,9,8,1_X(3)}.

The general solution to \eqref{HigherDGMaxwell} with the Ans\"atze 
\eqref{eq:Fansatz8,1} and \eqref{tildeUdualGravity}, according to the 
generalised Poincar\'e lemma, is 
\begin{multline}
    U_{11,9^n,7} = -\frac1{11} \varepsilon_{11} \tilde{U}_{9^n,7} + D^{n+1}_{(1)} Z_{11,8^n,6} + D^{n+1}_{(2)} Z_{10,9,8^{n-1},6} + D^{n+1}_{(n+2)} Z_{10,8^n,7} \,, \hfill
\end{multline}
\vspace{-8mm}
\begin{multline}
    U_{11,9^{n-1},8,8} = -\frac1{11} \varepsilon_{11} \tilde{U}_{9^{n-1},8,8} + D^{n+1}_{(1)} Z_{11,8^{n-1},7,7} + D^{n+1}_{(2)} Z_{10,9,8^{n-2},7,7} + D^{n+1}_{(n+1)} Z_{10,8^n,7} \,, \hfill
\end{multline}
\vspace{-8mm}
\begin{multline}
    U_{11,9^{n-1},8,7,1} = - \frac1{11} \varepsilon_{11} \tilde{U}_{9^{n-1},8,7,1} + D^{n+2}_{(1)} Z_{11,8^{n-1},7,6} + D^{n+2}_{(2)} Z_{10,9,8^{n-2},7,7} + D^{n+2}_{(n+1)} Z_{10,8^n,6} \\
    + D^{n+2}_{(n+2)} Z_{10,8^{n-1},7,7} + D^{n+2}_{(n+3)} Z_{10,8^{n-1},7,6,1} \,,
\end{multline}
\vspace{-8mm}
\begin{multline}
    U_{11,9^{n-2},8,8,8,1} = - \frac1{11} \varepsilon_{11} \tilde{U}_{9^{n-2},8,8,8,1} + D^{n+2}_{(1)} Z_{11,8^{n-2},7,7,7} + D^{n+2}_{(2)} Z_{10,9,8^{n-3},7,7,7} \\
    + D^{n+2}_{(n)} Z_{10,8^{n-1},7,7,1} + D^{n+2}_{(n+3)} Z_{10,8^{n-2},7,7,7,1} \,,
\end{multline}
where again it is understood that the right-hand sides are projected onto the symmetry types of the left-hand sides.

This concludes the proof of equations \eqref{LinearChi} and \eqref{LinearisedRicciU} for all levels $\ell=6+3n$, assuming the Ricci-flat equation \eqref{M=ddU} is satisfied for some $\tU$ field at level $\ell=3+3n$\,.

\subsection{Ricci-flat equations}
\label{sec:BI_sols_inhomog}

To complete the induction we still need to prove that, assuming \eqref{LinearChi} is satisfied for all levels $3+3n$ with $n\geqslant0$, equation 
\eqref{M=ddU} holds for some $U$ field at level $\ell\in\{4+3n,5+3n,6+3n\}$\,.

\subsubsection*{\texorpdfstring{Level $\ell=4+3n$ for $n\geqslant 0$}{Level l=4+3n for n>-1}}

Let us start with the first higher dual three-form.
Taking the curl of the duality equations, one finds that $X_{10\seco2}$ drops out and
\begin{equation}
    4\,\partial_b \cE_{a[10];b[3]} = 40\,\partial_a \partial_b A_{a[9],b[3]} - \varepsilon_{a[10]c}\,\partial^c F_{b[4]} \approx 0 \,.
\end{equation}
Taking a single trace of this equation gives that the Labasitida tensor of $A_{9,3}$ is proportional to the gradient of the four-form field strength, i.e.
\begin{equation}
    L[A]_{a[9],b[3]} \coloneqq M[A]_{a[9],b[3]} + 27\,\partial_a \partial_b A_{a[8]c,b[2]}{}^c \approx - \varepsilon_{a[9]}{}^{cd} \partial_c F_{b[3]d} \,.
\end{equation}
Note that the Labastida tensor does not vanish on-shell for mixed symmetry fields with at 
least one column of $D-2=9$ antisymmetric indices.
Although the right-hand side is not manifestly a double $\partial_a\partial_b$ curl, 
one checks using the Maxwell equation that it is closed under the $\partial_b$ and 
$\partial_c$ curls, and thus it is (on-shell) locally the double curl of an irreducible 
$[8,2]$ field $\tU_{b[8],c[2]}$, i.e.
\begin{equation}
\label{eq:ddY=ddA+dd*A}
    27\,\partial_b \partial_c \tU_{b[8],c[2]} \approx -27\,\partial_b \partial_c A_{b[8]}{}^d{}_{,c[2]d} -\varepsilon_{b[9]}{}^{de} \partial_d F_{c[3]e} \,.
\end{equation}
Note that the $\tU$ field is non-local in the gauge field $A_{3}$\,.

More generally one has using \eqref{LinearChi} that
\begin{equation}
    4\cdot10^{n} \partial_c \partial_{b^1} \cdots \partial_{b^n} \cE_{a[10];b[9]|_{i=1}^n,c[3]}
    = K[A]_{a[10],b^i[10]|_{i=1}^n,c[4]} - \varepsilon_{a[10]d}\,\partial^d K_{b^i[10]|_{i=1}^n,c[4]} \approx 0 \,,
\end{equation}
and therefore in particular 
\begin{equation}
    \partial^d K_{da[9],b^i[10]|_{i=1}^n,c[4]} \approx 0 \,.
\end{equation}
One computes that 
\begin{equation}
    \partial^d K_{da[9],b^i[10]|_{i=1}^n,c[4]} =  4\cdot10^{n} \partial_c \partial_{b^1} \cdots \partial_{b^n} M[A]_{a[9],b^i[9]|_{i=1}^n,c[3]} \approx 0 \,,
\end{equation}
from which it follows that the Maxwell tensor $M[A]_{9^{n+1},3}$ is annihilated by 
$D^{n+1}_{(i)}$ for all $i$.
By the generalised Poincar\'e lemma 
\cite{Bekaert:2002dt}, it is locally a double curl, as we wanted to 
prove.\footnote{Note that the same line of reasoning with the 
generalised Poincar\'e lemma was used in \cite{Bekaert:2015fwa}.}

To see this result slightly more explicitly, it is useful to consider the linear combination of duality equations 
\begin{multline}
    \partial^d \cE_{a[9]d;b^i[9]|_{i=1}^n,c[3]} - 9 \sum_{i=1}^n \partial_{b^i} \cE_{a[9]}{}^d{}_{;b^j[9]|_{j\ne i},b^i[8]d,c[3]} - 3\,\partial_c \cE_{a[9]}{}^d{}_{;b^i[9]|_{i=1}^n,c[2]d} \\
    = \partial^d F_{a[9]d;b^i[9]|_{i=1}^n,c[3]} - 9 \sum_{i=1}^n \partial_{b^i} F_{a[9]}{}^d{}_{;b^j[9]|_{j\ne i},b^i[8]d,c[3]} - 3\,\partial_c F_{a[9]}{}^d{}_{;b^i[9]|_{i=1}^n,c[2]d} \\
    - \varepsilon_{a[9]}{}^{de} \partial_d \Big( 9 \sum_{i=1}^n \partial_{b^i} A_{b^j[9]|_{j\ne i},b^i[8]e,c[3]} + 3\,\partial_c A_{b^i[9]|_{i=1}^n,c[2]e} \Big) \approx 0 \,.
\end{multline}
From this, one computes
\begin{multline}
\label{eq:Inhom_9...93}
    M[A]_{a[9],b^i[9]|_{i=1}^n,c[3]} \\
    \approx -81 \sum_{i\ne j} \partial_{b^i} \partial_{b^j} \Big( X^{\ord{1}}_{a[9]}{}^d{}_{;b^k[9]|_{k\ne i,j},b^i[8]d,b^j[8],c[3]} + \frac{3}{n+7} X^{\ord{2}}_{a[9]}{}^d{}_{;b^k[9]|_{k\ne i,j},b^i[8]d,b^j[8]c,c[2]} \Big) \\
    - 27 \sum_{i=1}^n \partial_{b^i} \partial_c \Big( X^{\ord{1}}_{a[9]}{}^d{}_{;b^j[9]|_{j\ne i},b^i[8],c[2]d} + \frac{8}{n+7} X^{\ord{2}}_{a[9]}{}^d{}_{;b^j[9]|_{j\ne i},b^i[8]d,c[2]} + \frac{2}{n+7} X^{\ord{2}}_{a[9]}{}^d{}_{;b^j[9]|_{j\ne i},b^i[8]c,cd} \Big) \\
    - 9\,\partial_a \Big( 9 \sum_{i=1}^n \partial_{b^i} A_{a[8]}{}^d{}_{,b^j[9]|_{j\ne i},b^i[8]d,c[3]} + 3\,\partial_c A_{a[8]}{}^d{}_{,b^i[9]|_{i=1}^n,c[2]d} \Big) \\
    - \varepsilon_{a[9]}{}^{de} \partial_d \Big( 9 \sum_{i=1}^n \partial_{b^i} A_{b^j[9]|_{j\ne i},b^i[8]e,c[3]} + 3\,\partial_c A_{b^i[9]|_{i=1}^n,c[2]e} \Big) \,.
\end{multline}
One sees that the first three lines are manifestly double curls, whereas the last line is annihilated by $D^{n+1}_{(1)}$\,.
Although the right-hand side is not manifestly in the irreducible representation with partition $[9^{n+1},3]$, the left-hand side is; hence the right-hand side must also be irreducible on shell.
It therefore follows that it must be annihilated on shell by $D^{n+1}_{(i)}$ for all $i$ and that the last line must be locally a double curl.

One observes moreover that the penultimate line of \eqref{eq:Inhom_9...93} reproduces some of the trace terms in the Labastida tensor, so for an appropriate field redefinition of the $X$ fields one can write down a Labastida-like equation of the form
\begin{multline}
\label{eq:Laba_9...93}
    L[A]_{a[9],b^i[9]|_{i=1}^n,c[3]} \approx 81 \sum_{i\ne j} \partial_{b^i} \partial_{b^j} \Big( \tfrac12\,W_{a[9];b^k[9]|_{k\ne i,j},b^i[8],b^j[8],c[3]} + \tfrac{3}{8}\,W_{a[9];b^k[9]_{k\ne i,j},b^j[8],c,b^i[8],c[2]} \Big) \\
    + 27 \sum_{i=1}^n \partial_{b^i} \partial_c W_{a[9];b^j[9]|_{j\ne i},b^i[8],c[2]}   \\
    + \varepsilon_{a[9]}{}^{de} \partial_d \Big( \partial_e A_{b^i[9]|_{i=1}^n,c[3]} - 9 \sum_{i=1}^n \partial_{b^i} A_{b^j[9]|_{j\ne i},b^i[8]e,c[3]} - 3\,\partial_c A_{b^i[9]|_{i=1}^n,c[2]e} \Big) \,,
\end{multline}
where the gauge transformation of the $W$ fields is understood to compensate for the non-gauge invariance of the Labastida tensor and the last line on the right-hand side.
This is consistent with the fact that one cannot write a free Lagrangian for the higher gradient duals alone without involving all the other gradient duals at lower levels.  
The $W$ fields are \emph{not} the same as the $\tilde{U}$ fields from before.
They are defined such that $\eqref{eq:Inhom_9...93}$ holds, and can be expressed in terms of the $\tilde{U}$ fields using \eqref{M=ddU,3form} and \eqref{eq:Laba_9...93}.

\subsubsection*{\texorpdfstring{Level $\ell=6+3n$ for $n\geqslant 0$}{Level l=6+3n for n>-1}}

One can proceed in the same way for higher dual gravitons.
Expanding the linear combination of duality equations
\begin{multline}
    \partial^e \mathcal{E}_{a[9]e;b^i[9]|_{i=1}^n,c[8],d} - 9 \sum_{i=1}^n \partial_{b^i} \mathcal{E}_{a[9]}{}^e{}_{;b^j[9]|_{j\ne i},b^i[8]e,c[8],d} \\
    + 8\,\partial_c  \mathcal{E}_{a[9]}{}^e{}_{;b^i[9]|_{i=1}^n,c[7]e,d} - \partial_d \mathcal{E}_{a[9]}{}^e{}_{;b^i[9]|_{i=1}^n,c[8],e} \approx 0 \,,
\end{multline}
one computes that
\begin{multline}
\label{eq:Inhom_9...981}
    M_{a[9],b^i[9]|_{i=1}^n,c[8],d} \\
    \approx -81 \sum_{i\ne j} \partial_{b^i} \partial_{b^j} \Big( X^{\ord{1}}_{a[9]}{}^e{}_{;b^k[9]|_{k\ne i,j},b^i[8]e,b^j[8],c[8],d} + \frac{8}{n+2} X^{\ord{2}}_{a[9]}{}^e{}_{;b^k[9]|_{k\ne i,j},b^i[8]e,b^j[8]c,c[7],d} \hspace{10mm} \\
    \hspace{20mm} + \frac{1}{n+10} X^{\ord{3}}_{a[9]}{}^e{}_{;b^k[9]|_{k\ne i,j},b^i[8]e,b^j[8]d,c[8]} - \frac{1}{9(n+10)} X^{\ord{3}}_{a[9]}{}^e{}_{;b^k[9]|_{k\ne i,j},b^i[8]e,c[8]d,b^j[8]} \Big)\\
    + 72 \sum_{i=1}^n \partial_{b^i} \partial_c \Big( X^{\ord{1}}_{a[9]}{}^e{}_{;b^j[9]|_{j\ne i},b^i[8],c[7]e,d} - \frac{3}{n+2} X^{\ord{2}}_{a[9]}{}^e{}_{;b^j[9]|_{j\ne i},b^i[8]e,c[7],d} \hspace{30mm} \\
    \hspace{10mm} + \frac{7}{n+2} X^{\ord{2}}_{a[9]}{}^e{}_{;b^j[9]|_{j\ne i},b^i[8]c,c[6]e,d} + \frac{1}{n+10} X^{\ord{3}}_{a[9]}{}^e{}_{;b^j[9]|_{j\ne i},b^i[8]d,c[7]e} \\
    \hspace{40mm} + \frac{10}{9(n+10)} X^{\ord{3}}_{a[9]}{}^e{}_{;b^j[9]|_{j\ne i},b^i[8]e,c[7]d} - \frac{1}{9(n+10)} X^{\ord{3}}_{a[9]}{}^e{}_{;b^j[9]|_{j\ne i},c[7]ed,b^i[8]} \Bigr) \\
    - \sum_{i=1}^n \partial_{b^i} \partial_d \Big( 9 X^{\ord{1}}_{a[9]}{}^e{}_{;b^j[9]|_{j\ne i},b^i[8],c[8],e} + \frac{72}{n+2} X^{\ord{2}}_{a[9]}{}^e{}_{;b^j[9]|_{j\ne i},b^i[8]c,c[7],e} \hspace{20mm} \\
    \hspace{40mm} + \frac{89}{n+10} X^{\ord{3}}_{a[9]}{}^e{}_{;b^j[9]|_{j\ne i},b^i[8]e,c[8]} - \frac{1}{n+10} X^{\ord{3}}_{a[9]}{}^e{}_{;b^j[9]|_{j\ne i},c[8]e,b^i[8]} \Big) \\
    - 8\,\partial_c \partial_d \Big(\,\frac{2}{n+2} X^{\ord{2}}_{a[9]}{}^e{}_{;b^i[9]|_{i=1}^n,c[7],e} - \frac{10}{n+10} X^{\ord{3}}_{a[9]}{}^e{}_{;b^i[9]|_{i=1}^n,c[7]e} \Big) \hspace{20mm} \\
    - 9\,\partial_a \Big( 9\,\sum_{i=1}^b \partial_{b^i} h_{a[8]}{}^e{}_{,b^j[9]|_{j\ne i},b^i[8]e,c[8],d} - 8\,\partial_c h_{a[8]}{}^e{}_{,b^i[9]|_{i=1}^n,c[7]e,d} + \partial_d h_{a[8]}{}^e{}_{,b^i[9]|_{i=1}^n,c[8],e} \Big) \\
    - \varepsilon_{a[9]}{}^{ef} \partial_e \Big( 9 \sum_{i=1}^n \partial_{b^i} h_{b^j[9]|_{j\ne i},b^i[8]f,c[8],d} - 8\,\partial_c h_{b^i[9]|_{i=1}^n,c[7]f,d} + \partial_d h_{b^i[9]|_{i=1}^n,c[8],f} \Big) \,.
\end{multline}
All the terms except the last line are manifestly double curls, and the last line is clearly annihilated by $D^{n+2}_{(1)}$\,.
Although the right-hand side is not manifestly in the irreducible representation with partition $[9^{n+1},8,1]$, the left-hand side is; hence the right-hand side must be irreducible on shell.
It follows that it must be annihilated by all $D^{n+2}_{(i)}$ on shell and the last line must be locally a double curl.

As for the three-form field, one can derive this directly from the divergence-free equation
\begin{equation}
    \partial^e K_{ea[9],b^i[10]|_{i=1}^n,c[9],d[2]} = 18\cdot10^{n} \partial_d \partial_c \partial_{b_1} \cdots \partial_{b_n} M[A]_{a[9],b^i[9]|_{i=1}^n,c[8],d} \approx 0 \,.
\end{equation}
The penultimate line of \eqref{eq:Inhom_9...981} involves the curl on the $a$ column of the Labastida tensor, and one can write a modified Labastida-like equation for $h_{9^{n+1},8,1}$ involving four irreducible $W$ fields.

This concludes the proof of \eqref{LinearisedRicciU} and \eqref{LinearChi} for all the propagating fields in $E_{11}$ exceptional field theory.
To summarise, we have proved that, although the Labastida tensor of a higher dual field with Young diagram $\rY$ does not vanish on-shell, it can be expressed locally as the sum of double curls of  fields in representations of Young diagrams $\rY_{ij}\llcurly\rY$.
These fields depend non-locally on the lower level fields and have been proved to exist thanks to the generalised Poincaré lemma.
The same is true of the Maxwell tensors of higher dual fields, i.e.~\eqref{M=ddU}, showing that the right-hand side of the Bianchi-type equation for the constrained fields 
is also the sum of double curls of fields in representations of Young diagrams 
$\rY_{ij}\llcurly\rY$.
This exhibits that the additional field $U^{\tilde{\Lambda}}$ in 
$L(\Lambda_4)$ -- necessary for the gauge invariance of $E_{11}$ exceptional 
field theory but absent in the tensor hierarchy algebra -- is different in nature 
from the fields $X^{\tilde{\alpha}}$ and 
$X^\Lambda$ in $R(\Lambda_2)\oplus L(\Lambda_{10})$ predicted by the tensor 
hierarchy algebra.
The field $U^{\tilde{\Lambda}}$ is determined as the double-curl source of the 
Maxwell tensor, while the fields $X^{\tilde{\alpha}}$ and $X^\Lambda$ are 
determined by a choice of gauge.

\section{First-order duality equations}
\label{sec:E11dualityrels}

In this section we study more closely the duality equations that follow from our 
Lagrangians as the Euler--Lagrange equations of the $\chi_{10\seco9^n,\,\rx}$ fields -- 
see \eqref{eq:dualityequation} and \eqref{eq:fieldstrength}.
In Section~\ref{sec:gauge_inv_E11}, we write the gauge transformations for the linearised 
fields and show their compatibility with the $E_{11}$ exceptional field theory up to additional 
gauge parameters that we show lie inside modules beyond $L(\Lambda_1)\supset R(\Lambda_1)$.
Then, in Section~\ref{sec:covariantdualities} we show that the first-order duality equations 
of the theory can be obtained directly from the usual covariant form of the higher gradient 
dualities.

\subsection{\texorpdfstring{Gauge invariance of the $E_{11}$ system}{Gauge invariance of the E11 system}}
\label{sec:gauge_inv_E11}

In the $\mathrm{GL}(11)$ decomposition, the gauge parameters $\xi^M$ in $R(\Lambda_1)$ are denoted by $\lambda$ in the three-form and six-form sectors, and by $\xi$ in the (dual) gravity sector:
\begin{equation}
    \xi^M = \big( \xi^m, \lambda_{m[2]}, \lambda_{m[5]}, \xi_{m[7];n}, \lambda_{m[8];n[3]}, \lambda_{m[9];n,p}, \dots \big) \,.
\end{equation}
Now that we have described the propagating equations for the higher gradient 
dual fields in $E_{11}$ exceptional field theory, we discuss the gauge invariance 
of these equations.
For fields at positive levels $\ell\geqslant1$, the linearised gauge transformation 
\eqref{eq:dxiJ} simplifies to
\begin{equation}
\label{eq:dxiphiPos}
    \delta_{(\xi,\alpha)} \phi{}^{\alpha_{(\ell)}} = T^{\alpha_{(\ell)} a}{}_{P_{(\ell+3/2)}} 
    \partial_a \xi^{P_{(\ell+3/2)}} \,, 
\end{equation}
while using the relations \eqref{LinearChi} that we have derived, 
the gauge transformation of the constrained fields can be written for the $X$ and 
$U$ fields as
\begin{align} 
    \delta_{(\xi,\alpha)} X^{\ta_{(\ell)}} &= T^{\ta_{(\ell)} a}{}_{P_{(\ell+3/2)}} \partial_a \xi^{P_{(\ell+3/2)}} + \Pi^{\ta_{(\ell)}}{}_{a P_{(\ell-3/2)}} \partial^a \xi^{P_{(\ell-3/2)}}+T^{\ta_{(\ell)} a}{}_{\tilde{P}_{(\ell+3/2)}} \partial_a \alpha^{\tilde{P}_{(\ell+3/2)}} \,,\nonumber\\
    \delta_{(\xi,\alpha)} X^{\Lambda_{(\ell)}} &= \Pi^{{\Lambda}_{(\ell)}}{}_{a P_{(\ell-3/2)}} \partial^a \xi^{P_{(\ell-3/2)}} + T^{\Lambda_{(\ell)} a}{}_{\tilde{P}_{(\ell+3/2)}} \partial_a \alpha^{\tilde{P}_{(\ell+3/2)}} \,,\label{eq:dxiChiPos}\\
    \delta_{(\xi,\alpha)} U^{\tilde{\Lambda}_{(\ell)}} &= \Pi^{\tilde{\Lambda}_{(\ell)}}{}_{a P_{(\ell-3/2)}} \partial^a \xi^{P_{(\ell-3/2)}} \,.\nonumber
\end{align}
In this section we will briefly describe the consistency between these 
$E_{11}$ formulas and the explicit gauge transformations in eleven dimensions. 

\subsubsection*{Higher dual three-form fields}

For the higher duals $A_{9^{n},3}$ we have the gauge transformations
\begin{equation}
    \delta_{(\xi,\alpha)} A_{a^i[9]_{i=1}^{n},b[3]} = 9\sum_{i=1}^{n} \partial_{a^i} \lambda_{a^j[9]_{j\ne i},a^i[8],b[3]} + 3\,\partial_{\langle b} \lambda_{a^i[9]_{i=1}^{n},b[2]\rangle} \,,
\end{equation}
in terms of parameters $\lambda_{9^{n-1},8,3}$ and $\lambda_{9^n,2}$ with 
the same projection as in \eqref{eq:Fansatz3}.
These irreducible representations are indeed in  $R(\Lambda_1)$.
We sketch a proof that the corresponding components of the representation matrices 
$T^{\alpha_{(\ell)} a}{}_{P_{(\ell+3/2)}}$ are all non-vanishing in 
Appendix~\ref{app:proof_of_spectrum}.
Upon recalling \eqref{eq:fieldstrength}, \eqref{chichi},  
\eqref{eq:Chi=dX3}, and \eqref{eq:Chi10981=dX}, 
we find that the duality equation 
\begin{equation}
    \cE_{a[10];b^i[9]_{i=1}^n,c[3]} \coloneqq F_{a[10];b^i[9]_{i=1}^n,c[3]} - \varepsilon_{a[10]}{}^d \partial_d A_{b^i[9]_{i=1}^n,c[3]} \approx 0 
\end{equation}
is gauge invariant, provided $X^{\ord{1}}_{10\seco9^{n-1},8,3}$ 
and $X^{\ord{2}}_{10\seco9^n,2}$ transform as
\begin{align}
\label{DeltaX3}
    \delta_{(\xi,\alpha)} X^{\ord{1}}_{a[10];b^i[9]_{i=1}^{n-1},c[8],d[3]} 
    &= -10\,\partial_a \lambda_{a[9],b^i[9]_{i=1}^{n-1},c[8],d[3]} 
    + \varepsilon_{a[10]}{}^e \partial_e \lambda_{b^i[9]_{i=1}^{n-1},c[8],d[3]} \nonumber\\
    &\hspace{15mm} + 8\,\partial_{\langle c} \alpha_{a[10];b^i[9]|_{i=1}^{n-1},c[7],d[3]\rangle } + 3\,\partial_{\langle d} \alpha_{a[10];b^i[9]|_{i=1}^{n-1},c[8],d[2]\rangle} \,,\\
    \delta_{(\xi,\alpha)} X^{\ord{2}}_{a[10];b^i[9]_{i=1}^{n},c[2]} &= -10\,\partial_a \lambda_{a[9],b^i[9]_{i=1}^{n},c[2]} + \varepsilon_{a[10]}{}^d \partial_d \lambda_{b^i[9]_{i=1}^{n},c[2]} \nonumber\\
    &\hspace{15mm} -9 \sum_{i=1}^n \partial_{\langle b^i} \alpha_{a[10];b^j[9]|_{j\ne i},b^i[8],c[2]\rangle} + 2\,\partial_{\langle c} \alpha_{a[10];b^i[9]|_{i=1}^n,c\rangle} \,.
\end{align}
Note that the $\rd_1\lambda$ terms only appear in the transformations of the irreducible fields $X^{\ord{1}}_{10\seco9^{n-1},8,3}$ and $X^{\ord{2}}_{10\seco9^n,2}$ that belong indeed to $R(\Lambda_2)$ and so are compatible with  $T^{\ta_{(\ell)}a}{}_{P_{(\ell+3/2)}}\partial_a\xi^{P_{(\ell+3/2)}}$ in \eqref{eq:dxiChiPos}.
In contrast, the $(\star\rd)_1\lambda$ terms appear in all irreducible components, in agreement with the ``$\partial\xi$\,'' terms in \eqref{eq:dxiChiPos}.

The gauge transformation terms with the $\alpha$ parameters have been defined 
such that they drop out in $\delta_\alpha \chi_{10\seco9^n,3}$ and therefore leave the 
field strength $F_{10\seco9^n,3}$ invariant. 
Note that one can determine the set of $\alpha$ parameters by 
viewing the $\chi^I$ fields given in \eqref{eq:dxiChiPos} 
-- recalling \eqref{LinearChi} and \eqref{chichi} --
as Labastida-like gauge potentials whose first column of height ten
plays an inert role. 
It turns out that the $X$ and $U$ fields can then be viewed as the gauge parameters for the would-be Labastida potential $\chi^I$, 
and the $\alpha$ parameters are, technically speaking, 
the reducibility parameters for $\chi^I\,$, i.e.~$\delta_\alpha X^{\tilde{\alpha}}$, $\delta_\alpha X^{\Lambda}$, and $\delta_\alpha U^{\tilde{\Lambda}}$ are such that $\delta_{\alpha}\chi^I=0$\,.
The set of reducibility parameters is given by the Labastida--Morris rule \cite{Labastida:1986gy,Bekaert:2002dt}.
However, they appear in this way in all irreducible components and are therefore not compatible with \eqref{eq:dxiChiPos} that predicts $\delta_\alpha U=0$\,.
The structure of the tensor hierarchy algebra and the exterior differential \eqref{dsuper} imply that the gauge parameters $(\xi^{M_{(\ell+3/2)}},\alpha^{\tilde{M}_{(\ell+3/2)}})$ featured in \eqref{eq:dxiChiPos} should be in $[8]\otimes[9^n,3]$, while we get four more irreducible representations in \eqref{DeltaX3}.
We will label the corresponding irreducible components of $\alpha$ that parametrise $\delta_\alpha U$ as $\beta_{11,9^{n-2},8,7,3}$\,, $\beta_{11,9^{n-2},8,8,2}$\,, $\beta_{11,9^{n-1},7,2}$\,, and $\beta_{11,9^{n-1},8,1}$ in order to distinguish them from the genuine $\alpha$ parameters in \eqref{eq:dxiChiPos}. 

We conclude that the gauge transformations \eqref{eq:dxiChiPos} do not exhaust all the gauge symmetries of the linearised equations and there are additional gauge parameters $\beta_{11,9^{n-2},8,7,3}$\,, $\beta_{11,9^{n-2},8,8,2}$\,, $\beta_{11,9^{n-1},7,2}$\,, and $\beta_{11,9^{n-1},8,1}$ that are not irreducible components of $\alpha^{\tilde{M}}$ in $\cT_1\ominus R(\Lambda_1)$.

\subsubsection*{Higher dual gravitons}

For higher dual gravitons $h_{9^n,8,1}$ we have, with the same projections, the gauge transformations
\begin{equation}
    \delta_{(\xi,\alpha)} h_{a^i[9]_{i=1}^n,b[8],c} = 9\sum_{i=1}^n \partial_{a^i} \xi_{a^j[9]|_{j\ne i},a^i[8],b[8],c} + 8\,\partial_{\langle b} \xi_{a^i[9]|_{i=1}^{n},b[7],c\rangle} + \partial_{\langle c} \xi_{a^i[9]|_{i=1}^n,b[8]\rangle} \,,
\end{equation}
in terms of the gauge parameters $\xi_{9^{n-1},8,8,1}$\,, $\xi_{9^n,7,1}$\,, and $\xi_{9^n,8}$\,.
The duality equation 
\begin{equation}
    \cE_{a[10];b^i[9]_{i=1}^n,c[8],d} \coloneqq F_{a[10];b^i[9]_{i=1}^n,c[8],d} - \varepsilon_{a[10]}{}^e \partial_e h_{b^i[9]_{i=1}^n,c[8],d} \approx 0
\end{equation}
is gauge invariant provided the fields $X^{\ord{1}}_{10\seco9^{n-1},8,8,1}$\,, $X^{\ord{2}}_{10\seco9^n,7,1}$\,, and $X^{\ord{3}}_{10\seco9^n,8}$ transform as
\begin{align}
\label{DeltaX81}
    \delta_{(\xi,\alpha)} X^{\ord{1}}_{a[10];b^i[9]|_{i=1}^{n-1},c^1[8],c^2[8],d} &= -10\,\partial_a \xi_{a[9],b^i[9]|_{i=1}^{n-1},c^1[8],c^2[8],d} + \varepsilon_{a[10]}{}^e \partial_e \xi_{b^i[9]|_{i=1}^{n-1},c^1[8],c^2[8],d} \\
    &\hspace{4.5mm} + 8\,\partial_{\langle c^1} \alpha_{a[10];b^i[9]|_{i=1}^{n-1},c^2[8],c^1[7],d\rangle} + 8\,\partial_{\langle c^2} \alpha_{a[10];b^i[9]|_{i=1}^{n-1},c^1[8],c^2[7],d\rangle} \nonumber\\
    &\hspace{25mm} + \partial_{\langle d} \alpha_{a[10];b^i[9]|_{i=1}^{n-1},c^1[8],c^2[8]\rangle}  \,,\nonumber\\
    \delta_{(\xi,\alpha)} X^{\ord{2}}_{a[10];b^i[9]|_{i=1}^{n},c[7],d} &= -10\,\partial_a \xi_{a[9],b^i[9]|_{i=1}^{n},c[7],d} + \varepsilon_{a[10]}{}^e \partial_e \xi_{b^i[9]|_{i=1}^{n},c[7],d} \\
    &\hspace{4.5mm} -9 \sum_{i=1}^{n} \partial_{b^i} \alpha_{a[10];b^j[9]|_{j\ne i},b^i[8],c[7],d} + 7\,\partial_{\langle c} \alpha_{a[10];b^i[9]|_{i=1}^n,c[6],d\rangle} \nonumber\\
    &\hspace{25mm} + \partial_{\langle d}  \alpha_{a[10];b^i[9]|_{i=1}^n,c[7]\rangle} \,,\nonumber\\
    \delta_{(\xi,\alpha)} X^{\ord{3}}_{a[10];b^i[9]|_{i=1}^{n},c[8]} &= -10\,\partial_a \xi_{a[9],b^i[9]|_{i=1}^{n},c[8]} + \varepsilon_{a[10]}{}^e \partial_e \xi_{b^i[9]|_{i=1}^{n},c[8]} \\
    &\hspace{4.5mm} - 9\sum_{i=1}^n \partial_{b^i} \alpha_{a[10];b^j[9]|_{j\ne i},b^i[8],c[8]} - 8\,\partial_{\langle c} \alpha_{a[10];b^i[9]|_{i=1}^n,c[7]\rangle} \,.\nonumber
\end{align}
Here as well the $\rd_1\xi$ terms only appear in the gauge transformations of the irreducible fields $X^{\ord{1}}_{10,9^{n-1},8,8,1}$\,, $X^{\ord{2}}_{10,9^{n},7,1}$\,, and $X^{\ord{3}}_{10,9^{n},8}$ that belong to $R(\Lambda_2)$ and are therefore compatible with  $T^{\ta_{(\ell)} a}{}_{P_{(\ell+3/2)}} \partial_a \xi^{P_{(\ell+3/2)}}$ in equation \eqref{eq:dxiChiPos}.
The $(\star\rd)_1\xi$ terms appear in all irreducible components, in agreement with \eqref{eq:dxiChiPos}.

The gauge transformations for the $\alpha$ parameter have been defined such that 
they drop out in $\delta_\alpha\chi_{10\seco9^n,8,1}$\,.
However, as they are defined they appear in all irreducible components and are therefore not compatible with equation \eqref{eq:dxiChiPos} that predicts $\delta_\alpha U=0$\,.
The structure of the tensor hierarchy algebra implies that the gauge parameters $(\xi^{M_{(\ell+3/2)}},\alpha^{\tilde{M}_{(\ell+3/2)}})$ featured in \eqref{eq:dxiChiPos} should be in $[8]\otimes[9^n,8,1]$, while we get seven more irreducible representations in \eqref{DeltaX81}.
We label the corresponding irreducible components of $\alpha$ that parametrise $\delta_\alpha U$ as $\beta_{11,9^{n-2},8,8,7,1}$\,, $\beta_{11,9^{n-2},8,8,8}$\,, $\beta_{11,9^{n-1},7,7,1}$\,, $\beta_{11,9^{n-1},8,6,1}$\,, $\beta^{\ord{1}}_{11,9^{n-1},8,7}$\,, $\beta^{\ord{2}}_{11,9^{n-1},8,7}$\,, and $\beta_{11,9^{n},6}$\,.

The $\beta$ parameters that appear in the gauge transformations $\delta_\alpha U$ of \eqref{DeltaX3} and \eqref{DeltaX81} and their equivalent for the six-form higher duals turn out to be inside the $E_{11}$ module $R(\Lambda_5)$: 
\begin{align}
    R(\Lambda_5) &\supset \big( [11,6] \big)_{\frac{15}{2}} \oplus \big( [11,7,2] \oplus [11,8,1] \big)_{\frac{17}{2}} \oplus \big( [11,7,5] \oplus [11,8,4] \big)_{\frac{19}{2}} \nonumber\\
    &\hspace{4.5mm} \oplus \big( [11,7,7,1] \oplus [11,8,6,1] \oplus 2\!\times\![11,8,7] \oplus [11,9,6] \big)_{\frac{21}{2}} \nonumber\\
    &\hspace{4.5mm} \oplus \bigoplus_{n\geqslant 2} \big( [11,9^{n-2},8,7,3] \oplus [11,9^{n-2},8,8,2] \oplus [11,9^{n-1},7,2] \oplus [11,9^{n-1},8,1] \big)_{3n+\frac{11}{2}} \nonumber\\
    &\hspace{4.5mm} \oplus \bigoplus_{n\geqslant 2} \big( [11,9^{n-2},8,7,6] \oplus [11,9^{n-2},8,8,5] \oplus [11,9^{n-1},7,5] \oplus [11,9^{n-1},8,4] \big)_{3n+\frac{13}{2}} \nonumber\\
    &\hspace{4.5mm} \oplus \bigoplus_{n\geqslant 2} \big( [11,9^{n-2},8,8,7,1] \oplus [11,9^{n-2},8,8,8] \oplus [11,9^{n-1},7,7,1] \oplus [11,9^{n-1},8,6,1] \nonumber\\
    & \hspace{45mm} \oplus 2\!\times\![11,9^{n-1},8,7] \oplus [11,9^{n},6] \big)_{3n+\frac{15}{2}} \,.
\end{align}
One checks that 
\begin{align}
\begin{aligned}
    R(\Lambda_1) \otimes R(\Lambda_4) &\;\cong\; R(\Lambda_1+\Lambda_4) \oplus R(\Lambda_5) \oplus R(\Lambda_3+\Lambda_{10}) \oplus \dots \,,\\
    R(\Lambda_1) \otimes R(\Lambda_{10}) &\;\cong\;  R(\Lambda_1+\Lambda_{10}) \oplus R(\Lambda_{11}) \oplus R(\Lambda_5) \oplus \dots \,,
\end{aligned}
\end{align}
whereas $R(\Lambda_5)$ does not appear in $R(\Lambda_1)\otimes R(\Lambda_2)$.
This suggests that there should exist some normalisations of the Clebsch--Gordan coefficients for
\begin{align}
    &R(\Lambda_5) \hookrightarrow R(\Lambda_1) \otimes R(\Lambda_4)&
    &R(\Lambda_5) \hookrightarrow R(\Lambda_1) \otimes R(\Lambda_{10})
\end{align}
such that the $\beta^{\vardbtilde{P}}\in L(\Lambda_5)\coloneqq R(\Lambda_5)\oplus\dots$ gauge transformations
\begin{align}
\label{eq:dbetaChiPos}
    \delta_{\beta} X^{\ta} &= 0 \,,&
    \delta_{\beta} X^{\Lambda} &= \Pi^{\Lambda M}{}_{\!\vardbtilde{P}} \partial_M \beta^{\vardbtilde{P}} \,,&
    \delta_{\beta} U^{\tilde{\Lambda}} &= \Pi^{\tilde{\Lambda} M}{}_{\!\vardbtilde{P}} \partial_M \beta^{\vardbtilde{P}} \,,
\end{align}
leave the field strength $F^I$ invariant.

\subsection{From covariant to first-order duality equations}
\label{sec:covariantdualities}

In Sections~\ref{sec:Lagrangian} and \ref{sec:GradientDuals} we have derived the covariant duality equations \eqref{FormHigherDuality} and \eqref{GravityHigherDuality}  from the $E_{11}$ exceptional field theory Lagrangians \eqref{cLc}.
In this section we shall do the reverse:~we will derive the first-order duality equations starting from the propagating mixed-symmetry field strength integrability conditions.

The gauge invariant curvature for the field $\phi_{9^n,\,\rx}$ is defined by 
\begin{equation}
\label{eq:K10nrx}
    K[\phi]_{10^n,\,\overline{\rx}} = \Bigg( \prod_{i=1}^{n+s(\rx)} \rd_i \Bigg) \phi_{9^n,\,\rx} = \rd_1 D^{n+s(\rx)-1}_{(1)} \phi_{9^n,\,\rx} \,,
\end{equation}
where $\overline{\rx}$ is obtained from the $\rx$ part of the diagram by adding one box to each column.
Using the (covariant) equation of motion for $\phi_{9^n,\,\rx}$ 
in \eqref{eq:Tr10K_Tr4K}--\eqref{eq:Tr10K_Tr9K_Tr2K}, 
one finds that  
\begin{equation}
    \rd_i (\star\rd)_1 K[\phi]_{0,10^n,\,\overline{\rx}} = 0\;,
\end{equation}
for all $i=1,\dots,n+1+s(\rx)$.
This implies by the generalised Poincar\'e lemma \cite{Bekaert:2002dt} that 
\begin{equation}
    (\star\rd)_1 K[\phi]_{10^n,\,\overline{\rx}} = K[{\phi}]_{10^{n+1},\,\overline{\rx}}
\end{equation}
for some higher dual field ${\phi}_{9^{n+1},\,\rx}$\,.
Therefore, one finds
\begin{equation}
    D_{(1)}^{n+s(\rx)} \big( (\star\rd)_1 \phi_{9^{n},\,\rx} - \rd_1 \phi_{9^{n+1},\,\rx} \big) = 0 \,.
\end{equation}
Using again the generalised Poincar\'e lemma, this gives 
\begin{equation}
  (\star \rd)_1 \phi_{9^{n},\,\rx} -\rd_1 \phi_{9^{n+1},\,{\rx}}
  = \sum_{\rY_i\prec[9^n,\,\rx]} \hspace{-3mm} \big( \rd_i X_{10\seco\rY_i} \big)_{[10\seco9^n,\,\rx]} \;,
\end{equation}
where we point the reader's attention to the $\mathrm{GL}(11)$ reducibility 
of both sides of the equation, with the first column of height ten not attached 
to the other columns. 
As a result, we see that the covariant duality equations imply the first-order duality 
equations in which the $X$ fields are `integration constants' in the generalised 
Poincar\'e lemma.
We therefore reproduce the linearised first-order duality equations in $E_{11}$ exceptional field theory.

\section{Summary of results and conclusion}
\label{sec:conclusion}

In this paper we have systematically studied the higher gradient dual fields in 
$E_{11}$ exceptional field theory \cite{Bossard:2021ebg} 
in the linearised approximation.
Our main result is to prove that the $E_{11}$ exceptional field theory 
pseudo-Lagrangian combined with appropriate components of the duality equation of the 
theory permits to derive the covariant field equations for all the higher gradient dual 
fields first identified in \cite{Boulanger:2012df}.
As a by-product, we propose parent Lagrangians for all these fields, which can either reproduce the original eleven-dimensional supergravity Lagrangian after integrating out all the St\"{u}ckelberg fields, or give new Lagrangians for the higher dual fields after eliminating the lower level fields in the linearised approximation.

We shall first summarise the technical results in more detail. Then we will discuss possible lines of research related to what we explored in this paper. 

\subsection*{Summary of results}

For the set of $E_{11}$ fields $\{\phi_{9^n,\,\rx}\}_{n\in \mathbb{N}}$\,, 
we have defined an infinite set of Lagrangians $\mathcal{L}^\ord{k}_{\rx}$ that depend on all the fields 
$\phi_{9^n,\,\rx}$ for $n\leqslant k+1$ with $\rx=[3]$, $[6]$, or $[8,1]$.
They are all obtained from the $E_{11}$ exceptional field theory 
pseudo-Lagrangian given in \cite{Bossard:2021ebg}, 
plus an infinite sum of squares of duality equations. 
The Euler--Lagrange equations for $\phi_{9^n,\,\rx}$ that follow from 
$\mathcal{L}^\ord{k}_{\rx}$ are the same for all $0\leqslant n\leqslant k$\,.
The difference of the Euler--Lagrange equations for $\phi_{9^n,\,\rx}$ from 
$\mathcal{L}^\ord{n}_{\rm x}$ and $\mathcal{L}^\ord{n-1}_{\rx}$ gives that the 
Maxwell-like tensor
\begin{equation}
    M[\phi]_{9^n,\,\rx} \;\coloneqq\; \Box \phi_{9^n,\,\rx} - \hspace{-3mm} \sum_{\rY_i\prec[9^n,\,\rx]} \hspace{-3mm} \rd_i \big( \partial \cdot \phi \big)_{\rY_i}
    \;=\; \Box \phi_{9^n,\,\rx} - \sum_{i=1}^{n+s(\rx)} \big( \rd_i \rd_i^\dagger \phi \big)_{[9^n,\,\rx]}
\end{equation}
is equal to an integrability condition for the gradient dual field strength of $\phi_{9^{n+1},\,\rx}$\,:
\begin{equation}
\label{Integrability}
    \rd_1 F_{10\seco9^n,\,\rx} - \hspace{-3mm} \sum_{\rY_i\prec[9^n,\,\rx]} \hspace{-3mm} \rd_i ( e_1 \cdot F )_{[11\seco\rY_i]}
    \;\approx\; \varepsilon_{11} \,M[\phi]_{9^n,\,\rx} \,,
\end{equation}
where $e_1$ takes one of the indices in $\rY$ and antisymmetrises it with the entire first column, see \eqref{eq:HoBianchi3} and \eqref{eq:HoBianchi8,1}.
For this integrability condition to be effective, one needs to show that the Maxwell-like tensor of $\phi_{9^n,\,\rx}$ is a second derivative on-shell, i.e.~that
\begin{equation}
\label{MaxwellDouble}
    M[\phi]_{9^n,\,\rx} \;\approx \sum_{\rY_{i,j}\llcurly[9^n,\,\rx]} \big( \rd_i \rd_j \tilde{U}_{\rY_{i,j}} \big)_{[9^n,\,\rx]} \,,
\end{equation}
for some fields $\tilde{U}_{\rY_{i,j}}$\,. 
For $n=0$ this is a direct consequence of the Labastida equation 
$L[\phi]_{\rx}\approx0$ with the Labastida tensor 
defined by \cite{Labastida:1986ft,Bekaert:2006ix} 
\begin{equation}
    L[\phi]_{9^n,\,\rx} \coloneqq M[\phi]_{9^n,\,\rx} + \hspace{-3mm} \sum_{\rY_{i,j}\llcurly[9^n,\,\rx]} \hspace{-3mm} \big( \rd_i \rd_j (\Tr_{i,j} \phi)_{\rY_{i,j}} \big)_{[9^n,\,\rx]} \,.
\end{equation}
It follows that the right-hand side of \eqref{Integrability} is a double curl on-shell for $n=0$ and we get that the St\"uckelberg field $\chi_{10\seco\rx}$ is a double curl.
For higher gradient duals the Labastida tensor $L[\phi]_{9^n,\,\rx}$ does not vanish on-shell, but we prove by induction that it is a double curl.

The induction works as follows.
After showing for some $n$ that $L[\phi]_{9^n,\,\rx}$ is a double curl on shell, one finds that the right-hand side of \eqref{Integrability} is a second derivative and so all the St\"uckelberg fields $\chi_{10\seco9^n,\,\rx}$ are total derivatives:
\begin{equation}
\label{chifromX}
    \chi_{10\seco9^n,\,\rx} \,\approx \sum_{\rY_i\prec[9^n,\,\rx]} \hspace{-3mm} \big( \rd_i X_{10\seco\rY_i} \big)_{[10\seco9^n,\,\rx]} \,.
\end{equation}
One can decompose the fields $X_{10\seco\rY_i}$ into those that belong to the tensor hierarchy algebra \eqref{eq:R2L10}, that are in the irreducible components of $\big([9]\otimes[9^n,\rx]\big)\ominus[9^{n+1},\rx]$, and the extra fields $U_{11\seco\rY_{i,j}}$ with $\rY_{i,j}\llcurly[9^n,\rx]$ from \eqref{eq:L4}, that must be present for the gauge invariance of the field equations.
The fields $X_{10\seco\rY_i}$ drop out in the integrability condition \eqref{Integrability} while the fields $U_{11\seco\rY_{i,j}}$ are identified with $\varepsilon_{11}\tilde{U}_{\rY_{i,j}}$ through \eqref{MaxwellDouble}, so
\begin{equation}
    \sum_{\rY_{i,j}\llcurly[9^n,\,\rx]} \hspace{-3mm} \big( \rd_i \rd_j U_{11\seco\rY_{i,j}} \big)_{[11\seco9^n,\,\rx]}
    \approx - \varepsilon_{11} M[\phi]_{9^n,\,\rx}
    \approx - \varepsilon_{11} \hspace{-3mm} \sum_{\rY_{i,j}\llcurly[9^n,\,\rx]} \hspace{-3mm} \big( \rd_i \rd_j \tilde{U}_{\rY_{i,j}} \big)_{[9^n,\,\rx]} \,.
\end{equation}
Inserting the result inside the divergence of the duality equation $\mathcal{E}_{10\seco9^n,\,\rx}\approx0$\,,
\begin{equation}
    (\partial \cdot \cE)_{9\seco9^n,\,\rx} -\hspace{-3mm} \sum_{\rY_i\prec[9^n,\,\rx]} \hspace{-3mm} \rd_i ( \Tr_{1,i} \cE )_{[9\seco\rY_i]} \approx 0 \,,
\end{equation}
one finds by integrability that the Labastida tensor $L[\phi]_{9^{n+1},\,\rx}$ is itself a double curl on-shell, which proves the induction.
Note however that $\tilde{U}_{\rY_{i,j}}$ are generally non-local in the propagating 
fields on shell.
They are determined via the generalised Poincar\'e lemma \cite{Bekaert:2002dt} 
from the conditions
\begin{equation}
    D_{(1)}^{n+s(\rx)} ( M[\phi]_{9^{n+1},\,\rx})
    \;\approx\; 0 \,.
\end{equation}
We find that the additional field $U^{\tilde{\Lambda}}\in L(\Lambda_4)$ plays a particular role in the dynamics:~they are directly related to the source terms in the Labastida equations of the higher dual fields and the dual graviton. 

Recall that the Bianchi-type equations that we solve to obtain \eqref{chifromX} are the covariant field equations of the theory \eqref{LENexpand}.
They split into the Ricci-flat equations and \eqref{BianchiE11}, the latter of which is just the condition $\rd F=0$ for the tensor hierarchy algebra differential complex \eqref{dTH} (when ignoring $\zeta_M{}^{\tilde{\Lambda}}$ \cite{Bossard:2017wxl}), 
which leads (via what one may call an ``exceptional geometry Poincar\'e lemma'') to $F=\rd\phi+\rd X$.
In contrast, one must use the field equations for all lower level fields to solve the Ricci-flat equations and arrive at $\zeta_M{}^{\tilde{\Lambda}}=\partial_M U^{\tilde{\Lambda}}$, as described above. 

Once we have shown that the St\"uckelberg fields are total derivatives according to \eqref{chifromX}, one obtains that the first-order duality equation $\mathcal{E}_{10\seco9^n,\,\rx}\approx0$ implies the duality equation for the covariant field strength as defined in \cite{Boulanger:2015mka}: 
\begin{equation}
    K[\phi]_{10^{n+1},\,\overline{\rx}} \approx (\star\rd)_1 K[\phi]_{10^n,\,\overline{\rx}} \,.
\end{equation}
Conversely, we have shown that the covariant duality equation above implies by the generalised Poincar\'e lemma that there exist $X_{10\seco\rY_i}$ as in \eqref{chifromX} such that the first-order duality equation 
\begin{equation}
    F_{10\seco9^n,\,\rx} \approx (\star\rd)_1 \phi_{9^n,\,\rx} \,,
\end{equation}
holds.

In this way we have proved that $E_{11}$ exceptional field theory provides parent Lagrangians for all higher dual potentials, where the equations of motions are equivalent to the covariant duality equations originally derived in \cite{Boulanger:2015mka}, up to the introduction of additional fields necessary to obtain gauge covariant first-order duality equations.

Although we have been working in the linearised approximation, it is a priori straightforward to generalise these parent Lagrangians to the non-linear level.
For this purpose one simply needs to consider
\begin{equation}
\label{cLcNL}
    \cL_c = \cL^{\scalebox{0.6}{sugra}} - \frac{1}{4} \sum_{\ell\,\geqslant\frac12} c_\ell\,\eta^{I_{(\ell)}J_{(\ell)}} \mathcal{E}_{I_{(\ell)}} \mathcal{E}_{J_{(\ell)}} \,,
\end{equation}
where $\cL^{\scalebox{0.6}{sugra}}$ is the non-linear (bosonic) supergravity Lagrangian in eleven dimensions \cite{Cremmer:1978km,Cremmer:1978ds}, and where $\partial_M \phi_\alpha$ is replaced by $J_{M \alpha}$ according to \eqref{JMQ} in the first-order duality equations.
At the non-linear level, the three Lagrangians $\mathcal{L}^\ord{k}_{\rx}$ 
for ${\rx}\in\{[3], [6], [8,1]\}$ 
do not decouple and must be considered together.
However, since there are extra St\"uckelberg fields associated with each higher dual field and with the dual graviton, the non-linearities in the Maurer--Cartan form can all be reabsorbed and the only important non-linear coupling comes from the metric and the Chern--Simons coupling for the three-form.
This reflects the fact that the only truly non-abelian gauge algebra is the algebra of space-time diffeomorphisms, in line with \cite{Boulanger:2008nd,Bergshoeff:2009zq}.

\subsection*{Conclusion and outlook} 

The $E_{11}$ pseudo-Lagrangian and the duality equations were shown to transform covariantly under generalised diffeomorphisms in \cite{Bossard:2021ebg}, but the complete set of gauge transformations of the theory was not derived there.
In particular, it was exhibited that the algebra of generalised diffeomorphisms only closes up to ancillary gauge transformations that can be interpreted in components as St\"{u}ckelberg gauge invariance of the higher duals and the dual graviton.
It would be interesting to derive the ancillary gauge transformation of the constrained fields and to analyse further the full gauge algebra of the theory.
As pointed out in \cite{Bossard:2021ebg}, it is a priori necessary to introduce certain 
`higher ancillary' transformations to define the full set of gauge transformations of the 
theory, i.e.~gauge parameters with more than one constrained index.
These gauge parameters should appear in the gauge transformations of the constrained fields that leave the field strength $F^I$ invariant.

In relation to the full gauge structure of the theory, it would also be good to obtain explicit expressions for the gauge-for-gauge and higher reducibility transformations of $E_{11}$ exceptional field theory and to understand the highest weight representations that enter.
Bianchi-type and Noether-type equations and reducibility parameters sit at certain grades in the tensor hierarchy algebra $\cT(\mf{e}_{11})$, and so far only the Bianchi-type equations have been worked out in any detail.
One could try to extend the analysis of \cite{Cederwall:2019bai} (see also \cite{Cederwall:2023xbj,Bossard:2024gry}) to the case of $E_{11}$ and other very-extended Kac--Moody symmetries.

As was discussed in the conclusion of \cite{Bossard:2021ebg}, it may actually be necessary to enlarge the set of fields and ancillary transformations to include fields in the infinite extension $\mathfrak{e}_{11}\oleft\bigoplus_{n\geqslant1}\overline{L(n\Lambda_2)}$ of the $\mathfrak{e}_{11}$ algebra to obtain a genuine Kac--Moody symmetry of the equations of motion.
This extension is the natural generalisation of the negative Virasoro extension $\mathfrak{e}_{9}\oleft\bigoplus_{n\geqslant1} \langle L_{-n}\rangle$ of the affine Kac--Moody algebra that is introduced in $E_9$ exceptional field theory in order to make the Geroch symmetry in the Maison--Breithenlohner linear system explicit \cite{Bossard:2021jix}. 
Hoping for integrability structures associated with Kac--Moody groups beyond $E_9$ might be wishful, the affine symmetry being itself only realised in the presence of commuting isometries.
One may nonetheless find an $E_9$ integrable structure in non-trivial backgrounds \cite{Cesaro:2024ipq,Cesaro:2025msv}.

Another motivation for introducing an extension of $E_{11}$ exceptional field theory with fields valued in $\mathfrak{e}_{11}\,\oleft\,\bigoplus_{n\geqslant1}\overline{L(n\Lambda_2)}$ is for the construction of the supersymmetric theory.
The supersymmetric $E_{11}$ exceptional field theory was considered in \cite{Bossard:2019ksx}, where it was established that the bilinear $\bar\epsilon\,\Psi$ that defines the supersymmetry transformation of the scalar fields cannot be in $\mathfrak{e}_{11}\ominus K(\mathfrak{e}_{11})$, but must at least include  $\mathfrak{e}_{11}\oleft\overline{L(\Lambda_2)} \ominus K(\mathfrak{e}_{11})$.
The more recent analysis of the supersymmetry transformation in $D=2$ supergravity shows  that the equivalent of the bilinear $\bar\epsilon\,\Psi$ transforms in $\mathfrak{e}_{9}\,\oleft\,\bigoplus_{n\geqslant1} \langle L_{-n}\rangle$ \cite{Paulot:2006zp,Konig:2025rlt}, so generalising to the $E_{11}$ case would certainly involve the infinite extension $\mathfrak{e}_{11}\oleft\bigoplus_{n\geqslant1}\overline{L(n\Lambda_2)}$. 

In this paper we have only discussed the eleven-dimensional solution to the section 
constraint.
We expect our results to generalise straightfowardly to the type IIB solution in ten 
dimensions.
One main difference is that the four-form admits a self-dual field strength, 
and so does not admit an action without breaking manifest Lorentz invariance or 
including auxiliary fields 
\cite{Henneaux:1987hz,Henneaux:1988gg,Schwarz:1993vs,Schwarz:1997mc,Hillmann:2009zf,Pasti:1996vs,DallAgata:1997gnw,DallAgata:1998ahf,Mkrtchyan:2019opf,Witten:1996hc,Witten:1999vg,Belov:2006jd,Sen:2015nph}.
The self-duality equation for the field strength of the four-form is the level zero 
component of the $E_{11}$ duality equation.
One obtains a Lagrangian of Henneaux--Teitelboim type \cite{Henneaux:1987hz} by 
adding the square of the electric component of the duality equation to the $E_{11}$ 
pseudo-Lagrangian. 
The higher dual fields corresponding to the chiral four-form are predicted from $E_{11}$ 
representations to have $[8^n,4]$ irreducible Young tableaux, and the chiral projection 
only appears for the original four-form  \cite{Bergshoeff:2016gub}. 

It would be interesting to extend the decomposition of the linearised $E_{11}$ 
exceptional field theory Euler--Lagrange equation that we considered in this paper
into a non-linear extension of the Bianchi-type equation \eqref{BianchiE11} and 
a non-linear extension of the
Ricci-flat equation \eqref{eq:generalisedRicci} in order to define a generalised 
Ricci tensor.
One initial difficulty is that the non-linear Euler--Lagrange equation involves infinitely many terms and cannot be analysed directly, see eq.~(5.16) in \cite{Bossard:2021ebg}.
Moreover, it is not obvious that the Euler--Lagrange equation splits into two vanishing terms  \eqref{BianchiE11} and \eqref{eq:generalisedRicci} at the non-linear level.
One may nevertheless consider defining a generalised geometry based on $E_{11}$, for which the Ricci tensor would vanish on-shell and would reproduce  \eqref{eq:generalisedRicci} in the linearised approximation.
A generalised Ricci tensor has been defined in $E_n\times\mathds{R}^+$ exceptional generalised geometry for $n\leqslant7$ \cite{Hull:2007zu,PiresPacheco:2008qik,Coimbra:2011ky,Coimbra:2011nw,Coimbra:2012af,Aldazabal:2013mya}.
Although the Levi--Civita connection is defined only up to some ambiguities, the Ricci tensor is unique and its vanishing is equivalent to the supergravity field equations.
The notion of connection and Ricci tensor has been generalised to exceptional field theory \cite{Park:2013gaj,Cederwall:2013naa,Blair:2014zba,Godazgar:2014nqa} and more recently to generalised Cartan geometry \cite{Hassler:2023axp,Hassler:2025rag}.
The linearised Ricci-flat equations we define suggest that an additional term in $\zeta_M{}^{\tilde{\Lambda}}$ should appear in the $E_n\times\mathds{R}^+$ Ricci tensor for $n\geqslant8$\,.
It would be interesting to see if such an additional field is indeed necessary in the definition of the Ricci tensor from these more geometric approaches.

One of the most exciting lines of research in $E_{11}$ exceptional field theory would be to study Belinsky--Khalatnikov--Lifshitz  (BKL) dynamics near a space-like singularity from this perspective.
Choosing a time direction in the $R(\Lambda_1)$ module, one can work out a Hamiltonian exceptional 
field theory with manifest $E_{10}$ invariance \cite{Bossard:2021ebg}. 
In principle this is well-suited to exhibit that the $E_{10}$ non-linear sigma model proposed in \cite{Damour:2002cu,Damour:2002et} describes the dynamics of supergravity in eleven dimensions near a space-like singularity.
The higher dualities described in this paper in the linearised approximation should reflect the property that the gradient expansion of the supergravity fields is described by the $\mathrm{GL}(10)$ components of the $E_{10}/K(E_{10})$ symmetric space coordinates with multiple columns of nine antisymmetrised indices \cite{Damour:2002cu,Damour:2004zy}.
It would be very interesting to see if $E_{11}$ exceptional field theory could shed some light on the problem of defining the conjectural Sugawara-like constraints necessary to retrieve supergravity from the $E_{10}$ sigma model \cite{Damour:2006xu,Damour:2007dt,Damour:2009ww}.

\section*{Acknowledgements}

We wish to thank Martin Cederwall, Axel Kleinschmidt, Jakob Palmkvist, and Zhenya Skvortsov for useful comments and discussions, and we are grateful to Axel Kleinschmidt for help with the proof in Appendix~\ref{app:proof_of_spectrum}.
J.O.~thanks G.B.~and G.B.~thanks N.B.~and J.O.~for hospitality at each other's institutions.
J.O.~would also like to thank Falk Hassler and David Osten for hosptiality at the University of Wroc{\l}aw, Poland, and Axel Kleinschmidt at the Max Planck Institute for Gravitational Physics, Potsdam, Germany, where parts of this work was completed.
The work of J.O.~was supported by the F.R.S.-FNRS (Belgium) grant number FC 43791 until September 2025, in November 2025 by the SONATA BIS grant 2021/42/E/ST2/00304 of the National Science Center (NCN), Poland, and in December 2025 by local funds of the Max Planck Institute for Gravitational Physics, Potsdam, Germany.
The work of N.B.~was partly supported by the F.R.S.-FNRS PDR grant number T.0047.24.

\appendix

\section{A brief review of Schur--Weyl duality}
\label{Appendix:Schur-Weyl}

\paragraph{Young diagrams and tableaux.}

Here we mostly use the content of \cite{Fulton_1996,Fulton:2004uyc} to which we 
refer for comprehensive expositions.
A Young diagram $\rY$ (or $\lambda$) is a collection of boxes, or cells, 
arranged in left-justified rows, with a non-increasing number of boxes in each row.
Listing the number of cells in each row gives a partition of the positive integer 
$|\rY|$ (or $|\lambda|$), the total number of cells.
Equivalently, a Young diagram is specified by a partition
\begin{equation}
    \rY
    = [h_1, \ldots, h_{\lambda_1}]
    = [\underbrace{D,\dots,D}_{\ell_D}, \underbrace{D-1, \ldots D-1}_{\ell_{D-1}}, \ldots, \underbrace{1,\ldots, 1}_{\ell_1}]
    = (\lambda_1,\ldots,\lambda_{h_1}) 
    = \lambda
\end{equation}
of $|\rY|=\sum_{i=1}^{\lambda_1}h_i=\sum_{i=1}^Di\,\ell_i=\sum_{i=1}^{h_1}\lambda_i$\,, i.e.~a non-increasing sequence of columns of heights $h_1\geqslant h_2\geqslant\cdots\geqslant h_{\lambda_1}>0$\,, or a non-increasing sequence of row lengths $\lambda_1\geqslant\lambda_2\geqslant\cdots\geqslant\lambda_{h_1}>0$\,, where $h_1\leqslant D$ for an integer $D$ that denotes the dimension of space-time in the present context.

As explained in \cite{Fulton_1996}, any way of assigning a positive integer 
-- be it from the set $\{1, \ldots, |\lambda|\}$ or from the set 
$\{1, \ldots, D\}$ -- in each cell of a Young diagram is called a \emph{filling} when the entries need not be different, and called a \emph{numbering} when the entries are distinct.
The result is generically called a \emph{tableau}.
Entries of tableaux can just as well be taken from any alphabet (totally ordered set), and in case there are not enough letters in the chosen alphabet, we will be choosing letters bearing indices.
For example, we could write
\ytableausetup{mathmode,boxsize=1.3em,centerboxes}
\begin{equation} 
    T_{[3,2,2]} = 
    \begin{ytableau}
    a_1 & b_1 & c_1 \\
    a_2 & b_2 & c_2 \\
    a_3
    \end{ytableau}
    \;.
\end{equation}
As we review below, in case the indices $\{a_1, a_2, \ldots, c_2\}$ take their values in the set $\{1,\ldots,D\}$\,, such a tableau is a convenient way to encode all the components of a $\mathrm{GL}(D,\mathds{R})$-irreducible tensor belonging to the Schur--Weyl module $\mathbb{S}_{\lambda}(\mathds{R}^D)$\,.
A \emph{semistandard} Young tableau $T_\lambda$ is a filling that is non-decreasing rightward across each row, and strictly increasing downward each column of the Young diagram.
A \emph{standard} tableau $T_\lambda$ is a semistandard tableau such that the entries are numbers from $1$ to $|\lambda|$, each occurring once.
The \emph{normal} Young tableau associated with $\lambda$ is the standard tableau in which the integers $\{1,2,\ldots,|\lambda|\}$ appear in order from left to right and from the top row to the bottom row of $\lambda$\,.

\paragraph{Schur--Weyl duality.}

For a given tensor field over space-time, $\mathrm{GL}(D,\mathds{R})$-irreducibility means that the over-antisymmetrisation constraints on the sets of antisymmetric indices of the tensor (i.e.~the columns in the Young diagram) are satisfied:~antisymmetrising an entire column of indices with one index from any column to its right vanishes identically if the columns are part of the same $\mathrm{GL}(D,\mathds{R})$-irreducible component.
This is part of the well-known Schur--Weyl construction, exposed e.g.~in Chapters 6 and 15 of \cite{Fulton:2004uyc}, where over-antisymmetrisation constraints appear from the quotient $\mathbb{S}^\bullet$ of the graded commutative ring $A^\bullet(V)$ over $V=\mathds{R}^D$,
\begin{align}
    A^{\bullet}(V) &= \mathrm{Sym}^\bullet (V \oplus \wedge^2 V \oplus \ldots \oplus \wedge^D V) \label{multiform1}\\
    &= \bigoplus_{\ell_1,\ell_2,\ldots,\ell_D} \mathrm{Sym}^{\ell_D} (\wedge^D V) \otimes \ldots \otimes \mathrm{Sym}^{\ell_2}(\wedge^2 V) \otimes \mathrm{Sym}^{\ell_1}(V)
    \eqqcolon \bigoplus_{\boldsymbol{\ell}}A^{\boldsymbol{\ell}}(V) \,,\label{multiform2} 
\end{align}
(the direct sum being over all the multi-indices ${\boldsymbol{\ell}}=(\ell_1,\ldots,\ell_D)\in\mathbb{Z}_{\geqslant 0}^D$\,), by the two-sided ideal $I^\bullet$ generated by all the elements of the form
\begin{multline}
\label{overantisymm}
    (v_1 \wedge \ldots \wedge v_p) \cdot (w_1 \wedge \ldots \wedge w_q) \\
    - \sum_{i=1}^p(v_1 \wedge \ldots v_{i-1} \wedge w_1 \wedge v_{i+1} \wedge \ldots \wedge v_p) \cdot (v_i \wedge w_2 \wedge \ldots \wedge w_q) \,,
\end{multline}
for all $p\geqslant q\geqslant1$ and all vectors $v_1,\ldots,v_p,w_1,\ldots,w_q$ in $V=\mathds{R}^D$.
The relations by which one quotients the ring $A^\bullet(V)$, i.e.~setting the above elements \eqref{overantisymm} to zero, are referred to as the Pl\"ucker relations in \cite{Fulton:2004uyc}. 
They are nothing but the over-antisymmetrisation constraints expressed in a coordinate-free way. 
The quotient $\mathbb{S}^\bullet=A^\bullet(V)/I^\bullet$ gives the direct sum of all 
contravariant rank-$h$ finite-dimensional $\mathrm{GL}(D,\mathds{R})$-irreducible 
representations, $\mathbb{S}^\bullet=\bigoplus_\lambda\mathbb{S}_\lambda V$\,, summing over 
all the non-negative integers, and for each such integer $h$, summing over the partitions 
$\lambda$ of $h$\,.

\underline{Remark:} As noticed in \cite{Fulton:2004uyc}, the symmetrised product appearing on the right-hand side of \eqref{multiform1} is not absolutely necessary, as the over-antisymmetrisation constraints \eqref{overantisymm} enforce the symmetrisation between all pairs of columns having the same height $p=q$\,.
One could therefore as well define the ring $\mathbb{S}^\bullet$ to be the full tensor algebra on $V\oplus\wedge^2V\oplus\ldots\oplus\wedge^DV$ modulo the ideal generated by all the elements of the form \eqref{overantisymm}, omitting all the symmetrisations in \eqref{multiform2}.

For a fixed non-negative integer $h$\,, one has the irreducible decomposition
\begin{equation}
    V^{\otimes h} = \underbrace{V \otimes V \otimes \ldots \otimes V}_{h~\text{factors}}
    \;\cong\,
    \bigoplus_\lambda\,(\mathbb{S}_\lambda V)^{\oplus m_\lambda} \,,
\end{equation}
where the sum is over all the partitions $\lambda$ of $h$\,, and $m_\lambda$ is the dimension of the irreducible representation $V_\lambda=\mathbb{C}\mathfrak{S}_h\cdot c_\lambda$ of the symmetric group $\mathfrak{S}_h$ associated with the partition $\lambda$ of $h$\,, where $c_\lambda\in\mathbb{C}\mathfrak{S}_h$ is the Young symmetriser corresponding to the partition $\lambda$ of $h$\,. 
If $R$ (resp.~$Q$) is the subgroup of $\mathfrak{S}_h$ preserving the rows (resp.~columns) of the normal (in fact, any numbering of the cells would do) Young tableau of $\lambda$, one has $c_\lambda=\big(\sum_{r\in R}e_r\big)\big(\sum_{q\in Q}(-1)^qe_q\big)$\,. 

The non-negative integer $m_\lambda=\mathrm{dim}\,V_\lambda$ corresponds to the number of standard Young tableaux associated with the partition (or Young diagram) $\lambda$ of $h$\,, and $\mathbb{S}_\lambda V=\mathrm{Im}(c_\lambda\vert_{V^{\otimes h}})$\,. 
One has $c_\lambda\cdot c_\lambda=n_\lambda\,c_\lambda$ with $n_\lambda=|\lambda|!/\mathrm{dim}\,V_\lambda=|\lambda|!/m_\lambda$\,, so that one defines the primitive idempotent, also called Young projector, $e_\lambda=c_\lambda/n_\lambda$\,.

On the space $V^{\otimes h}$ of contravariant tensors of rank $h$, the symmetric group $\mathfrak{S}_h$ acts from the right, say, as $(v_1\otimes v_2\otimes\ldots\otimes v_h)\cdot\sigma=v_{\sigma(1)}\otimes v_{\sigma(2)}\otimes \ldots\otimes v_{\sigma(h)}$\,, while $\mathrm{GL}(D,\mathds{R})$ acts from the left on $V^{\otimes h}$.
These two actions commute, which is at the basis of the Schur--Weyl duality between irreducible representations of $\mathfrak{S}_h$ and $\mathrm{GL}(D,\mathds{R})$ on $V^{\otimes h}$.
We also recall the dimension of the Schur module $\mathbb{S}_\lambda(V)$\,:
\begin{equation}
    \mathrm{dim}\,\mathbb{S}_\lambda(V)
    = \frac{m_\lambda}{h!}\,\prod_{(i,j)\in\lambda} (D-i+j)
    = \prod_{(i,j)\in\lambda} \frac{(D-i+j)}{h_{ij}} \,,
\end{equation}
where $h_{ij}$ is the hook length of the cell situated at the $i^\text{th}$ row and $j^\text{th}$ column of $\lambda$\,.

The quotient by the over antisymmetrisation relations can be enforced on every summand $A^{\boldsymbol{\ell}}(V)$ in \eqref{multiform2} by acting with the corresponding Young symmetriser $c_\lambda\,$, where $\lambda$ corresponds to 
the multi-index $\boldsymbol{\ell}=(\ell_D,\ldots,\ell_1)$ giving $\mathbb{S}^\bullet=\oplus_{\boldsymbol{\ell}}\,\mathbb{S}^{\boldsymbol{\ell}}(V)$\,.
We recall the fact (Proposition~15.15 of \cite{Fulton:2004uyc}) that the semistandard tableaux of shape $\lambda$ form a basis of the Schur--Weyl module 
$\mathbb{S}^{\boldsymbol{\ell}}(V)$\,.
Pictorially, the components $T^{a_1[D],\ldots,a_{\ell_D}[D],b_1[D-1],\ldots,b_{\ell_{D-1}}[D-1],\ldots\ldots,e_1[2],\ldots,e_{\ell_2}[2],f_1,\ldots,f_{\ell_1}}$ of a tensor 
$T_{\lambda}\in\mathbb{S}^{\boldsymbol{\ell}}(V)$ 
are represented by the tableau
\ytableausetup{boxsize=2.4em}
\begin{equation}
\begin{ytableau}
    a_1^1&\cdots&a_{\ell_D}^1&b_1^1&\cdots&b_{\ell_{D-1}}^1&\cdots&e_1^1&\cdots&e_{\ell_2}^1&f_1&\cdots&f_{\ell_1}\\
    a_1^2&\cdots&a_{\ell_D}^2&b_1^2&\cdots&b_{\ell_{D-1}}^2&\cdots&e_1^2&\cdots&e_{\ell_2}^2\\
    \tvdots&\tddots&\tvdots&\tvdots&\tddots&\tvdots&\none[\reflectbox{$\tddots$}]\\
    a_1^{D-1}&\cdots&a_{\ell_D}^{D-1}&b_1^{D-1}&\cdots&b_{\ell_{D-1}}^{D-1}\\
    a_1^D&\cdots&a_{\ell_D}^D
\end{ytableau}
\end{equation}

\section{Proof of the spectrum of extra fields}
\label{app:proof_of_spectrum}

We have verified \eqref{eq:R2L10} explicitly up to level 13 and \eqref{eq:L4} up to level 14\,.
To prove these equations at all levels, it is useful to consider the level decomposition with respect to $H_2=(\Lambda_2,H)$ under representations of $\mathrm{SL}(2)\times E_9$\,, the Levi subgroup that commutes with $H_2$\,.
One has
\begin{align}
    &H_2 = \frac{6}{11} H_{11} + H_2^{A_{10}} \,,&
    &D = - \frac{8}{33} H_{11} - \frac{19}{18} H_2^{A_{10}} \,,
\label{ChangeWeightBasis}
\end{align}
where $H_2^{A_{10}}$ is the Cartan generator of the $\mathrm{SL}(11)$ Levi subgroup commuting with $H_{11}$\,, $H_2$ is the central charge of the affine algebra $E_9$ commuting with $H_2$\,, and $D$ is the derivation acting on $E_9$\,.
The level decomposition with respect to $H_2$ gives 
\begin{align}
    R(\Lambda_2) &\cong {\bf 1}_1 \oplus {\bf 2} \otimes R(\Lambda_0)^{E_9}_{-\frac12} \oplus \Big( R(\Lambda_1)^{E_9}_{-2+\frac1{16}} \oplus R(\Lambda_7)^{E_9}_{-2+\frac12} \oplus R(\Lambda_1)^{E_9}_{-1+\frac1{16}} \oplus \dots \Bigr) \oplus \dots \,,\nonumber\\
    R(\Lambda_{10}) &\cong R(\Lambda_7)^{E_9}_{-2+\frac12} \oplus \dots \,,\\
    R(\Lambda_4) &\cong R(\Lambda_1)^{E_9}_{-1+\frac1{16}} \oplus \dots \,,\nonumber
\end{align}
where ${\bf 2}$ is the fundamental of $\mathrm{SL}(2)$\,.
The subscript $w$ in the $E_9$ representations $R(\lambda)^{E_9}_w$ is the shift of the eigenvalue of the derivation $D=L^\ord{k}_0+w$ such that $L^\ord{k}_0$ is the Sugawara Virasoro\footnote{One defines $L^\ord{k}_n=\frac{1}{2k+60}\sum_n\kappa_{AB}:T^A_nT^B_{m-n}:$ with a normal ordering prescription that preserves the $\mathrm{SL}(9)$ acting on ${[8]}_{4/9} $ instead of $E_8$\,.} generator acting on the level $k$ representations of $E_9$ that commute with $\mathrm{SL}(9)\subset E_9$\,.
We label the fundamental weights of $E_9$ such that the fundamental weight $\Lambda_i$ associated to a node $i$ of the $E_9$ Dynkin diagram corresponds to the node $i+3$ of the $E_{11}$ Dynkin diagram.
The basic representation of level $k=1$ decomposes into eigenspaces of $L^\ord{1}_0$ as
\begin{align}
    R(\Lambda_0)^{E_9} \supset \null& \big( [8] \big)_{\frac{4}{9}} \oplus \big( [2] \big)_{\frac{7}{9}} \oplus \big( [5] \big)_{\frac{10}{9}} \oplus \big( [7,1] \oplus [8] \big)_{\frac{13}{9}} \oplus \bigoplus_{n\,\geqslant\,1} \big( [9^{n-1},8,3] \oplus [9^n,2] \big)_{\frac{7}{9}+n} \label{E9L0}\\
    & \oplus \bigoplus_{n\,\geqslant\,1} \big( [9^{n-1},8,6] \oplus [9^n,5] \big)_{\frac{10}{9}+n} \oplus \bigoplus_{n\,\geqslant\,1} \big( [9^{n-1},8,8,1] \oplus [9^n,7,1] \oplus [9^n,8] \big)_{\frac{13}{9}+n} \,,\nonumber
\end{align}
where the subscripts $\ell_0$ are the eigenvalues of $L_0^\ord{1}$ and the partitions label the $\mathrm{SL}(9)$ irreducible representations.
Since all the eigenvalues are strictly positive, the generators $L_n^{\ord{1}}$ with $n\geqslant1$ act faithfully on all $\mathrm{SL}(9)$ irreducible representations with lowest $L_0^\ord{1}$ eigenvalue.
This allows to prove that the representations above are indeed present in $R(\Lambda_0)^{E_9}$ for all $n\geqslant 1$\,.
In fact, they appear with a multiplicity that grows polynomially with $n$\,.

The same argument applies to the level decomposition of the representations at $k=2$ into eigenvectors of $L^\ord{2}_0$.
Explicitly, we have 
\begin{align}
\begin{aligned}
    R(\Lambda_1)^{E_9} \supset \null& \big( [7] \big)_{\frac{119}{144}} \oplus \big( [8,2] \big)_{\frac{167}{144}} \oplus \big( [8,5] \big)_{\frac{215}{144}} \oplus \big( [8,7,1] \oplus [8,8] \oplus [9,7] \big)_{\frac{263}{144}} \\
    & \oplus \bigoplus_{n\geqslant2} \big( [9^{n-2},8,8,3] \oplus [9^{n-1},8,2] \big)_{\frac{23}{144}+n} \\
    & \oplus \bigoplus_{n\geqslant2} \big( [9^{n-2},8,8,6] \oplus [9^{n-1},8,5] \big)_{\frac{71}{144}+n}  \label{E9L1} \\
    & \oplus \bigoplus_{n\geqslant2} \big( [9^{n-2},8,8,8,1] \oplus [9^{n-1},8,7,1] \oplus [9^{n-1},8,8] \oplus [9^n,7] \big)_{\frac{119}{144}+n} \,, 
\end{aligned}
\end{align}
and
\begin{align}
\begin{aligned}
    R(\Lambda_7)^{E_9} \supset \null& \big([1]\big)_{\frac{13}{18}}
    \oplus \big([4]\big)_{\frac{19}{18}} \oplus \big( [6,1] \oplus [7] \big)_{\frac{25}{18}} \\
    & \oplus \bigoplus_{n\geqslant1} \big( [9^{n-1},7,3] \oplus [9^{n-1},8,2] \oplus [9^n,1] \big)_{\frac{13}{18}+n} \\
    & \oplus \bigoplus_{n\geqslant1} \big( [9^{n-1},7,6] \oplus [9^{n-1},8,5] \oplus [9^n,4] \big)_{\frac{19}{18}+n} \label{E9L7} \\
    & \oplus \bigoplus_{n\geqslant1} \big( [9^{n-1},8,7,1] \oplus [9^{n-1},8,8] \oplus [9^{n},6,1] \oplus [9^{n},7] \big)_{\frac{25}{18}+n} \,.
\end{aligned}
\end{align}
Under the bi-grading with respect to $H_{11}$ and $H_2^{A_{10}}$ we find 
\begin{equation}
    R(\Lambda_2) \supset [9]_3 = {\bf 1}_1 \oplus {\bf 2} \otimes {[8]}_{\frac{4}{9}} \oplus [7]_{\frac{119}{144}} \subset {\bf 1}_1 \oplus {\bf 2} \otimes R(\Lambda_0)^{E_9}_{-\frac12} \oplus R(\Lambda_1)^{E_9}_{-2+\frac1{16}} \,,
\end{equation}
and all the other $\mathrm{SL}(9)$ irreducible representations in $R(\Lambda_0)^{E_9}_{-1/2}$ correspond to lowest $H_2^{A_{10}}$ level components of irreducible representations of $\mathrm{SL}(11) \subset E_{11}$\,.
One can check that the condition of $\mathrm{SL}(11)$ lowest weight for a representation $[10,p(9\ell_0-5)]$ with the partition $p(9\ell_0-5)$ of $9\ell_0-5$ with no number exceeding nine
\begin{equation}
    H_2^{A_{10}} \big( {\bf 2} \otimes [p(9\ell_0-5)] \big) = - \frac{2}{11}(9 \ell_0 - 5) - \frac{9}{11} \,,
\end{equation}
matches the $H_2^{A_{10}}$ level of the representation of $L_0^\ord{1}$ value $\ell_0$ in $R(\Lambda_0)^{E_9}_{-1/2}$ using \eqref{ChangeWeightBasis}, namely $H_2^{A_{10}}=-\frac{18}{11}D-\frac{8}{11}H_2$ and 
\begin{equation}
    - \frac{18}{11} ( \ell_0 - \tfrac12 ) - \frac{8}{11} (1) = - \frac{2}{11}(9 \ell_0 - 5) - \frac{9}{11} \,.
\end{equation}

In the exact same way, the $\mathrm{SL}(9)$ irreducible representations in $R(\Lambda_7)^{E_9}_{-3/2}$ and $R(\Lambda_1)^{E_9}_{-15/16}$ correspond to lowest $H_2^{A_{10}}$ level components of irreducible representations of $\mathrm{SL}(11) \subset E_{11}$\,.
One checks the condition of lowest weight for a representation $[11,p(9\ell_0-\frac{11}{2})]$ with the partition $p$ of $9\ell_0-\frac{11}{2}$ with no number exceeding nine
\begin{equation}
    H_2^{A_{10}} \big( [p(9\ell_0-\tfrac{11}{2})] \big) = - \frac{2}{11} (9 \ell_0 - \tfrac{11}{2}) \,,
\end{equation}
matches the $ H_2^{A_{10}}$ level of the representation of  $L_0^\ord{2}$ value $\ell_0$ in $R(\Lambda_7)^{E_9}_{-\frac32}$ using \eqref{ChangeWeightBasis},
\begin{equation}
    - \frac{18}{11} (\ell_0 - \tfrac{3}{2}) - \frac{8}{11} ( 2) =  - \frac{2}{11} (9 \ell_0 - \tfrac{11}{2}) \,,
\end{equation}
and so too does the lowest weight for a representation $[11,p(9\ell_0-\frac{7}{16})]$ with the partition $p$ of $9\ell_0-\frac{7}{16}$
\begin{equation}
    H_2^{A_{10}} \big( [p(9\ell_0-\tfrac{7}{16})] \big) = - \frac{2}{11} ( 9 \ell_0 - \tfrac{7}{16} ) \,,
\end{equation}
match the $H_2^{A_{10}}$ level of the representation of $L_0^\ord{2}$ value $\ell_0$ in $R(\Lambda_1)^{E_9}_{-\frac{15}{16}}$
\begin{equation}
    - \frac{18}{11} ( \ell_0 - \tfrac{15}{16} ) - \frac{8}{11} (2) = - \frac{2}{11} ( 9 \ell_0 - \tfrac{7}{16} ) \,.
\end{equation}

This proves \eqref{eq:R2L10} and \eqref{eq:L4}, and moreover that 
\begin{align}
\begin{aligned}
\label{eq:R10underSL11}
    R(\Lambda_{10}) \supset \null& \big([11,1]\big)_4 \oplus \big([11,4]\big)_5 \oplus \big( [11,6,1] \oplus [11,7] \big)_6 \\
    & \oplus \bigoplus_{n\,\geqslant\,1} \big( [11,9^{n-1},7,3] \oplus [11,9^{n-1},8,2] \oplus [11,9^n,1] \big)_{4+3n} \\
    & \oplus \bigoplus_{n\,\geqslant\,1} \big( [11,9^{n-1},7,6] \oplus [11,9^{n-1},8,5] \oplus [11,9^n,4] \big)_{5+3n} \\
    & \oplus \bigoplus_{n\,\geqslant\,1} \big( [11,9^{n-1},8,7,1] \oplus [11,9^{n-1},8,8] \oplus [11,9^{n},6,1] \oplus [11,9^{n},7] \big)_{6+3n} \,,
\end{aligned}
\end{align}
and 
\begin{align}
\label{eq:R2Big}
    R(\Lambda_2) \supset \null& \big([9]\big)_3 \oplus \big( [10,2] \oplus [11,1] \big)_4 \oplus \big( [10,5] \oplus [11,4] \big)_5 \nonumber\\
    & \oplus \big( [10,7,1] \oplus [10,8] \oplus [11,6,1] \oplus 2{\times} [11,7] \big)_6 \nonumber\\
    & \oplus \big( [10,8,3] \oplus [10,9,2] \oplus [11,7,3] \oplus 2{\times} [11,8,2] \oplus [11,9,1] \big)_{7} \nonumber\\
    & \oplus  \big( [10,8,6] \oplus [10,9,5] \oplus [11,7,6] \oplus 2{\times} [11,8,5] \oplus [11,9,4] \big)_{8} \nonumber\\
    & \oplus \big( [10,8,8,1] \oplus [10,9,7,1] \oplus [10,9,8] \nonumber\\
    &\hspace{20mm} \oplus 2{\times} [11,8,7,1] \oplus 2{\times} [11,8,8] \oplus [11,9,6,1] \oplus 2{\times} [11,9,7] \big)_{9} \\
    &\hspace{-10mm} \oplus \bigoplus_{n\geqslant2} \big( [10,9^{n-1},8,3] \oplus [10,9^n,2] \oplus [11,9^{n-1},7,3] \oplus 2{\times} [11,9^{n-1},8,2] \nonumber\\[-3mm]
    &\hspace{70mm} \oplus [11,9^n,1] \oplus  [11,9^{n-2},8,8,3] \big)_{4+3n} \nonumber\\
    &\hspace{-10mm} \oplus \bigoplus_{n\geqslant2} \big( [10,9^{n-1},8,6] \oplus [10,9^n,5] \oplus [11,9^{n-1},7,6] \oplus 2{\times} [11,9^{n-1},8,5] \nonumber\\[-3mm]
    &\hspace{70mm} \oplus [11,9^n,4] \oplus [11,9^{n-2},8,8,6] \big)_{5+3n} \nonumber\\
    &\hspace{-10mm} \oplus \bigoplus_{n\geqslant2} \big( [10,9^{n-1},8,8,1] \oplus [10,9^n,7,1] \oplus [10,9^n,8] \oplus 2{\times} [11,9^{n-1},8,7,1] \nonumber\\[-3mm]
    &\hspace{5mm} \oplus 2{\times} [11,9^{n-1},8,8] \oplus [11,9^{n},6,1] \oplus 2{\times}  [11,9^{n},7] \oplus [11,9^{n-2},8,8,8,1] \big)_{6+3n} \,.\nonumber
\end{align}
Note that all required $\mathrm{SL}(9)$ representations already appear in the $R(\Lambda_2)$ module, but the fields $\zeta_a{}^\Lambda$ in $L(\Lambda_{10})$ and $\zeta_a{}^{\tilde{\Lambda}}$ in $L(\Lambda_{4})$ are necessary for the theory to be gauge invariant.
Moreover, $R(\Lambda_{10})$ and $R(\Lambda_4)$ alone most probably do not include all the relevant extra fields.
For example, $R(\Lambda_2+\Lambda_{10})\supset [11,9,1]_7$ and $R(2\Lambda_3)\supset [11,8,8]_9$\,, so one expects most irreducible components of $\zeta_a{}^\Lambda$ and $\zeta_a{}^{\tilde{\Lambda}}$ to contribute at arbitrarily high levels.

Note that the same argument can be used to derive that 
\begin{align}
\begin{aligned}
\label{eq:R1underSL11}
    R(\Lambda_{1}) \supset \null& \big(\overline{[1]}\big)_\frac32 \oplus \big([2]\big)_\frac52 \oplus \big([5]\big)_\frac72 \oplus \big( [7,1] \oplus [8] \big)_\frac92\oplus \bigoplus_{n\,\geqslant\,1} \big( [9^{n-1},8,3] \oplus [9^{n},2]  \big)_{\frac52+3n} \\
    & \oplus \bigoplus_{n\,\geqslant\,1} \big( [9^{n-1},8,6] \oplus [9^{n},5]   \big)_{\frac72+3n} \oplus \bigoplus_{n\,\geqslant\,1} \big(  [9^{n-1},8,8,1] \oplus [9^{n},7,1] \oplus [9^n,8] \big)_{\frac92+3n} \,,
\end{aligned}
\end{align}
using the branching of $R(\Lambda_1)$ under $\mathrm{SL}(2)\times E_9$\,:
\begin{equation}
    R(\Lambda_1) \,\cong\, \mathbf{2}_\frac12 \oplus R(\Lambda_0)^{E_9}_{-1} \oplus \dots \,.
\end{equation}
Using Appendix~E.1 of \cite{Bossard:2023jid}, one obtains that the restriction of $T^{\alpha_{(\ell)} a}{}_{P_{(\ell+3/2)}}$ to $\alpha_{(\ell)}$ in the $[9^n,\rx]$ with $\ell=3n+\ell(\rx)$ and $P_{\ord{\ell+3/2}}$ in the representations in \eqref{eq:R1underSL11} with the same $\ell$, is non-vanishing for the lowest $H_2^{A_{10}}$ levels and must therefore be non-vanishing for the corresponding irreducible representations of $\mathrm{GL}(11)$.

The infinite sequence of irreducible $\mathrm{GL}(11)$ representations with an increasing number of columns of height nine was here derived from the decomposition of $E_{11}$ under $\mathrm{SL}(2)\times E_9$ and the action of the positive Virasoro generator $L_{1}$ on affine Kac--Moody modules.
We presume that one should be able to understand this directly from the action of the generator $t^{\tilde{\alpha}}$ on $\mathfrak{e}_{11}$ within the tensor hierarchy algebra \eqref{eq:T0CR}.
This reasoning has been applied to $E_{10}$ in \cite{Cederwall:2025muh} and should generalise to $E_{11}$\,.

\section{Comments on non-propagating field strengths}
\label{sec:CountingIR}

Let us briefly describe the lowest-level  field strengths for the non-propagating $E_{11}$ fields with one column of ten indices.
The first example arises at level four and was described in \cite{Bossard:2021ebg}.
One has 
\begin{align}
\begin{aligned}
\label{F11,1,1}
    F_{a[10];b[3]} &= 10\,\partial_a A_{a[9],b[3]} + 3\,\chi_{b;a[10],b[2]} + \frac32 \big( \chi_{b;a[10]b,b} + \zeta_{b;a[10]b,b} \big) \,,\\
    F_{a[11],b,c} &= 11\,\partial_a B_{a[10],b,c} + 2\,\partial_{(b|} C_{a[11],|c)} + \frac52 \chi_{(b|;a[11],|c)} + \frac12  \zeta_{(b|;a[11],|c)} \,,
\end{aligned}
\end{align}
where $A_{9,3}$, $B_{10,1,1}$, and $C_{11,1}$ are the level four fields in $E_{11}/K(E_{11})$, and we have distinguished the field components of $\chi_M{}^{\tilde{\alpha}}$ and $\zeta_M{}^\Lambda$ explicitly.
Although $\chi_{b;a[10]b,b}$ and $\chi_{(b|;a[11],|c)}$ are distinct irreducible components of the same constrained field, we must take into account that the components of $\chi_M{}^{\tilde{\alpha}}$ associated with propagating $E_{11}$ fields are argued to be total derivatives in the linearised approximation -- see \eqref{LinearChi}.
Writing instead
\begin{align}
\begin{aligned}
\label{F11,1,1Lin}
    F_{a[10];b[3]} &= 10\,\partial_a A_{a[9],b[3]} + 3\,\partial_b X_{a[10],b[2]}  + \frac32 \partial_b \big( X_{a[10]b,b} + Y_{a[10]b,b} \big) \,,\\
    F_{a[11],b,c} &= 11\,\partial_a B_{a[10],b,c} + 2\,\partial_{(b|} C_{a[11],|c)} + \frac52 \partial_{(b|} X_{a[11],|c)} + \frac12  \partial_{(b|} Y_{a[11],|c)} \,,
\end{aligned}
\end{align}
one exhibits that two distinct fields $X_{11,1}$ and $Y_{11,1}$ are needed to define the two gauge invariant field strengths.
One observes that, although $R(\Lambda_2)$ provides all the necessary irreducible $\mathrm{GL}(11)$ representations at level four to include the field $\chi_{a[10];b[3]}$ in the field strength $F_{a[10];b[3]}$, one needs an extra field in the $[1\,;11,1]$ in $R(\Lambda_{10})$ to obtain the extra St\"{u}ckelberg field in the $[11,1,1]$   necessary to write a covariant field strength for the non-propagating field $B_{10,1,1}$\,.
One finds therefore that $R(\Lambda_2)\oplus R(\Lambda_{10})$ provides precisely the necessary number of $\mathrm{GL}(11)$ irreducible representations.
Moreover, note that the $[11,1]$ fields appearing in $F_{10\seco3}$ and $F_{11,1,1}$ are linear combinations mixing the corresponding fields $X_{11,1}$ and $Y_{11,1}$ in $R(\Lambda_2)$ and $R(\Lambda_{10})$. 

The same analysis applies at level five.
The fields in $E_{11}/K(E_{11})$ are $A_{9,6}$, $B_{10,4,1}$, $C_{11,3,1}$, and $C_{11,4}$.
The field strength $F_{10\seco6}$ requires St\"{u}ckelberg fields in the $[1\,;10,5]$ and the $[1\,;11,4]$, and the field strength $F_{11,4,1}$ requires St\"{u}ckelberg fields in the $[1\,;11,3,1]$ and the $[1\,;11,4]$.
The two St\"{u}ckelberg fields in the $[1\,;11,4]$ are linear combinations of the corresponding components of $\chi_M{}^{\tilde{\alpha}}$ in $R(\Lambda_2)$ and $\zeta_M{}^\Lambda$ in $R(\Lambda_{10})$.
In this case as well, $R(\Lambda_2)\oplus R(\Lambda_{10})$ at level five provides precisely the necessary irreducible representations to obtain all the St\"{u}ckelberg fields.

At level six one finds for the first time the need to introduce the additional field $\zeta_M{}^{\tilde{\Lambda}}$ in  $R(\Lambda_4)$.
We have explained in this paper that this can be verified by looking at the propagating fields only using equation \eqref{LEN}.
Here we will observe that one can find this result by simply counting irreducible representations when one includes the non-propagating fields.
The fields in $E_{11}/K(E_{11})$ at level six are $h_{9,8,1}$, $B_{10,6,2}$, $B_{10,7,1}$, $B_{10,8}$, $C_{11,4,3}$, $C_{11,5,1,1}$, $C_{11,7}$, and $C^\ord{i}_{11,6,1}$  for $i=1,2$\,.
The field strength $F_{10\seco8,1}$ requires one St\"{u}ckelberg field in the $[1\,;10,7,1]$, one in the $[1\,;10,8]$, one in the $[1\,;11,6,1]$, and two St\"{u}ckelberg fields in the $[1\,;11,7]$.
The reducible field strength $F_{11\seco6\seco2}=F_{11,6,2}+F_{11,7,1}+F_{11,8}$ requires one St\"{u}ckelberg field in the $[11,5,2]$, two in the $[11,6,1]$, and two in the $[11,7]$.
In order to obtain all these fields, one needs to include all the field components of $X^{\tilde{\alpha}}$, $X^\Lambda$ and $U^{\tilde{\Lambda}}$ at level six in $R(\Lambda_2)$, $R(\Lambda_{10})$, and $R(\Lambda_4)$\,:
\begin{align}
\begin{aligned}
    R(\Lambda_2)|_{\ell=6} &= [10,8] \oplus [10,7,1] \oplus [11,5,2] \oplus 2\times[11,6,1] \oplus 2\times[11,7] \,,\\
    R(\Lambda_{10})|_{\ell=6} &= [11,6,1] \oplus [11,7] \,,\\
    R(\Lambda_{4})|_{\ell=6} &= [11,7] \,. 
\end{aligned}
\end{align}

Note that this sharp matching for the irreducible representations in $R(\Lambda_2)\oplus R(\Lambda_{10})\oplus R(\Lambda_4)$ does not extend to higher levels.
At level seven the fields in  $E_{11}/K(E_{11})$ include 
\begin{equation}
    A_{9,9,8,1}\,,\,
    B_{10,9,1,1}\,,\,
    C_{11,9,1}\,,\,
    B_{10,10,1}\,,\,
    B_{10,9,2}^\ord{1}\,,\,
    B_{10,9,2}^\ord{2}\,,\,
    B_{10,8,2,1}\,,\,
    B_{10,8,3}\,,\,
    B_{10,7,4}\,,\,
    C_{11,10}\,,\,
    \label{ListLevel7}
\end{equation}
and other `$C$' fields with at least one column with eleven indices and none of height nine or ten.
It is natural to interpret the non-propagating fields $B_{10,9,1,1}$ and $C_{11,9,1}$ as higher gradient duals of the non-propagating fields $B_{10,1,1}$ and $C_{11,1}$ at level four.
The structure of the tensor hierarchy algebra indeed implies equation \eqref{All9!}, such that $B_{10,9,1,1}$ and $C_{11,9,1}$ must have field strengths 
\begin{align}
\begin{aligned}
    F_{a[10];b[10],c,d} &= 10\,\partial_a B_{b[10],a[9],c,d} + \chi_{a[10];b[10],c,d} \,,\\
    F_{a[10];b[11],c} &= 10\,\partial_a C_{b[11],a[9],c} + \chi_{a[10];b[11],c} \,.
\end{aligned}
\end{align}
All the other fields in \eqref{ListLevel7} have a column of ten indices and a corresponding field strength with a column of eleven indices.\footnote{It is not obvious that $C_{11,10}$ should have a field strength $F_{11,11}$, and it could only appear in the field strength $F_{11,10,1}$ of $B_{10,10,1}$. In counting irreducible representations, we have assumed instead that there is a field strength $F_{11,11}$. In the second eventuality one would have one additional field component of $(X,Y,U)$ in the $[11,10]$ that would not be needed.} 
Writing the field strengths for all these fields and introducing the corresponding St\"{u}ckelberg fields, one obtain the set of components of $X^{\tilde{\alpha}}$, $X^\Lambda$ and $U^{\tilde{\Lambda}}$ at level seven in $R(\Lambda_2)$, $R(\Lambda_{10})$, and $R(\Lambda_4)$ that one needs.
We find that there are two components that are not needed, one $[11,8,1,1]$ among three and one $[11,9,1]$ among eight.
We find therefore an almost perfect match, but there are two irreducible components of $X^{\tilde{\alpha}}$, $X^\Lambda$ and $U^{\tilde{\Lambda}}$ at level seven that must drop out in the field strengths.
In general, $L(\Lambda_2)\oplus L(\Lambda_{10})\oplus L(\Lambda_4)$ will provide more irreducible representations than is required to write all the gauge invariant field strengths and we expect the excess number of fields to grow exponentially at high level.
It therefore seems that there must exist an additional shift gauge invariance that can be used to set them to zero.
Such additional gauge invariance with gauge parameters with two constrained indices is expected to be necessary to close the gauge algebra on constrained fields.
Defining them in $E_{11}$ exceptional field theory is beyond the scope of this paper.

The counting performed in this appendix is similar to, but not identical to, the counting that was carried out in \cite{Boulanger:2024lwk}, where representations of $\mf{e}_{11}$ were compared against irreducible $\mathrm{GL}(11)$ components of first-order variables in the unfolded formulation of mixed-symmetry gauge fields in $E_{11}/K(E_{11})$.
These variables do not have the same symmetry types, i.e.~Young diagrams, as the field strengths $F^I$ in $E_{11}$ exceptional field theory, and a possible relationship between them is unclear.

\addcontentsline{toc}{section}{References}

\bibliographystyle{JHEP}
\bibliography{biblio}

\end{document}